\documentclass[1p]{elsarticle}
\usepackage{hyperref}
\usepackage{mathtools}
\usepackage{booktabs}
\usepackage{rotating}
\usepackage{listings}
\usepackage{algorithm}
\usepackage[noend]{algpseudocode}
\usepackage{amsmath,amsfonts,amssymb,amscd,amsthm,xspace}
\usepackage[centerlast,small,sc]{caption}
\usepackage{graphicx}
\usepackage{epstopdf}
\usepackage{mathtools}
\usepackage{booktabs}
\usepackage{rotating}
\usepackage{algpseudocode}
\usepackage{subcaption}

\graphicspath{{./}}
\begin{document}

\begin{frontmatter}

\title{A Distributed Real-Time Recommender System for Big Data Streams}
%
\author[NU]{Heidy Hazem}
\ead{h.hazem@nu.edu.eg}

\author[UT,CU]{Ahmed Awad\corref{mycorrespondingauthor}}
\ead{ahmed.awad@ut.ee}

\author[NU]{Ahmed Hassan}
\ead{ahassan@nu.edu.eg}

%
\cortext[mycorrespondingauthor]{Corresponding author}

\address[NU]{Nile University, Giza, Egypt}
\address[UT]{University of Tartu, Tartu, Estonia}
\address[CU]{Cairo University, Giza, Egypt}
\begin{abstract}

In today's data-driven world, recommender systems (RS) play a crucial role to support the decision making process. They serve both end users to make better choices as well as organizations to better understand their customers. As users become continuously connected to the internet, they become less patient and less tolerant to obsolete recommendations made by RS underpinning services they consume such as movie recommendations on Netflix or books to read on Amazon. This, in turn, requires a continuous training of the RS systems to cope with both the \emph{online} fashion of data and the changing nature of user tastes and interests, known as \emph{concept drift}.

In general, a streaming (online) RS has to address three requirements: continuous training and recommendation, handling concept drifts and ability to scale. In literature, approaches for streaming recommender systems have been proposed. Mostly, they address the first two requirements and do not consider scalability. That is because, they run the training process on a single machine. Such a machine, no matter how powerful it is, will eventually fail to cope with the volume of the data, a lesson learned from big data processing. 


To tackle the third challenge, we propose a \emph{Splitting} and \emph{Replication} mechanism for building distributed streaming recommender systems. Our mechanism is inspired by the the successful \emph{shared-nothing} architecture that underpins contemporary big data processing systems. We have applied our mechanism on two well-known approaches for online recommender systems, namely, matrix factorization using Incremental Stochastic Gradient Descent (ISGD) and item-based collaborative filtering using modified incremental cosine similarity. We have implemented our mechanism on top of Apache Flink, a distributed stream processing system. We conducted experiments comparing the performance of the baseline (single machine) approach with our distributed approach. By evaluating on different data sets, not only improvement in processing latency and throughput but also in accuracy has been observed. Our experiments show online recall improvement by $40\%$ with more than $50\%$ less memory consumption.

\end{abstract}


\end{frontmatter}


\section{Introduction}
\label{intro}

Today, we are living in a world driven by decisions resulting from the analysis of big data. Data driven decisions are not only limited to optimizing strategic and business related findings, but also day-to-day choices can be improved by data collected from daily transactions and decisions, such as what restaurant to have lunch in, which vacation destination to travel to, and which product to buy. Recommender systems ~\cite{recommenderSystemsBook16} have arisen to predict and recommend items that could meet user's preferences. Recommender systems are the backbone supporting the business of different business verticals like, e-commerce, e.g. Amazon, entertainment, e.g. Netflix, tourism and transportation, e.g. suggestions in Google maps, etc.

In principle, recommender systems \emph{learn} some \emph{similarity} from users' feedback, whether it is an explicit rating like stars ratings or implicit ratings like listening to a song many times. This feedback updates a rating matrix $R^{n \times m}$ where the rows are representing $n$ users and the columns are the $m$ items and the rating of user $u$ towards item $i$ is the intersection between the user's row and item's column is represented as $r_{ui}$. Traditionally, this learning, like other machine learning jobs, takes place in an \emph{offline} mode. That is, finite sets of user-item feedback are collected and a model iterates over this data trying to fill the empty cells of $R$ reducing some error function~\cite{recommenderSystemsBook16}. Among the different types of recommender systems: content-based, collaborative filtering, and knowledge-based, collaborative filtering (CF) is generally the most applicable~\cite{recommenderSystemsBook16}. The one million dollars Netflix Prize is considered a paradigm shift towards using CF.

As users become more continuously connected to the internet, they become less patient and less tolerant to obsolete recommendations made by RS underpinning services they consume. In return, an RS has to train more frequently and on web-scale volumes of data arriving at high speed. That, in practice, shifts the task of training an RS from offline on a finite set of data where the learning model can iterate several time to an online setting where the model has to process the user-item feedback once and has to continuously alternate between recommending and learning. Moreover, the learning has to be done in a distributed setup to cope with the large volumes of the data. 

To develop a web-scale online recommender system, the following requirements have to be addressed:
\begin{enumerate}
     \item The learning model must be able to produce a result and be updated after each feedback has been received without passing over all the past data (latency)
    \item Concept drifts ought to be taken care of by adjusting the model with each feedback to adapt to the change in users' tastes and preferences.
    \item Learning from big data must be processed in a distributed streaming environment (scalability)
\end{enumerate}

In literature, techniques have been proposed~\cite{albertBifetMLDataStreamsBook18} to handle mainly the first requirement and partially the second one. Very little work has been proposed to address the third requirement, we discuss that in more detail in Section~\ref{sec:related:work}. The work in this paper has been mainly designed to handle the third requirement of distributing the streaming  recommendation algorithms by adopting the shared-nothing architecture. Shared-nothing architecture is the distribution architecture behind the success of the different big data processing systems from the early days of Hadoop to the latest development of stream processing systems like Apache Flink~\cite{Flink2015} and Beam~\cite{BEAM}. We propose a novel technique of \emph{Splitting} and \emph{Replication} to allow distribution with replication of user-item feedback across different machines in the computing cluster. We apply our technique to two different streaming recommendation algorithms. Preliminary results of this work have been reported in our previous work~\footnote{Reference has been removed due to double-blind restriction}. In this paper, we extensively extend our work. In particular, the main contributions of this paper can be summarized as follows: 

\begin{enumerate}
   \item Proposing the splitting and replication technique by adopting the shared-nothing architecture for distributing the streaming recommender systems
   \item Two online streaming algorithms (ISGD~\cite{vinagre2014fast}, Cosine similarity~\cite{huang2015tencentrec}) have been adapted in distributed versions utilizing splitting and replication mechanism
\item  To account for state growth and to handle concept drifts, two different techniques from cache management, least recently used LRU and least frequently used LFU, have been applied over the distributed versions of the streaming algorithms. 
 \item A comparative evaluation with the baseline of the two algorithms on two data sets has been conducted showing the superiority of our approach not just in the processing speed and memory consumption but also in the improved online recall (accuracy). Our experiments show online recall improvement by $40\%$ with more than $50\%$ less memory consumption.

\end{enumerate}

 The rest of this paper is organized as follows. Background about the problem formulation and CF is given in Section~\ref{sec:background}. Related work on distributed recommender systems is discussed in Section~\ref{sec:related:work}. Our contribution of the splitting and replication technique along with adaptation of ISGD~\cite{vinagre2014fast} and item Cosine similarity ~\cite{huang2015tencentrec} are detailed in Section~\ref{sec:splitting:replication}. Section~\ref{sec:evalution} describes the evaluation setup and details the experimental results of our technique. Finally, conclusions and outlook to future work is communicated in Section~\ref{sec:conclusion}.

\section{Background and Preliminaries}
\label{sec:background}

In general, recommender systems can be categorized according to the type of data used to train them. There are three broad classes of such systems: content-based~\cite{contentBasedRecommenderSystem19}, collaborative filtering~\cite{collaborativeFilteringRecommenderSystemDS13}, and knowledge-based~\cite{knowledgeBasedRecommenderSystem18} systems. Nevertheless, some systems can use hybrid approaches combining two or more of the basic models~\cite{recommenderSystemsBook16}. In this paper, we are interested in collaborative filtering approaches as they utilize users and community ratings without the overhead of representing items nor the knowledge background of the users.

Collaborative filtering Assumes that the system has $n$ users and $m$ items and the rating matrix representing the rating from user $u$ to an item $i$ is $R^{n \times m}$.  Rows in $R$ represent the users and the columns are the items while the intersection between them is the rating $r_{ui}$. $R$ is very sparse matrix as users usually rate a small subset of the items and likewise, an item is rated by a very small subset of users~\cite{recommenderSystemsBook16}. CF goal is to predict the ratings of all the missing items to then recommend the items with high predicted ratings. CF can be further classified into \emph{neighborhood-based} and model-based methods. Neighborhood-based methods depend on finding similarity between users or items. Model-based methods apply machine learning techniques to learn a model from the rating matrix. In general, neighborhood methods outperform model-based methods when learning from sparse rating matrices~\cite{recommenderSystemsBook16}. Moreover, neighborhood methods are simpler to implement and this is more convenient in an online (streaming) setup. Thus, in the rest of this paper we are focusing on neighborhood methods.

The problem of learning from the rating matrix for neighborhood CF methods can be reduced to a common optimization problem of low-rank matrix factorization~\cite{ALS08,matrixFactorization09}. Starting with the low-rank matrix factorization, a loss function $L$ is characterized to minimize the cost of errors which implies to minimize the square of the differences between the real ratings $R_{nm}$and the expected ratings $\hat{R}_{nm}$


\begin{equation}
L = min_{U.,I.}\sum_{(u,i)\in D} (R_{ui} - U_{u}.I_{i}^T )^2 + \lambda(\|U_u\|^2 + \|I_i\|^2) 
\label{lossFunction}
\end{equation}

The second term in the loss function,  \autoref{lossFunction}, is the $L2$ regularization term that is added to eliminate overfitting, whereas $\lambda$ is the regularization parameter. Since the optimization problem to solve \autoref{lossFunction} is non convex, traditional convex optimization methods do not work. Many approaches have been proposed to solve this problem like alternating least squares(ALS) which minimizes the loss function by alternating between fixing the users' vectors $U_{u}$'s or items' vectors $I_{i}$'s. While one of them is fixed the other one can be solved using the least-squares problem. Stochastic gradient descent SGD \cite{le1998lecun} is another approach to tackle the minimization of the loss function $L$. SGD iterates over each rating $r_{ui}$ seeking to estimate the predicted error $\hat{r}_{ui}$ by firstly initializing the matrix by random numbers. Then, SGD calculates the error resulting from the prediction:
    
    \begin{equation}
    \hat{e_{ui}}\ = r_{ui} - U_{n}.I_{m}^T 
    \label{dotProduct}
    \end{equation}
    
Iteratively, SGD iterates to minimize the prediction error companion to its sequences of updates to the users vectors (\autoref{eq:1}) and items vectors (\autoref{eq:2}) with learning rate $\eta$ until computation converges. 
    
    \begin{equation}
    U_{u} := U_{u}+\eta(err_{ui}.I_{i} - \lambda U_{u})
    \label{eq:1}
    \end{equation}
    \begin{equation}
      I_{i} := I_{i}+\eta(err_{ui}.U_{u}- \lambda I_{i} )
      \label{eq:2}
   \end{equation}

In a streaming setup, iterating over all data points until convergence is not possible to due both the unbounded nature of the stream as well as the huge latency implied by iteration. Vinagre et al.~\cite{vinagre2014fast} proposed ISGD as an incremental matrix factorization algorithm that is based on SGD. For every received user-item rating, ISGD updates the model. So this algorithm overcomes the first two challenges mentioned in Section~\ref{intro}. 

Apart from matrix factorization, the core of neighborhood-based methods is to define a similarity measure among users or items. One of the most widely used similarities metrics is Cosine similarity in the range $[0,1]$ where values are proportional to similarity (\autoref{eq:cosine:similarity}). Vectors $\vec{i}$ and $\vec{j}$ can represent users or items respectively.

    \begin{equation}
    Sim (\vec{i},\vec{j}) = \dfrac{\vec{i}.\vec{j}}{\vert\vert{\vec{i}}\vert\vert \times \vert\vert{\vec{j}}\vert\vert} 
    \label{eq:cosine:similarity}
    \end{equation}
     
where $\vert\vert{\vec{i}}\vert\vert$ is the Euclidean norm of vector $i$

 The incremental Cosine similarity algorithm,  which is proposed in~\cite{huang2015tencentrec}, has been adopted in this work. The authors adapted the Cosine similarity metric for calculating the items' similarities and the estimated rate for introducing incremental recommendation for implicit feedback. The authors divided the calculation of similarity score of item pairs that can be naturally incremented as shown in Equation~\ref{sim_eq}. 

\begin{equation}
      Sim(i_{p},i_{q})' \leftarrow \frac{\sum_{u\in{U}} min(r_{up},r_{uq}) + \Delta{min(r_{up},r_{uq})}}  {\sqrt{\sum{r_{up}+\Delta{r_{up}}}} \sqrt{\sum{r_{uq}+\Delta r_{uq}}}}
  \label{sim_eq}
\end{equation}

where $sim(i_{p},i_{q})'$ is the similarity after considering a new received rating interaction, e.g. $r_{up}$. To estimate the rating for the unseen items for each user which is given by the weighted average rating of the top $k$ similar items rated before by the user (item's neighbors), the estimated rating can be calculated as per Equation~\ref{r} where the $N^k(i_p)$ is the item's neighbors.

\begin{equation}
\hat{r}_{u,p} = \frac{\sum_{i_q\in{N^k(i_p)}} sim(i_p,i_q)r_{u,i_q} }{\sum_{i_q\in{N^k(i_p)}} sim(i_p,i_q)}
\label{r}
\end{equation}

Cosine similarity and ISGD have been adopted in our work to make them scalable, more details are given in Section~\ref{sec:splitting:replication}.

\section{Related Work}
\label{sec:related:work}

Many distributed (scalable) approaches have been proposed for batch (offline) recommendation algorithms. Low-rank matrix factorization is the preponderant among them. A core challenge in such distributed approaches is how to split the enormous matrix $R$. Gemulla et al ~\cite{gemulla2011large} proposed a way to partition $R$ by dividing it into $d$ blocks called stratum where their column and rows do not overlap and permute over the entire matrix by \begin{math}\pi\end{math} permutations. Sebastian Schelter et. al~\cite{schelter2014factorbird} have proposed another approach to distribute the computation of SGD; the authors adapted the parameter server architecture~\cite{li2014scaling} to run SGD in that setting. In that approach, factorized matrices $U$ and $I$ are divided over computing nodes in a cluster, naming them the parameter nodes and the same for the input data $R$ is distributed over machines representing the learner machines in the architecture. For each record in $R$, each learning machine is responsible for fetching the needed vectors from the parameter machine to update them and write them back on the parameters machines. The authors adapted the architectures to minimize the communication between nodes by partitioning $R$ first by the key of user.  So, when updating $U$ or $I$, one of them turns to be local. The parameter server architecture suffers from the overhead of managing locking made on vectors when being updated by the learners. To avoid this overhead, a parallelization scheme called HOGWILD! is proposed in~\cite{recht2011hogwild}. HOGWILD!'s authors proved that SGD can be run in parallel without locking if most updates affect non-overlapping parts of the model, i.e. different cells of $R$. 

In the literature, there is no much work on online (streaming) RS. Agrarwal et al.~\cite{offlineInitializationOnlineUpdate2010} partially address the problem of online training recommender systems by first initializing the model in an offline setting and then use it for prediction and updating the model in an online setting. For fully online training and prediction, Papagelis et al.~\cite{papagelis2005incremental} have adapted the Pearson correlation similarity metric to work incrementally by updating the value between users when a new rating comes in the user-user similarity matrix while the last values of similarities are kept in the memory. Another~approach \cite{miranda2008incremental} computes user similarity based on the tally of items rated by each user kept in matrix $F$, which can be incremented for streaming scenarios. The square matrix $F$ describes user-user similarity. StreamRec is proposed in~\cite{streamRec2011} as an online recommender system using Cosine similarity between user vectors. Zaouali et al.~\cite{zaouali2018distributed} utilize Jaccard similarity between users which can be incrementally computed by nature. All of the mentioned approaches can be modified to calculate item-item similarity. Online low-rank matrix factorization approaches have also been proposed. Most of them utilize SGD as it can be adapted to be incrementally calculated. Vinagre et al.~\cite{vinagre2014fast} propose ISGD, we have discussed ISGD in Section~\ref{sec:background}. Xiangna et al.~\cite{he2016fast} propose an online matrix factorization algorithm eALS capitalizing on the Alternating Least square ALS matrix factorization algorithm~\cite{ALS08}. In addition, the authors propose optimization techniques to tackle the expensive computations by reducing the missing ratings depending on item's popularity. These approaches maintain the model on a single machine, i.e., no-distributed processing.


Offline (batch) distribution techniques of the rating matrix $R$  are not applicable in an online (streaming) context, as they assume that the size of $R$ is known beforehand. In the streaming context, the overall matrix $R$ is changing as new ratings may introduce new rows, columns, or both in $R$. Muqeet et al.~\cite{ali2011parallel} propose an approach to tackle the challenge of the changing size of $R$ by utilizing the master-slave architecture. The master node doles out blocks from $R$ dynamically to the free workers and locks all the other blocks dependent on rows or columns related to the working block. After the worker finishes the processing, the worker conveys back to the master node to free the locked blocks. Another responsibility of the master is to route the user-item ratings to the workers that need them. Consequently, the training samples are distributed over workers. Additionally, the matrices $U$ and $I$ are distributed over data stores called matrix stores. Obviously, embracing locking can lead to deadlocks in addition to the processing and memory overhead of maintaining locks.

TencentRec~\cite{huang2015tencentrec} is an online distributed approach for item-based collaborative filtering that is developed by Tencent corporation. TencentRec is implemented on top of Apache Storm~\cite{Storm2015}. TencentRec adapts Cosine similarity metric to work incrementally. However, the user-item rating history is maintained in a shared centralized state (memory). We follow a similar approach for computing the similarity in a distributed way. Yet, we also distribute and replicate the state to faithfully follow a shared-nothing architecture. We give more details in the next section.

\section{Splitting and Replication of User-Item Rating}
\label{sec:splitting:replication}

\begin{figure}[t]
    \centering
	\includegraphics[width=\linewidth]{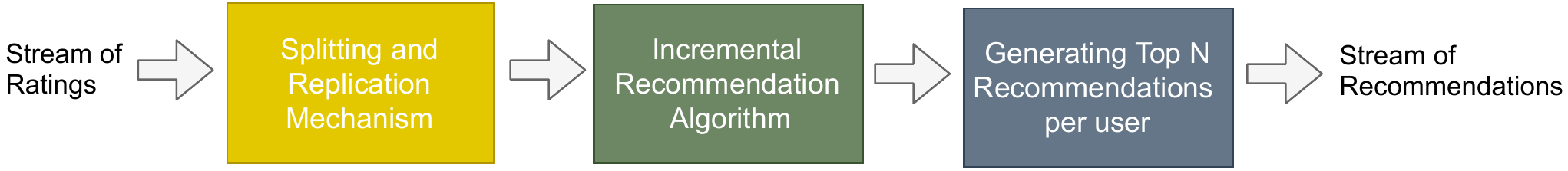}
	\caption{A pipeline of online recommender system with splitting and replication as the first operator}
	\label{Pipeline}
\end{figure}

In this section, we describe our approach to address the scalability requirements for an online recommender system. We benefit from the shared-nothing architecture on which modern distributed data processing system, i.e. big data systems, are built. The scalability feature of such systems is to partition data into disjoint \emph{splits} where each split is processed by a separate node (worker). However, for recommender systems, it is necessary to learn from the ratings of \emph{similar} users or to learn about \emph{similar} items. In this case, a pure partitioning limits the learning opportunities for a recommender system and thus reduces the chance of offering interesting items to users. To combat this limitation, we introduce a factor of replication of item and user vectors. This replication will allow some of the vectors to be present on more than one worker. In this case, each worker can still learn from local information without need to synchronize with other workers. We call our approach \emph{Splitting and Replication}. Our approach is further motivated by the observation that incremental stochastic gradient descent can run in parallel without locking if the predominance of the updates undergoes minor modification to the model. This has been analyzed and proved in~\cite{recht2011hogwild}. Figure~\ref{Pipeline}, shows how our approach fits in the processing pipeline of a stream of ratings.

\begin{figure}[h!]
    \centering
	\includegraphics[width=\linewidth]{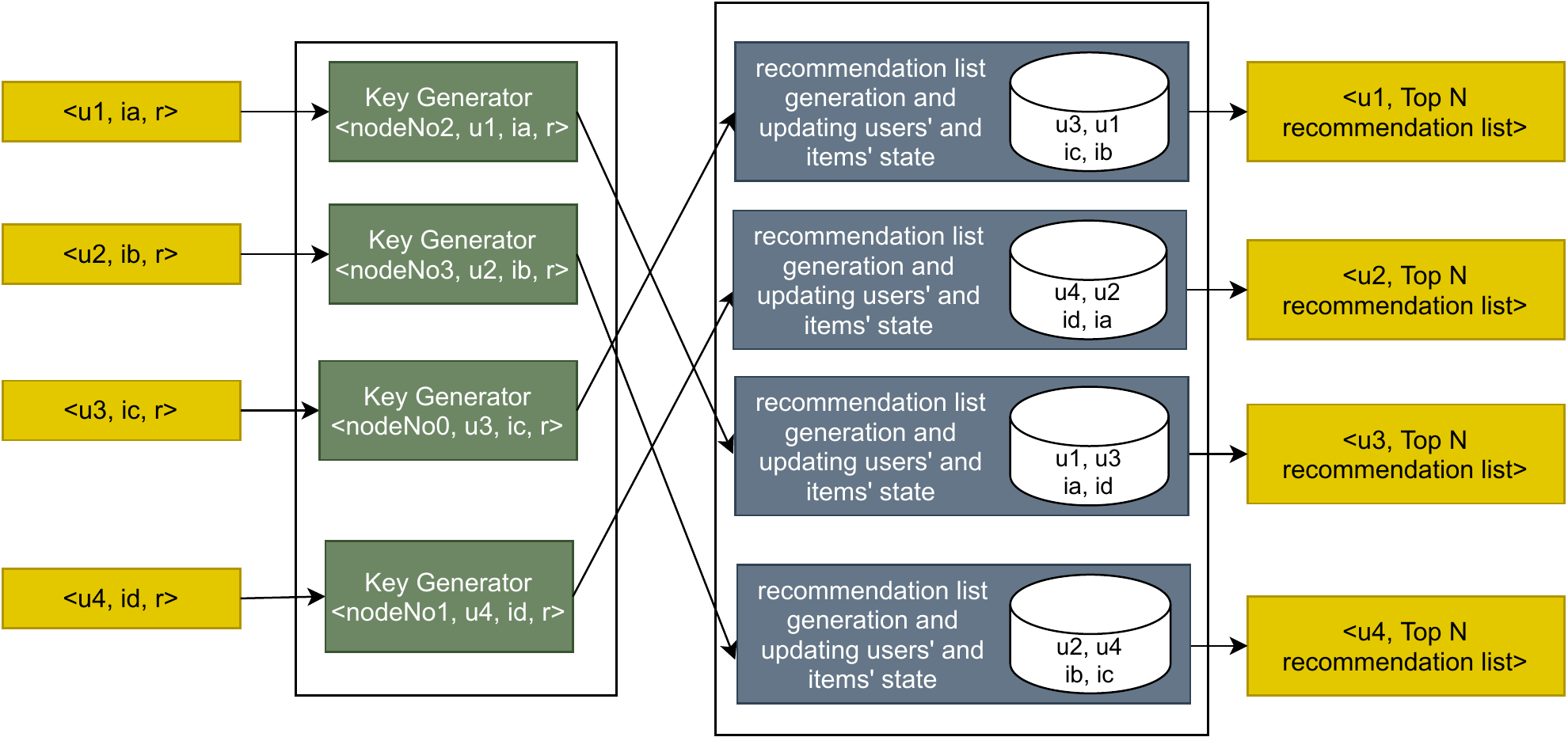}
	\caption{Overview of splitting and replication mechanism working with streaming recommender system}
	\label{fig:splliting:replication:inner:working}
\end{figure}


Figure~\ref{fig:splliting:replication:inner:working} shows the inner working of our approach. The input for the Splitting and Replication mechanism is a stream of ratings' interactions in tuples format $\langle user, item,\\ rating\rangle$. To control the number of splits and replications, we use the hyperparameter $n_{c}$ which is the number of the workers in the cluster and replication factor $n_{i}$. Partitioning based on either the user or the item only is not possible, because streaming recommendation algorithms should be updated incrementally with the new coming records, and a user's representation i.e. vector of every user and item's representation (vector) should be both accessible to the worker to apply the learning algorithm. If the stream is partitioned by the user only, all the items that are being rated can reside on a single worker in the cluster, and likewise, if the partitioning was based on items only. The splitting and replication mechanism comes over the aforementioned problem by partitioning by the pair of the user and item $\langle user, item \rangle$ as the partitioning key with allowing replication of user and items without synchronization. The main objectives of the mechanism are guaranteeing that each user-item pair hits only one node and controlling the size of partitions to guarantee scalability. Replication is a by-product as several items can still appear on the same worker as they will be associated with other users. Likewise, several users will appear on the same worker. Once the user rating has been directed to a worker, it is used by an incremental recommender algorithm to both generate a recommendation and to update the recommender model maintained by the algorithm.

\begin{algorithm}

  \caption{Rating routing}
  \label{algSRM}
  \textbf{Input:} a tuple of $\langle u,i,r\rangle$ of ratings, $n_c$ the number of nodes,\\
  \hspace*{\algorithmicindent} $n_i$, $w$ control the replication\\
  \textbf{Output:} $key$ the worker identifier to which the rating tuple is routed\\
  

\begin{algorithmic}

    \State $n_{ciw} \gets \dfrac{n_c}{n_i}+w $
    \State  $itemsWorkersCandidatesList \gets \phi$
    \State  $usersWorkersCandidatesList \gets \phi$
    \State $itemHash \gets i~modulo~n_i$
    \State $userHash \gets u~modulo~n_{ciw}$
    \For {$x=0\dots n_{ciw}$}
        \State $itemsWorkersCandidatesList  = itemsWorkersCandidatesList \cup \{ itemHash*n_{ciw} + x\}$
    \EndFor
    \For {$y=0\dots n_{i}$}
        \State $usersWorkersCandidatesList  = usersWorkersCandidatesList \cup \{ userHash+(y*n_{c})+w\}$
    \EndFor
    
    \State $key \gets itemsWorkersCandidates\cap usersWorkersCandidates$\\
    \Return $key.first()$
  \end{algorithmic}
\end{algorithm}

 The splitting and replication mechanism routes the user-item ratings to worker nodes. First, it distributes the stream over $n_{c}$  workers in the cluster while the pair can reside on just a single node. The mechanism replicates the representation of users and items over cluster nodes based on replication factor $n_{i}$. The replication here does not imply that they share the same values nor are they synchronized. But, it means that the replication of belonging, for instance, two vectors on different nodes can represent the same user. Yet, each vector of them is calculated separately based on the neighborhood of interactions within the same node as shown in figure \ref{fig:splliting:replication:inner:working}.

In many cases, the number of items is considerably smaller compared to the number of users. This encourages us to manage the splitting by replicating the items' representation over the nodes more than or equal to users' representation, and that can be controlled through mechanism's parameters. Hence, the items matrix can be divided into $n_{i}$ splits. As a constraint for gaining the best possible scalability from the algorithm to utilize all the available workers, the number of workers in the cluster $n_c$ should be equal to $n_{i}^2+w.n_{i}$ where $w\in \mathbb{N}_0$. For example, if $n_i=2$, the items' state ought to be split into two halves and each half should be stored on half of the nodes while the users' state is partitioned over $n_c/2$ of the nodes. The $n_c$, $n_i$, and $w$ are parameters for splitting and replication mechanisms so, they are the knobs to control the replication factor with respect to the nature of every use case. For accomplishing the previously mentioned partitioning, a generated key is assigned to the received tuple of the stream to ensure that the representation of the user/item pair is directed to only one worker,  the key is produced through three primary steps, firstly every item is hashed using modulo $n_i$ and each user is hashed using modulo $n_c/n_i + w$, the yield of the hashing is utilized as a seed to generate the candidates' workers ids that the user or item may reside. The candidates' workers are generated through the accompanying formulas shown in Algorithm ~\ref{algSRM}. The last step to get the key, which is utilized for directing the tuples to the workers, is to get the common key in the generated lists and how guaranteeing a single worker by pair coming from two main considerations.

In the following subsections, we adapt two incremental algorithms to work with our splitting and replication mechanism. Namely, in Section~\ref{sec:disgd}, we adapt ISGD for matrix factorization and make it distributed. Similarly, in Section~\ref{sec:distributed:similarity}, we develop a distributed version of the incremental Cosine similarity.

\subsection{Distributed Matrix Factorization}
\label{sec:disgd}
\renewcommand{\algorithmicreturn}{\textbf{emit}}
\begin{algorithm}
  \caption{DISGD}
  \label{alg:parallel:isgd}
 \textbf{Input:} a stream $S$ of $\langle u,i,r\rangle$ ratings,
  
  $N$ size of recommendation list, $\lambda, \eta$ to control learning for SGD, $n_i$ replication factor, and  $k$ the number of latent features\\
  \textbf{Output:}Top $N$ recommendation list
  \begin{algorithmic}
    \State $n_c \gets n_i^2+w$
    \For{ \textbf{each} $s \in S$}
        \State $key \gets generateKey (s, n_i, n_c, w )$
        \State Forward $s$ to worker $key$
        \State On the receiving worker $key$ of $s$ do: 
        
        \For{\textbf{each} $p \in I$}
            \If{p $\not\in$ user's rated items }
                \State $\hat{r_up} = U_{u}.I_{p}^T$
                \State \textbf{emit} $\langle u$,Top N recommendation list$\rangle$
            \EndIf

            \If{$s.u \not\in Rows(U)$ }
                \State $U_{u} \gets Vector(size : k)$
                \State $U_{u} \sim \mathcal{N}(0,0.1)$
            \EndIf
            \If{$s.i \not\in Rows(I)$ }
                \State $I_{i} \gets  Vector(size : k)$
                \State $I_{i} \sim \mathcal{N}(0,0.1)$
            \EndIf
            \State $err_{ui} \gets 1-U_{u}.I_{i}^T$
            \State $U_{u} \gets U_{u}+\eta(err_{ui}.I_{i}-  \lambda U_{u})$
            \State $I_{i} \gets  I_{i}+\eta(err_{ui}.U_{u} - \lambda I_{i})$
        \EndFor
    \EndFor  
\end{algorithmic}
\end{algorithm}

DISGD is the proposed algorithm utilizing the Splitting and Replication mechanism for accomplishing the scaling of ISGD. Coming to ISGD, it is the adapted version of the SGD algorithm, it works incrementally, and updates itself using only the recent event coming from the stream of ratings. Compared to SGD, ISGD has two primary differences. First, it proceeds with a single pass over the available data. Second, it manages just a positive Boolean rating matrix. Thus, the error is measured by $err_{ui}=1-\hat{ R_{ui}}$. For every received user-item rating, ISGD re/calculates user and item vectors as per equations~\ref{eq:1} and~\ref{eq:2} where $\eta$ is the gradient step size. DISGD parallelizes ISGD as shown in Algorithm~\ref{alg:parallel:isgd}. Basically, the newly received rating tuple is routed to a worker where the respective user and item vectors on that worker are updated in an incremental fashion.


\subsection{Distributed Incremental Cosine Similarity}
\label{sec:distributed:similarity}

\begin{algorithm}
  \caption{DICS}
  \label{alg:parallel:TenRec}
  \textbf{Input:} a stream $S$ of $\langle u,i,r\rangle$ ratings,
$N$ size of recommendation list, $n_i$ replication factor, and  $k$ the number of latent features\\
  \textbf{Output:}Top $N$ recommendation list

  \begin{algorithmic}
  
    \State $n_c \gets n_i^2+w$
    \For{ \textbf{each} $s \in S$}
        \State $key \gets generateKey (s, n_i, n_c, w )$
        \State Forward $s$ to worker $key$
        \State On the receiving worker $key$ of $s$ do:
        
        \For{\textbf{each} $p \in I$}
            \If{p $\not\in$ user's rated items }
                \State calculate $\hat{r_{up}}$ as per Equation~\ref{r}
                \State \textbf{emit}  $\langle u$,Top N recommendation list$\rangle$
            \EndIf
            
            \For{\textbf{each} $item~pair \in items'~pairs$}
                \State \textbf{update} {$sim(i_p,i_q)'$ as per Equation~\ref{sim_eq}}
            \EndFor
        \EndFor
    \EndFor
\end{algorithmic}
\end{algorithm}

Distributed Incremental Cosine Similarity (DICS) has utilized the modified Cosine similarity metric proposed by TencentRec \cite{huang2015tencentrec} as described in Section~\ref{sec:background}. The application of our Splitting and Replication mechanism on top of the incremental Cosine similarity is described in Algorithm~\ref{alg:parallel:TenRec}. Each worker memory (state) contains $pairCount$ of every pair of items on the worker and the user-item rating history is saved in the form of a hash table where the key is the user identifier and the value is the list of rated items per user. Upon receiving a user-item rating, it is first routed to the target worker. On that worker, based on the saved user's history rated items, the estimated ratings for unrated items by user are calculated as per Equation~\ref{r}. This estimate is used in the next step to generate the top-N recommendation list for the user.  Next, the algorithm updates all the related rated items by the user and all the items' pairs similarities which contain the current item.

\section{Evaluation}
\label{sec:evalution}

In this section, we discuss the evaluation of the Splitting and Replication mechanism against baseline centralized approaches. The justification of system selection for our approach is given in Section~\ref{sec:implementation}. The experimental setup and evaluation metrics are discussed in Section~\ref{sec:experimental:setup}. Finally, evaluation results are discussed in Section~\ref{sec:evaluation:results}.

\subsection{System Selection and Implementation}
\label{sec:implementation}
To develop a proof of concept of our approach, we have investigated the top three open-source distributed stream processing systems: Apache Storm~\cite{Storm2015}, Apache Spark~\cite{sparkStructuredStreaming18}, and Apache Flink~\cite{Flink2015}. As indicated in~\cite{huang2015tencentrec}, Storm is built with no support for stateful computations. In our case, state is holding the rating matrix $R$. State has to be partitioned across nodes in the distributed system. On the other hand, both Spark and Flink support distributed stateful computations on data streams. However, Spark has two limitations. First, it has limited support for application developers to work with stateful computations outside the predefined APIs. Second, it works with so-called micro-batches. That is, it emulates stream processing by processing a small number of received items at a time. On the other hand, Flink is a native stream processing engine that allows element-by-element processing. Moreover, it gives a rich set of APIs to manage the distributed state by application developers~\cite{stateManagementFlink17}. Flink streaming engine outperforms the other engines in terms of latency~\cite{karimov2018benchmarking,ShahverdiAS19}. For these reasons, we realized our splitting and replication mechanism on top of Apache Flink, namely version $1.8.1$.

\subsection{Setup and Evaluation Metrics}
\label{sec:experimental:setup}

Our evaluation is based on comparing the baseline (single machine) implementation of the ISGD and Cosine similarity with the distributed approach achieved by applying our splitting and replication mechanism. For the distributed setup, we deployed Flink on a cluster that comprises three nodes with $96$ cores each. Each core is running at $2.3~GHz$ with $30~GB$ of main memory on each node. All the single-node solutions are implemented on top of Apache Fink with a configuration to run on a single core. For reproducibility purposes, all the implementation and experiments are publicly available on GitHub~\footnote{\url{https://github.com/heidyhazem/DSRS}}.

\begin{algorithm}[h!]
  \caption{Prequential online evaluator using incremental recall }
  \label{alg:evaluateion}
  
  \hspace*{\algorithmicindent} \textbf{Input:} $\langle user, item, r\rangle$, user-item feedback element on the stream,\\ \hspace*{\algorithmicindent}$N$: the size of the topmost recommendations.\\
  \hspace*{\algorithmicindent} \textbf{Output:} $Recall@N \in \{0,1\}$
  \begin{algorithmic}[1]
\State $RecList \gets getRecommendations(\langle user, item, r\rangle, N)$
\If { $item \in RecList $} 
    \State $Recall@N \gets 1$
  \Else    
  \State $Recall@N \gets 0$
  \EndIf
  
   \State $UpdateModel(\langle user, item, r\rangle)$\\
   \Return $Recall@N$
  \end{algorithmic}
  \end{algorithm}

The traditional approach of evaluating batch recommender systems is by splitting the dataset into training and testing parts~\cite{recommenderSystemsBook16}. Collaborative filtering algorithms are trained on the training part of the data, whereas the unseen testing part is used to update the model until conversion. Unfortunately, the aforementioned evaluation scenario is not applicable in the (online) streaming scenario. That is because, data are theoretically unbounded. Hence, the splits into training and testing cannot be applied. Additionally, each data element on the stream, a user-item rating in our case, should be processed once to update the model and to give recommendations. Moreover, it is quite challenging to compensate for the real bias due to the recommendation list presented to the users as the user's behaviors are affected directly by what is recommended and it is more probable to interact with what the user sees. For these challenges, we adapt prequential evaluation proposed in~\cite{bifet2015efficient, albertBifetMLDataStreamsBook18}. Prequential evaluation is used in evaluating online machine learning algorithms. We adapt it to evaluate the performance of online recommender systems. Namely, we adapt it to measure the \emph{recall} of recommendations (Algorithm~\ref{alg:evaluateion}). The prequential evaluation using recall works as follows: for every coming rating $\langle user,item, rate\rangle$, it is used first for generating top $N$ recommendations. Online recall returns $1$ if the item is within the top-N recommendations, otherwise it is $0$. Then, the item is used for training.  

In addition to using recall for measuring the accuracy of the recommender algorithms, we compare latency, throughput and memory overhead of the different configurations. For the prequential evaluation we set $N=10$. We compute a moving average of recall over a window of $5000$ elements, i.e., user-item ratings.  All the algorithms' hyper-parameters have been selected by optimizing the first $20\%$ of the datasets (handling the cold startup problem~\cite{coldStartup05,coldStartup2012}).

For our distribution approach, we configured the replication factor $n_{i}\in\{2,4,6\}$ and the number of the workers is the least possible $n_{c}=n_{i}^2$  configuration to obtain the benefit from the splitting and replication technique. Hence, the corresponding $n_{c}\in\{4,16,36\}$ workers.

To cope with the unboundedness of data streams, we have employed techniques from cache management~\cite{cacheManagement2016} to control the growth of the partitions of the rating matrix $R$ held by each node in the cluster. Namely, we employ least frequently used (LFU) and least recently used (LRU) techniques. We refer to them as forgetting techniques. We measure their effect on the memory size, the number of entries in $R$. We do not measure the memory in bytes, as this is tricky with Flink~\cite{ShahverdiAS19}. Rather, we provide two visualizations. The first shows the distribution of count frequencies for items and users. The second shows the evolution of users/items count over the time. The forgetting techniques have two main parameters the first one is the trigger threshold parameter that defines when the scanning should be triggered and the second parameter is the controller parameter to define which items/users should be deleted. The trigger threshold parameter in LFU in the experiments has been defined by a certain count of records $c$ so, after processing every $c$ records the scan starts while in LRU has been defined by a certain time $t$ so, after $t$ time the scan starts. The controller parameter in LFU is the frequency of item is requested or count of users' requests towards items. The controller parameter in LRU is a threshold for the difference between the current time and last timestamp for user/item. For the sake of studying the effects of the forgetting techniques, LRU's parameters are chosen to get the best recall while LFU's parameters are optimized to get the least memory consumption.

In this work, two datasets have been used for evaluation purposes. The first dataset is the Movielens 25M \footnote{\url{https://grouplens.org/datasets/movielens/}}. The second dataset is the Netflix Prize dataset \footnote{\url{https://www.netflixprize.com/leaderboard.html}} which is widely used in benchmarks~\cite{RSBenchmark2012,RSBenchmark2020}. Both datasets' ratings are on an ordinal scale from $1$ to $5$ stars. Preprocessing steps have been applied to the datasets. First, the datasets have been ordered ascending based on the timestamp for emulating the streaming data. Second, since the central online algorithms are working only on binary ratings, positive feedback only is considered from the data sets. For this, we have filtered out the feedback of fewer than $5$ stars. The datasets' characteristics after filtering are reported in Table~\ref{tab:data:sets} 

\begin{table}[h!]
\centering
\begin{tabular}{|c|c|c|c|c|c|c|}
\hline
\textbf{Dataset} & \textbf{Ratings} & \textbf{Users} & \textbf{Items}& \textbf{avg}&\textbf{avg } &\textbf{Spar-}  \\ 
                 &                  &                &               & \textbf{ratings/user} &   \textbf{ratings/item}      &\textbf{sity}\\
\hline
MovieLens-25M & 3612474 & 155002 & 27133 &23.3 &133& 99.91\% \\ \hline
Netflix & 4086048 & 394106 & 3001 &10.6 & 1361.5 & 99.65\%                        \\ \hline
\end{tabular}
\caption{Datasets characteristics after filtering}
\label{tab:data:sets}
\end{table}

\subsection{Results}
\label{sec:evaluation:results}




\subsubsection{DISGD}
\label{sec:disgd:results}

In this section, we discuss the evaluation of DISGD compared to the central version ISGD as the baseline. The D/ISGD hyperparameters values used in equations \ref{eq:1}and~\ref{eq:2} are as follows, $\lambda=0.01$, $\mu=0.05$, and the latent feature $k=10$.

\begin{figure}[h!]
    \centering 
    
\begin{subfigure}[t]{0.5\textwidth}
  \includegraphics[width=\linewidth]{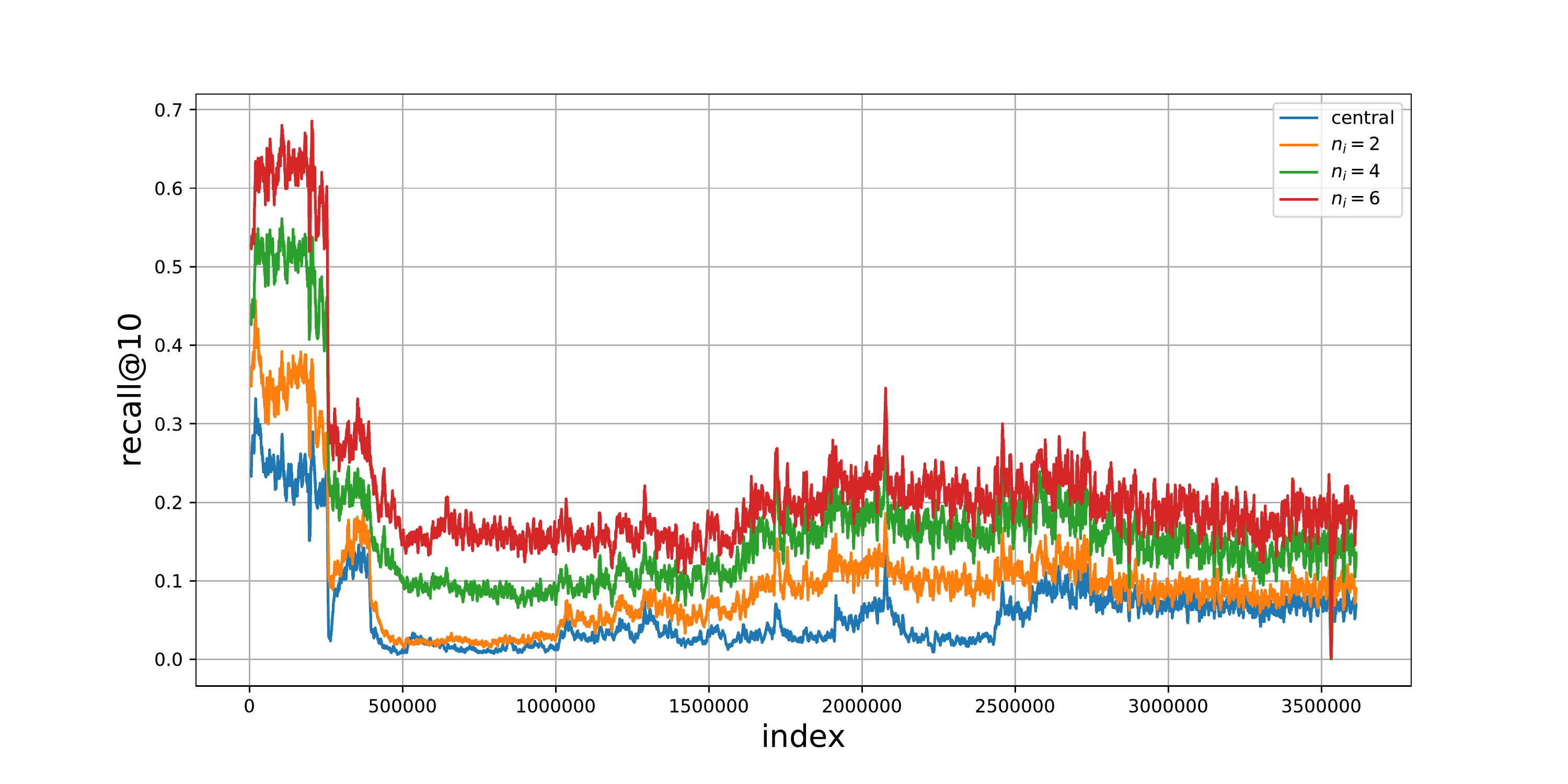}
  \caption{Movielens-25M}
  \label{fig:disgd:recall:no:forget:ML25}
\end{subfigure}~
\begin{subfigure}[t]{0.5\textwidth}
  \includegraphics[width=\linewidth]{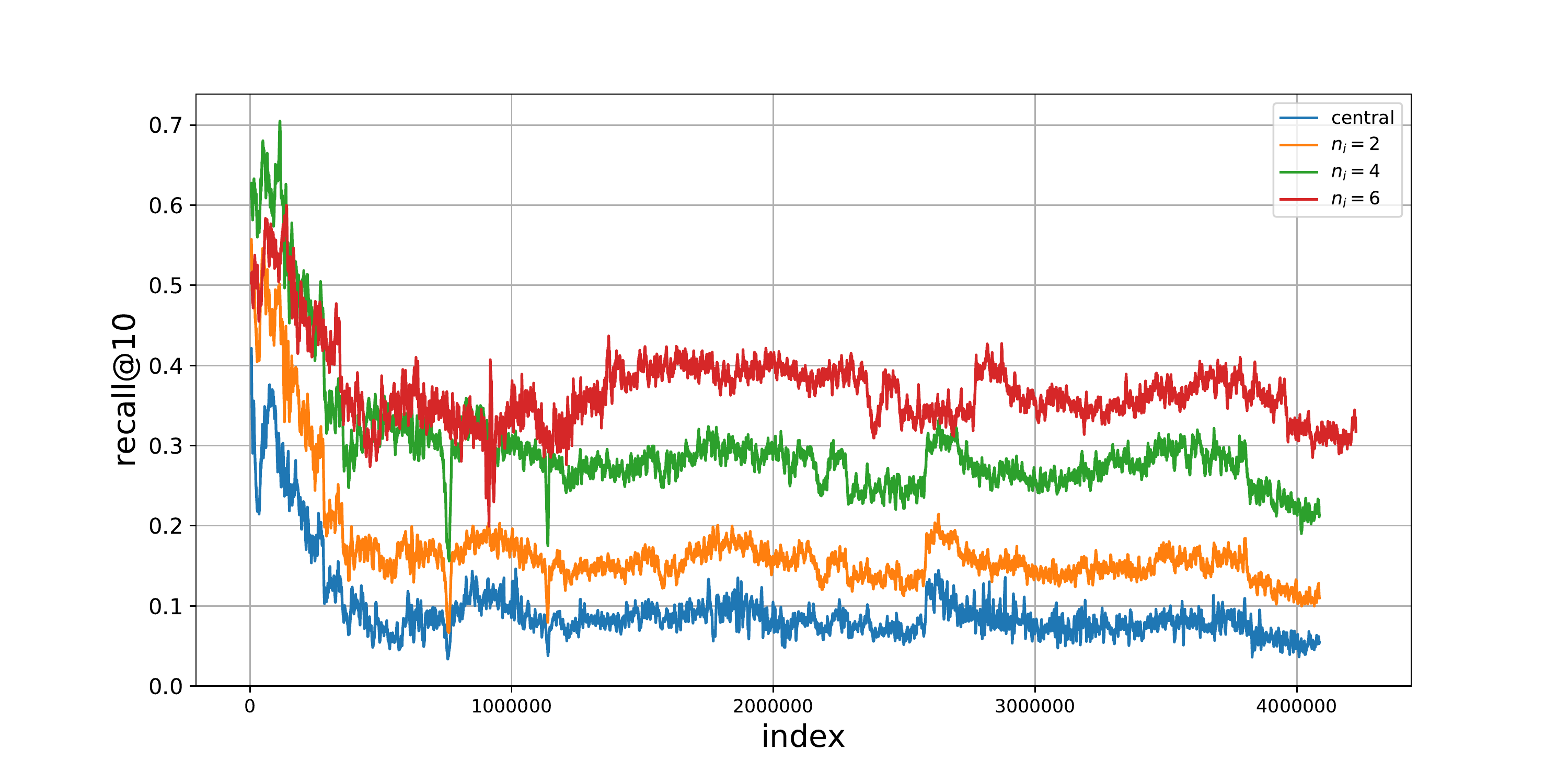}
  \caption{Netflix}
  \label{fig:disgd:recall:no:forget:NFlix}
\end{subfigure}
\caption{Moving average Recall@10 for ISGD (central) and DISGD for different $n_i$}
\label{fig:disgd:recall:no:forget}
\end{figure}

Figure~\ref{fig:disgd:recall:no:forget} captures the recall performance over time for the different $n_i$ configurations. Obviously, the splitting and replication mechanism dramatically improves the recall for both datasets over the centralized ISGD. The average recall is enhanced from $3\%$ to more than $40\%$ based on the $n_{i}$ and the dataset. Empirically, there is a direct positive correlation between the recall and $n_{i}$.

\begin{figure}[h!]
    
\begin{subfigure}{0.5\textwidth}
  \includegraphics[width=\linewidth]{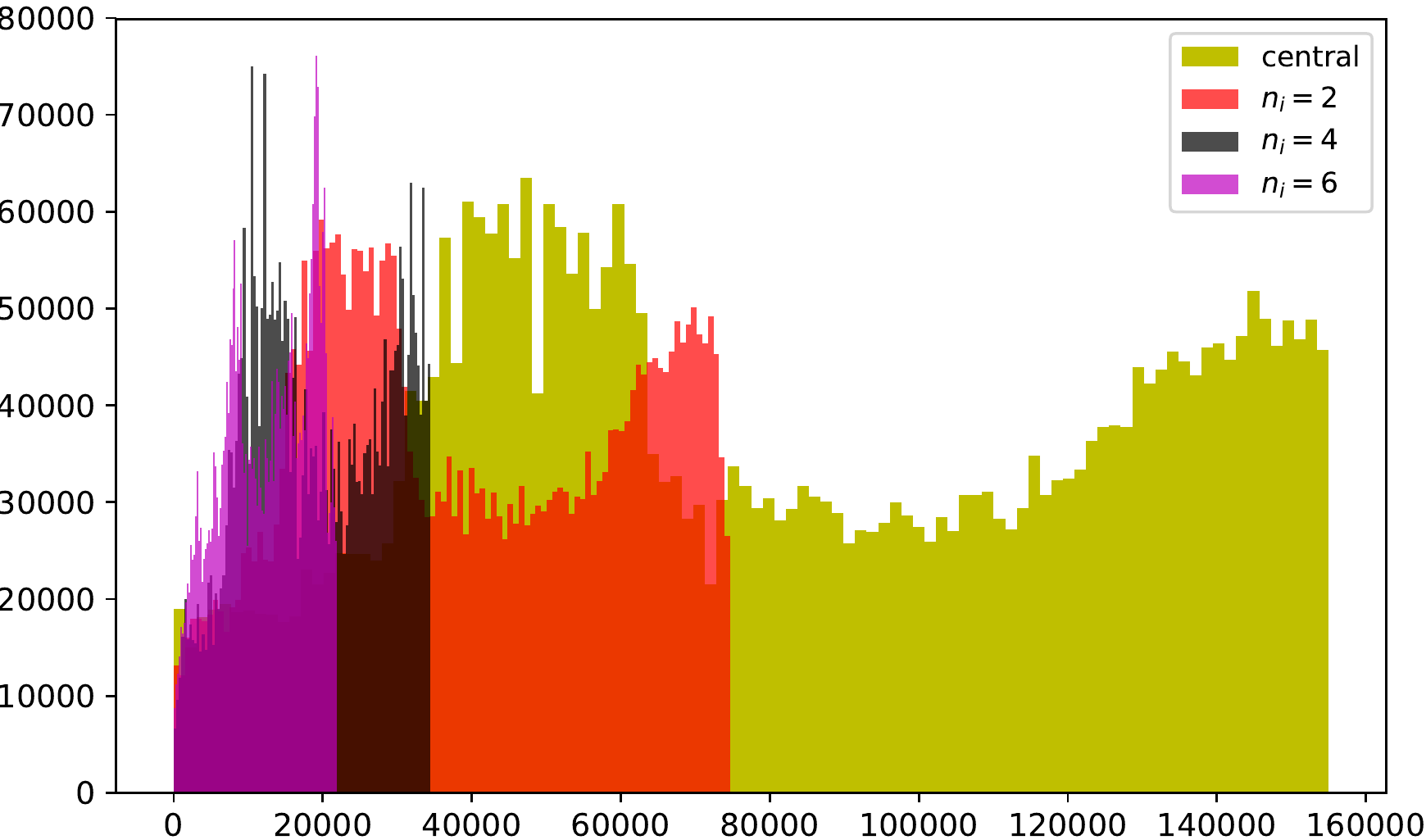}
  \caption{Movielens-25M users' state distribution}
  \label{fig:disgd:memory:no:forget:ML25:users}
\end{subfigure}
~
\begin{subfigure}{0.5\textwidth}
  \includegraphics[width=\linewidth]{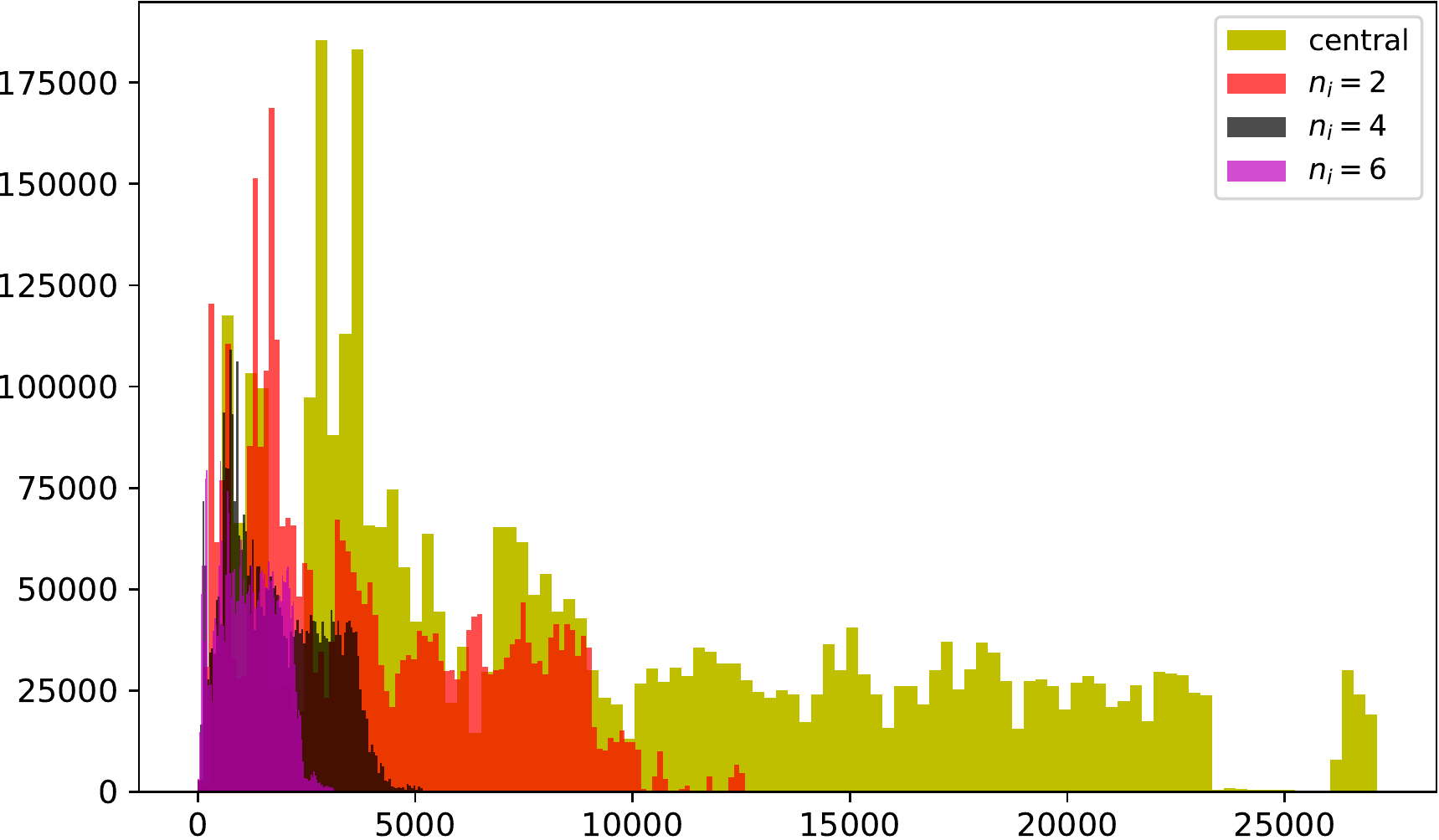}
  \caption{Movielens-25M items' state distribution}
  \label{fig:disgd:memory:no:forget:ML25:items}
\end{subfigure}
\qquad
\begin{subfigure}{0.5\textwidth}
  \includegraphics[width=\linewidth]{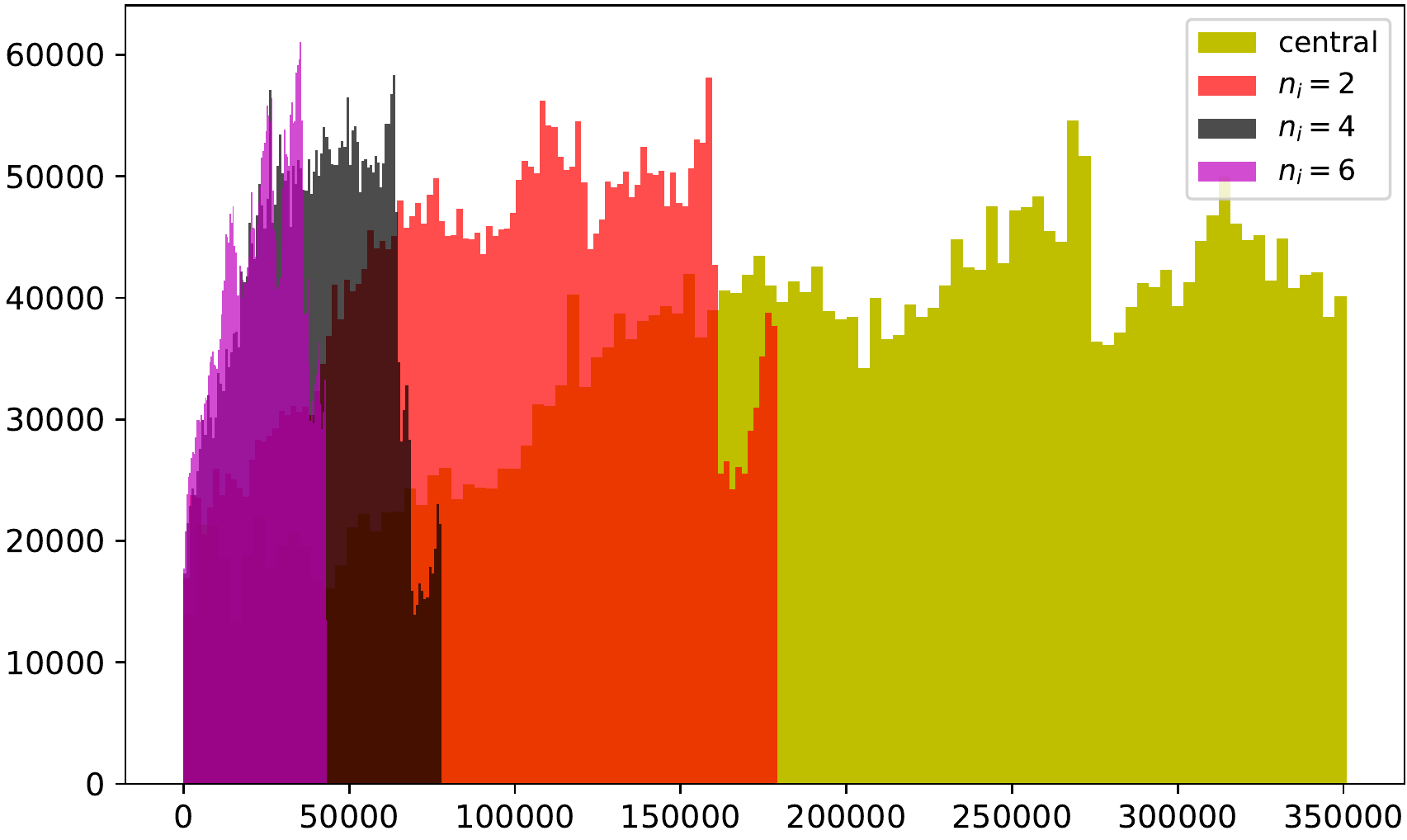}
  \caption{Netflix users' state distribution}
  \label{fig:disgd:memory:no:forget:NFlix:users}
\end{subfigure}~ 
\begin{subfigure}{0.5\textwidth}
  \includegraphics[width=\linewidth]{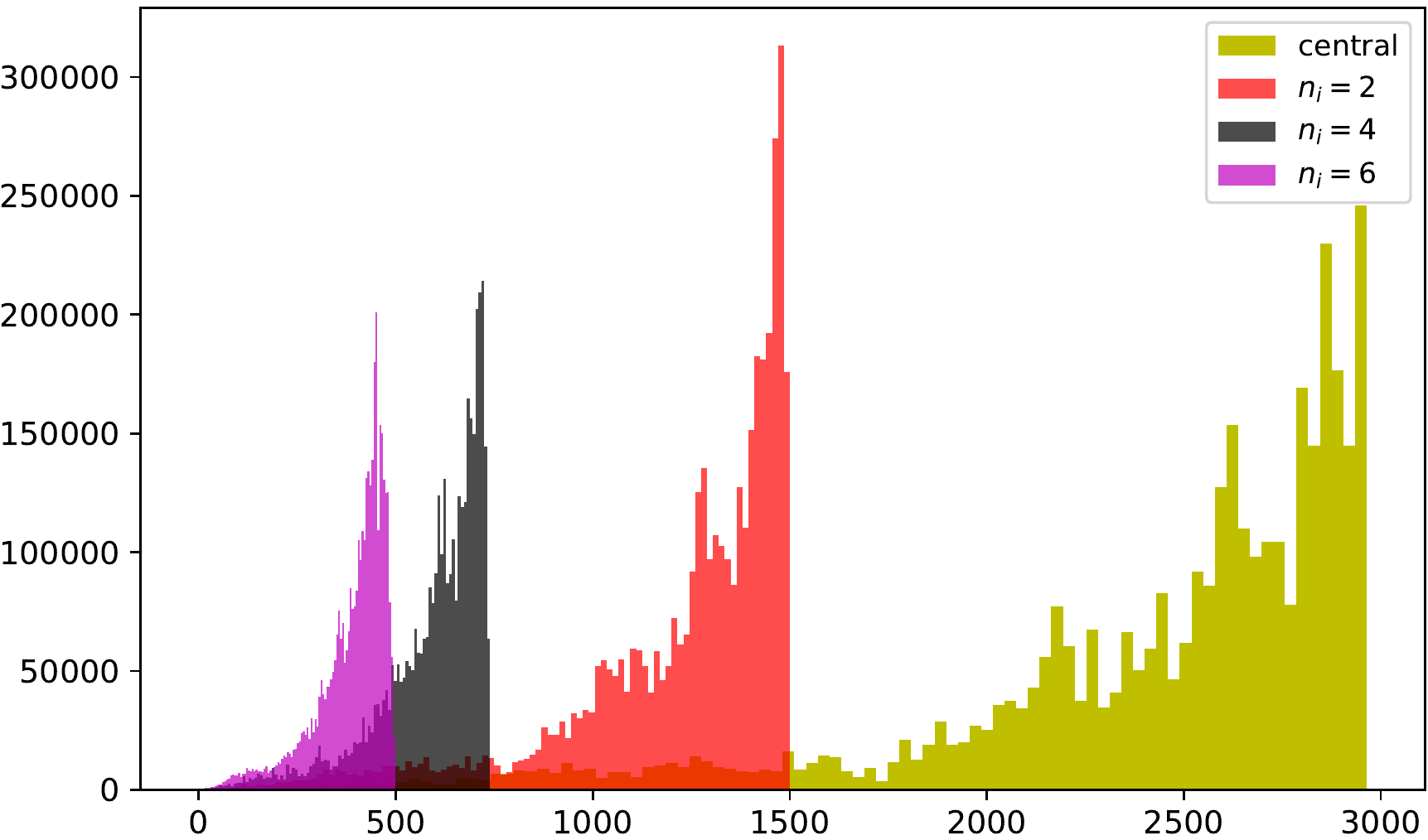}
  \caption{Netflix items' state distribution}
  \label{fig:disgd:memory:no:forget:NFlix:items}
\end{subfigure}
\caption{Memory distribution for ISGD (central) and DISGD for the different $n_i$ values. X-axis shows the size number of entries and Y-axis shows the frequency.}
\label{fig:disgd:memory:no:forget}
\end{figure}

The DISGD memory distribution for different $n_i$ is plotted in Figure~\ref{fig:disgd:memory:no:forget}. Although our approach admits some replication of values across nodes, the overall size of respective user and items states are almost linearly reduced as we distribute the work over more nodes. For instance, the case of the MovieLens data set, the state for users representation (Figure~\ref{fig:disgd:memory:no:forget:ML25:users}) reaches the maximum size of $~155K$ as reported in Table~\ref{tab:data:sets}. With our splitting and replication mechanism, The user's state size is appreciably lessened while increasing $n_i$, the mean of the distribution for $n_i=2$ is dropped to around half, and for $n_i=4$, $n_i=6$, the mean equals $23\%$ and $15\%$ of central configuration users' state mean respectively. Regarding items' s state reduction (Figure~\ref{fig:disgd:memory:no:forget:ML25:items}), it also drops obviously with increasing  $n_i$. The mean of items' state distribution drops to $43\%$, $19\%$, $12.9\%$ of the mean of the central configuration items' state for $n_i = 2, 4, 6$ respectively which indicates the great gain obtained from the splitting and replication mechanism and this can be extended with increasing $n_i$ or enlarging $w$ to better distribute the users' state. The same pattern of state size drop can be observed for the Netflix users state (Figure~\ref{fig:disgd:memory:no:forget:NFlix:users}) and items state (Figure~\ref{fig:disgd:memory:no:forget:NFlix:items}) respectively.

\begin{figure}[h!]
    
\begin{subfigure}{0.5\textwidth}
  \includegraphics[width=\linewidth]{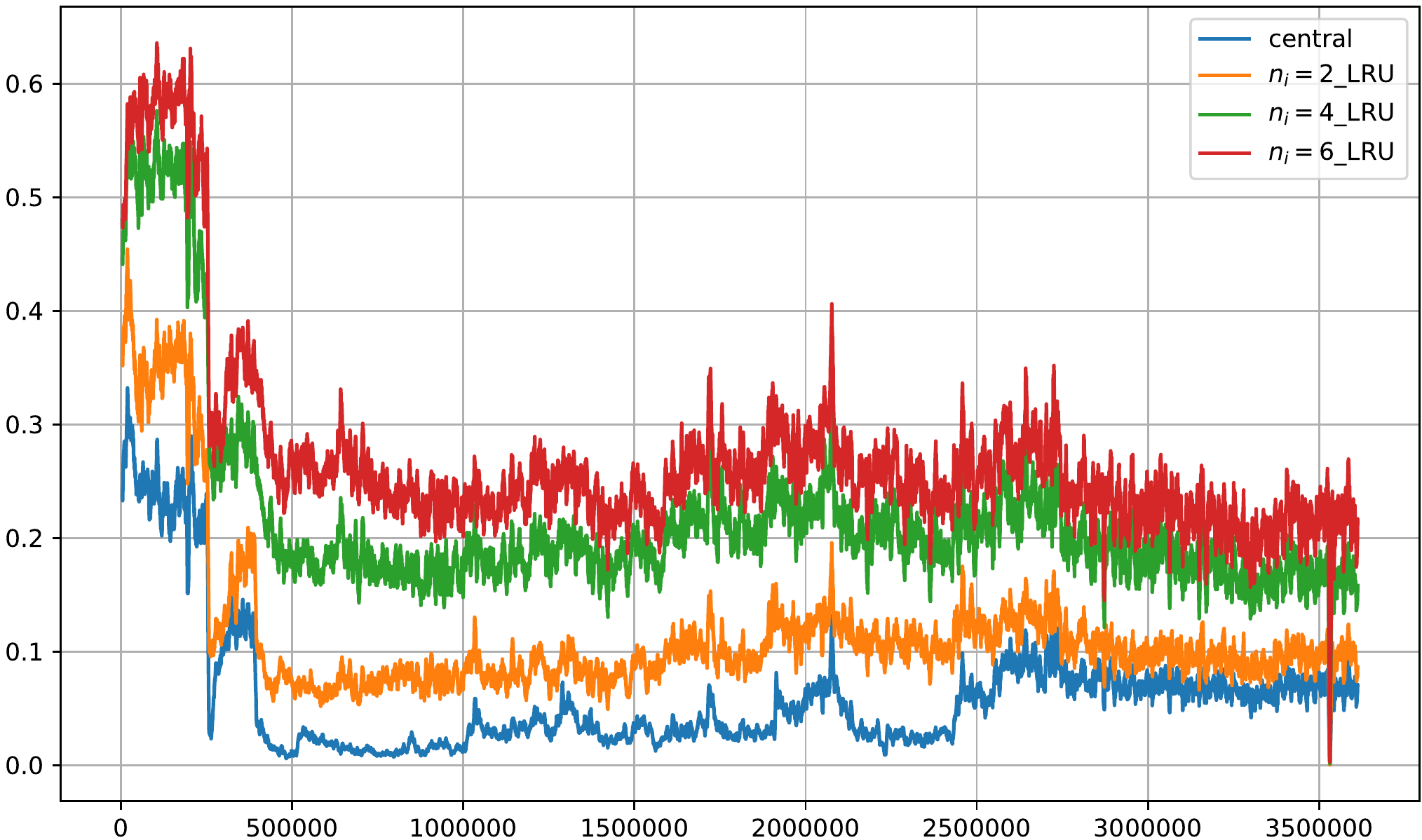}
  \caption{Movielens-25M with LRU Forgetting}
  \label{fig:disgd:recall:forget:ML25:LRU}
\end{subfigure}~ 
\begin{subfigure}{0.5\textwidth}
  \includegraphics[width=\linewidth]{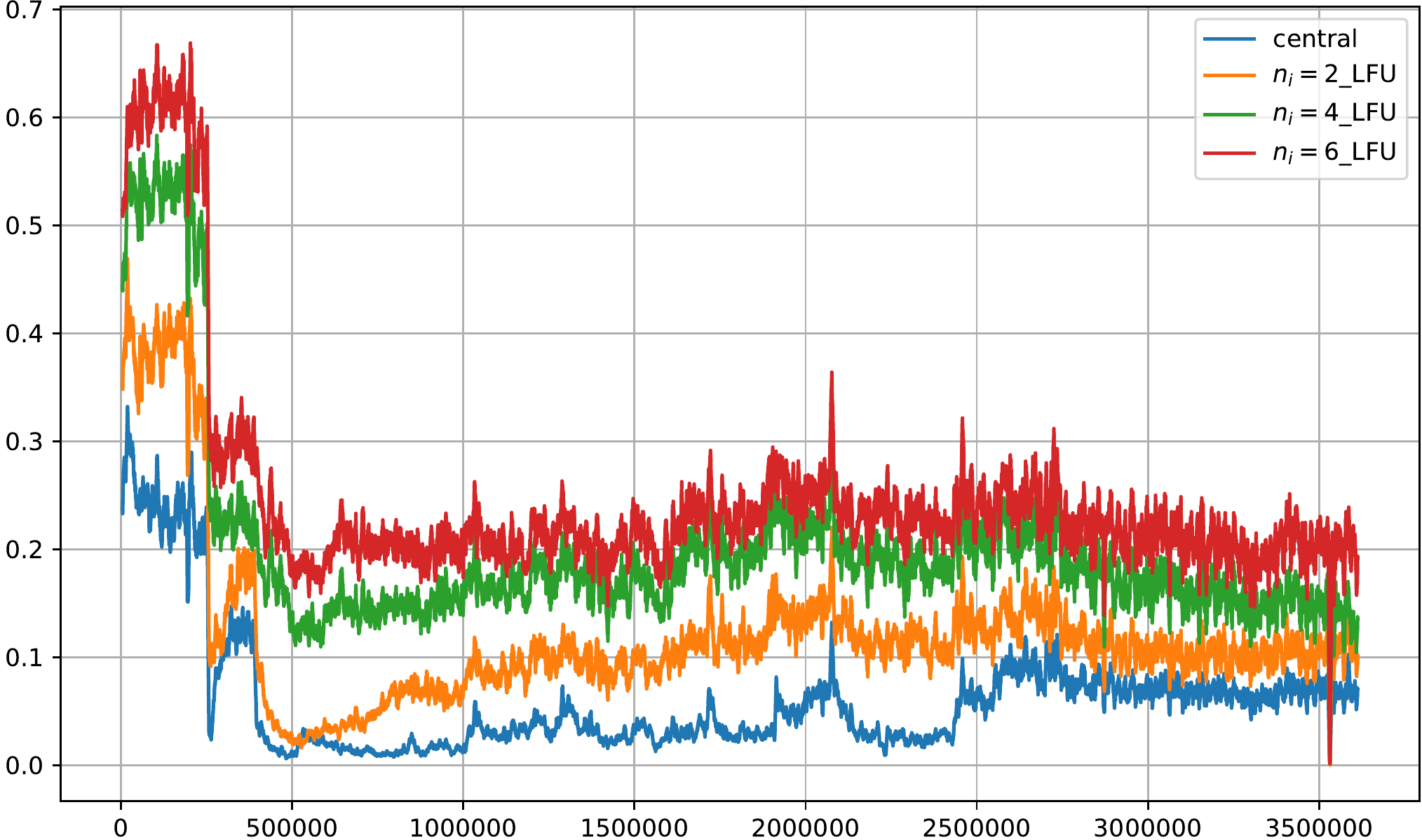}
  \caption{Movielens-25M with LFU Forgetting}
  \label{fig:disgd:recall:forget:ML25:LFU}
\end{subfigure} 
\qquad
\begin{subfigure}[t]{0.5\textwidth}
  \includegraphics[width=\linewidth]{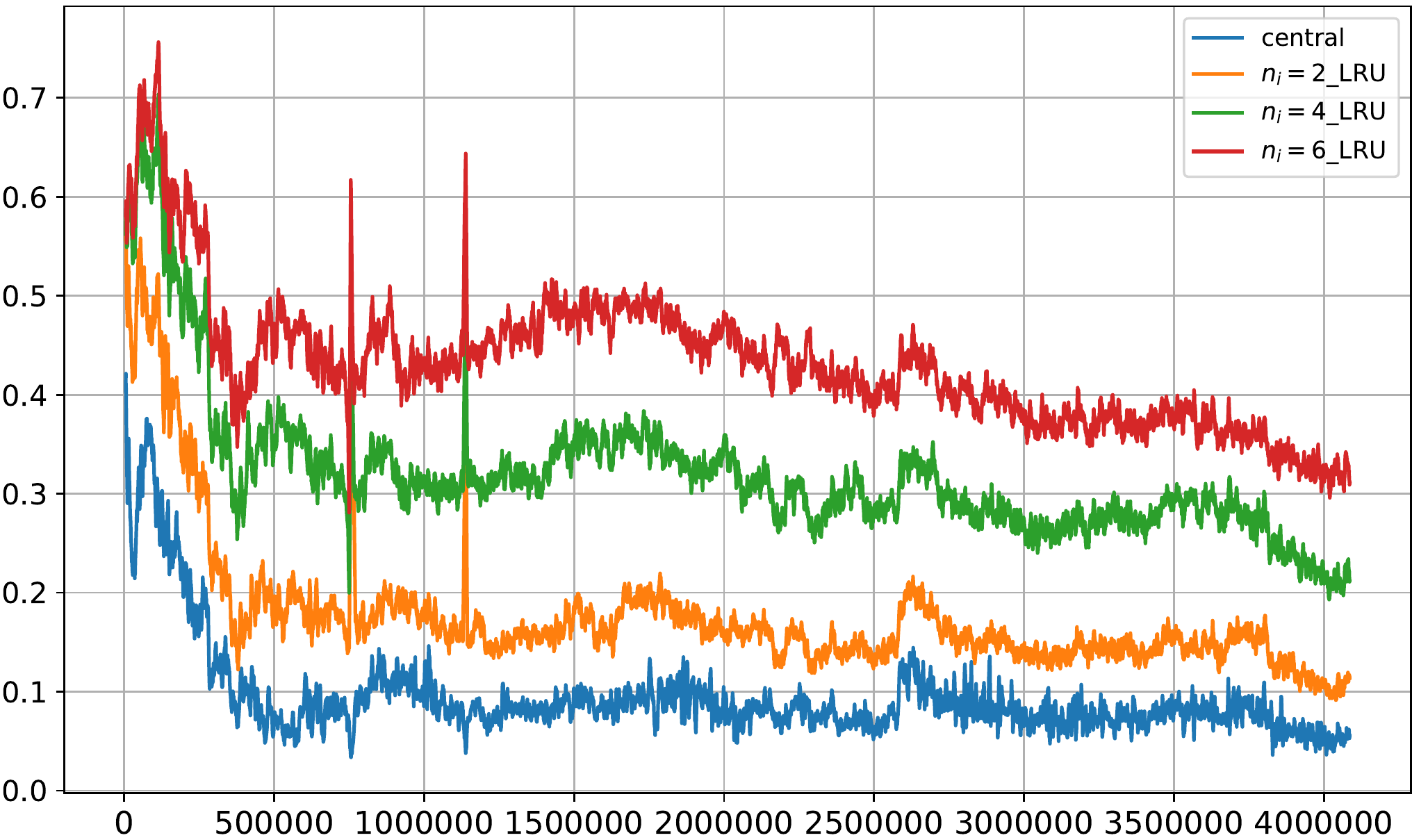}
  \caption{Netflix with LRU Forgetting}
  \label{fig:disgd:recall:forget:NFlix:LRU}
\end{subfigure}~ 
\begin{subfigure}[t]{0.5\textwidth}
  \includegraphics[width=\linewidth]{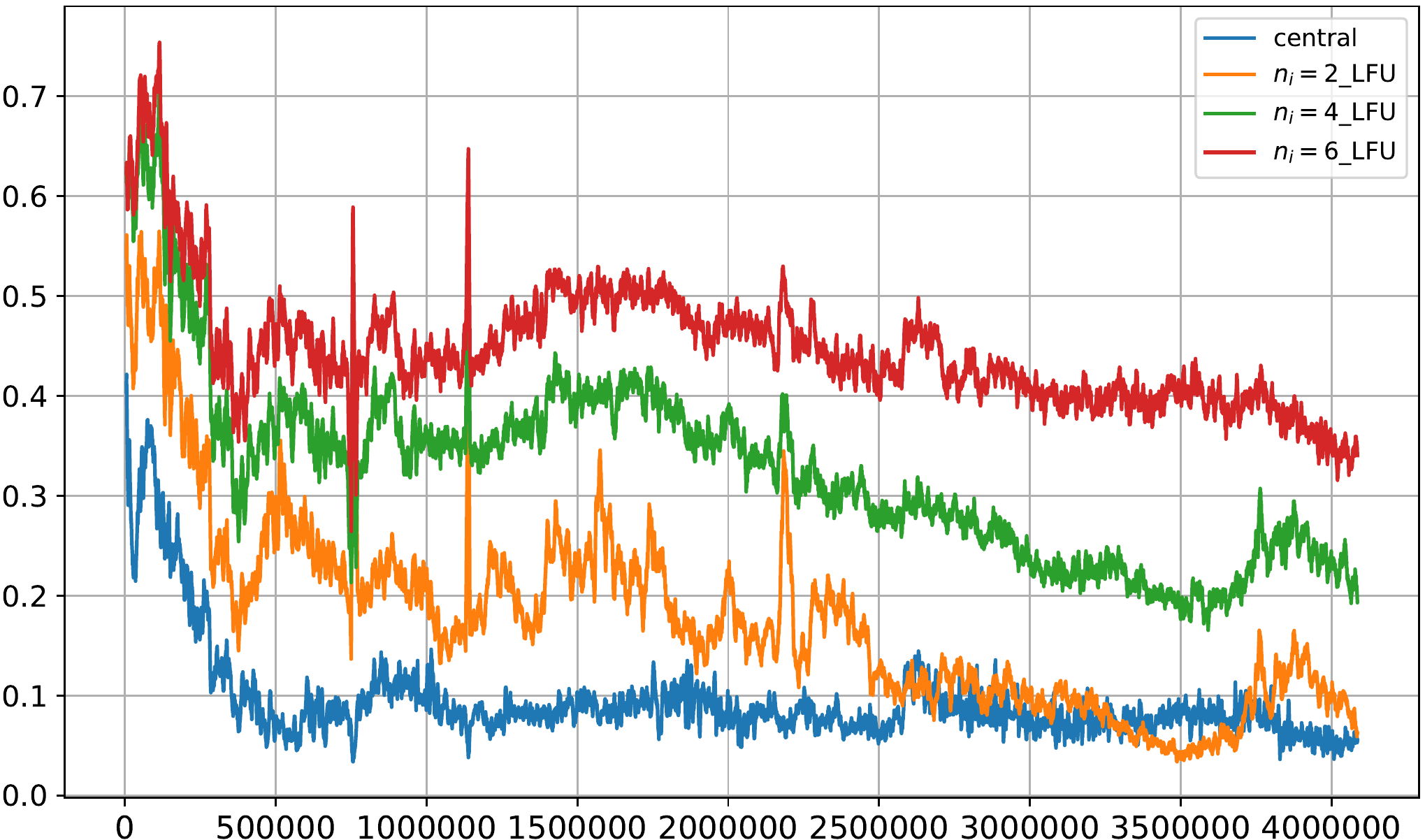}
  \caption{Netflix with LFU Forgetting}
  \label{fig:disgd:recall:forget:NFlix:LFU}
\end{subfigure}
\caption{The effect of applying forgetting techniques on Recall@10 for ISGD (central) and DISGD for the different $n_i$ values}
\label{fig:disgd:recall:forget}
\end{figure}

As stated earlier, to control the expected unlimited growth of users and items states in a production setting, we employed LFU and LRU as means to forget respective user/item vectors from the state (memory) of the workers. Figure~\ref{fig:disgd:recall:forget} shows the effect of the respective techniques on the recall for the two data sets. Obviously, the recall is still better than the centralized approach. Surprisingly, by comparing respective figures to the non-forgetting configuration, e.g. comparing figures~\ref{fig:disgd:recall:forget:ML25:LRU} and~\ref{fig:disgd:recall:forget:ML25:LFU} to Figure~\ref{fig:disgd:recall:no:forget:ML25}, we can observe that the forgetting techniques have enhanced the recall. Albeit being a slight enhancement, this shows how forgetting old user taste to items can enhance the prediction, possibly because of concept drift occurrence, in addition to the control of memory growth. In particular, LRU leads to better recall compared to LFU. The same can be observed for the Netflix data set, figures~\ref{fig:disgd:recall:forget:NFlix:LRU} and~\ref{fig:disgd:recall:forget:NFlix:LFU}.


\begin{figure}[h!]
\begin{subfigure}[t]{0.5\textwidth}
  \includegraphics[width=\linewidth]{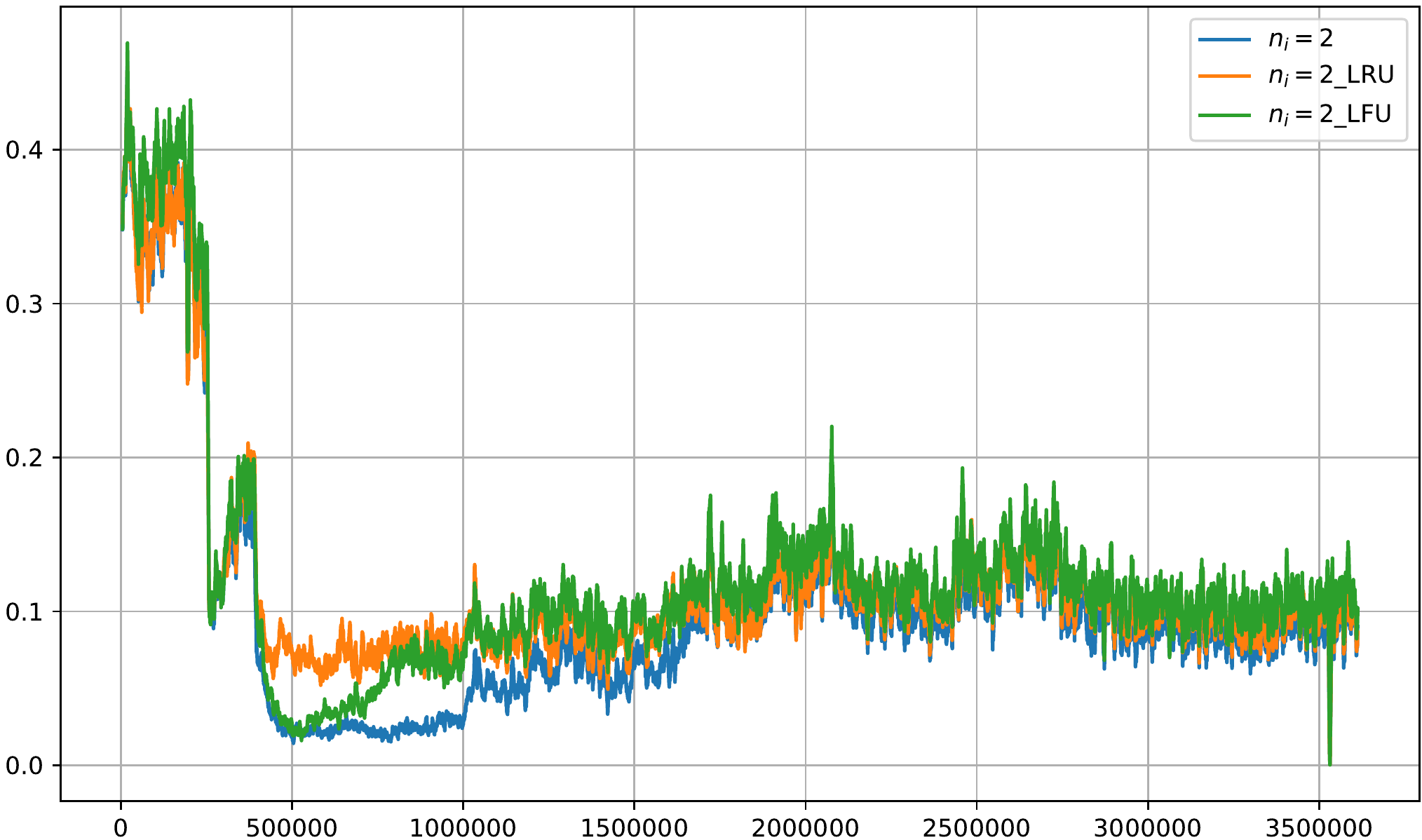}
  \caption{Movielens-25M with $n_{i}=2$}
  \label{ISGD_ML_using_ni_2_LRU}
\end{subfigure}
~
\begin{subfigure}[t]{0.5\textwidth}
  \includegraphics[width=\linewidth]{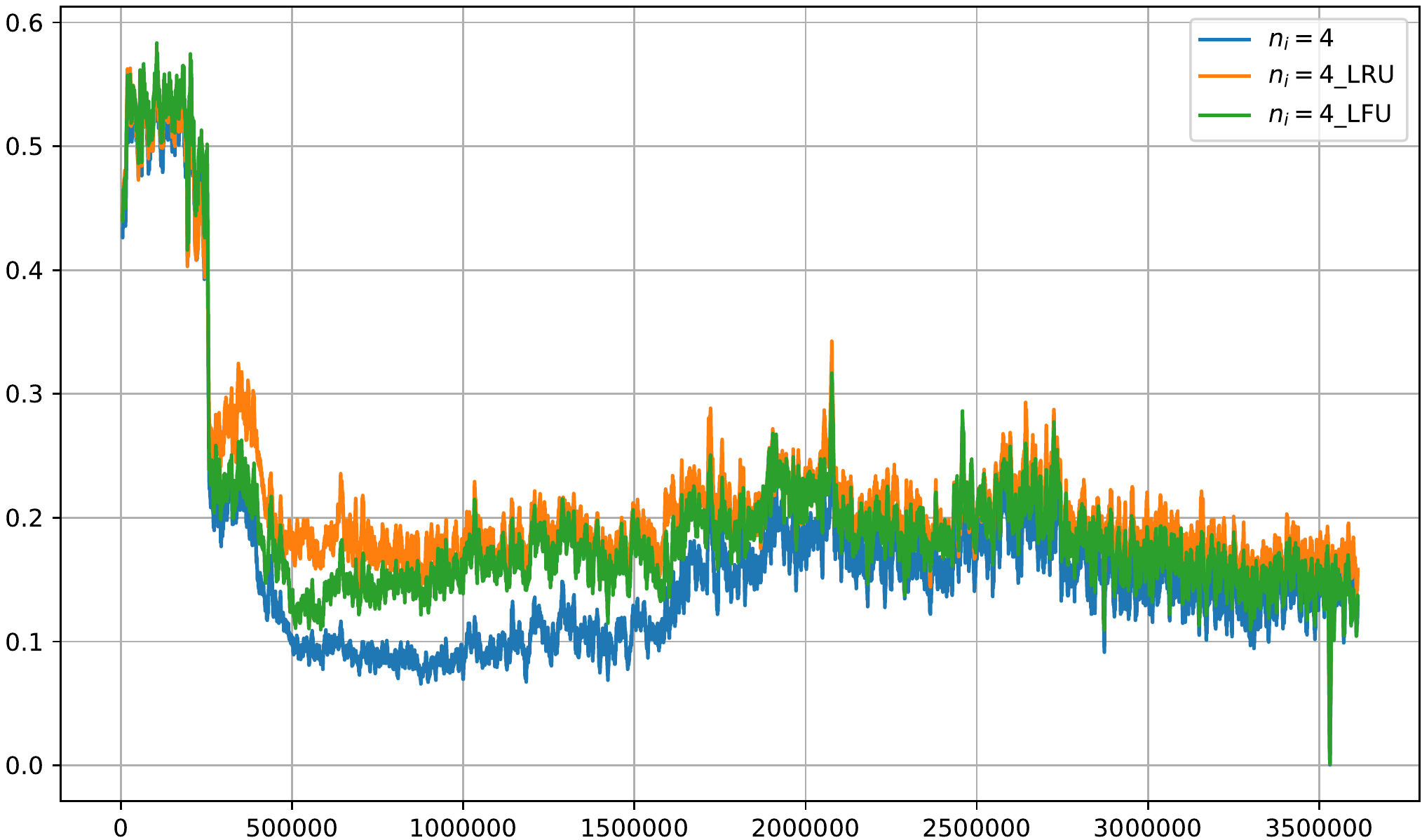}
  \caption{Movielens-25M with $n_{i}=4$}
  \label{ML_using_ni_4}
\end{subfigure}
\caption{Comparison of the effect of LFU and LRU on Recall@10 for ISGD (central) and DISGD for the different $n_i$ values}
\label{fig:images_DISGD_closer_look}
\end{figure}
\begin{figure}[h!]\ContinuedFloat
\begin{subfigure}[t]{0.5\textwidth}
  \includegraphics[width=\linewidth]{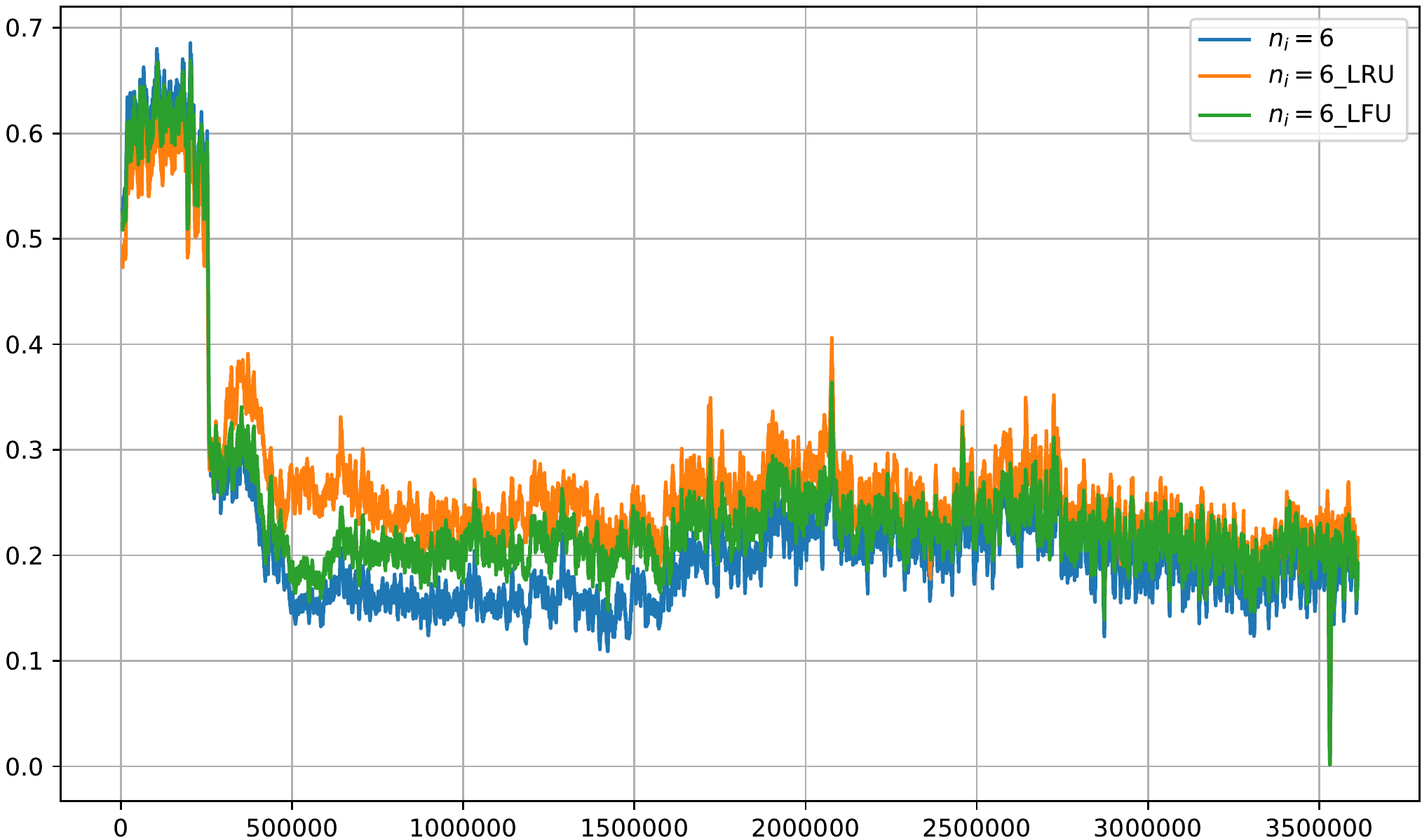}
  \caption{Movielens-25M with $n_{i}=6$}
  \label{ML_using_ni_6}
\end{subfigure}
~
\begin{subfigure}[t]{0.5\textwidth}
  \includegraphics[width=\linewidth]{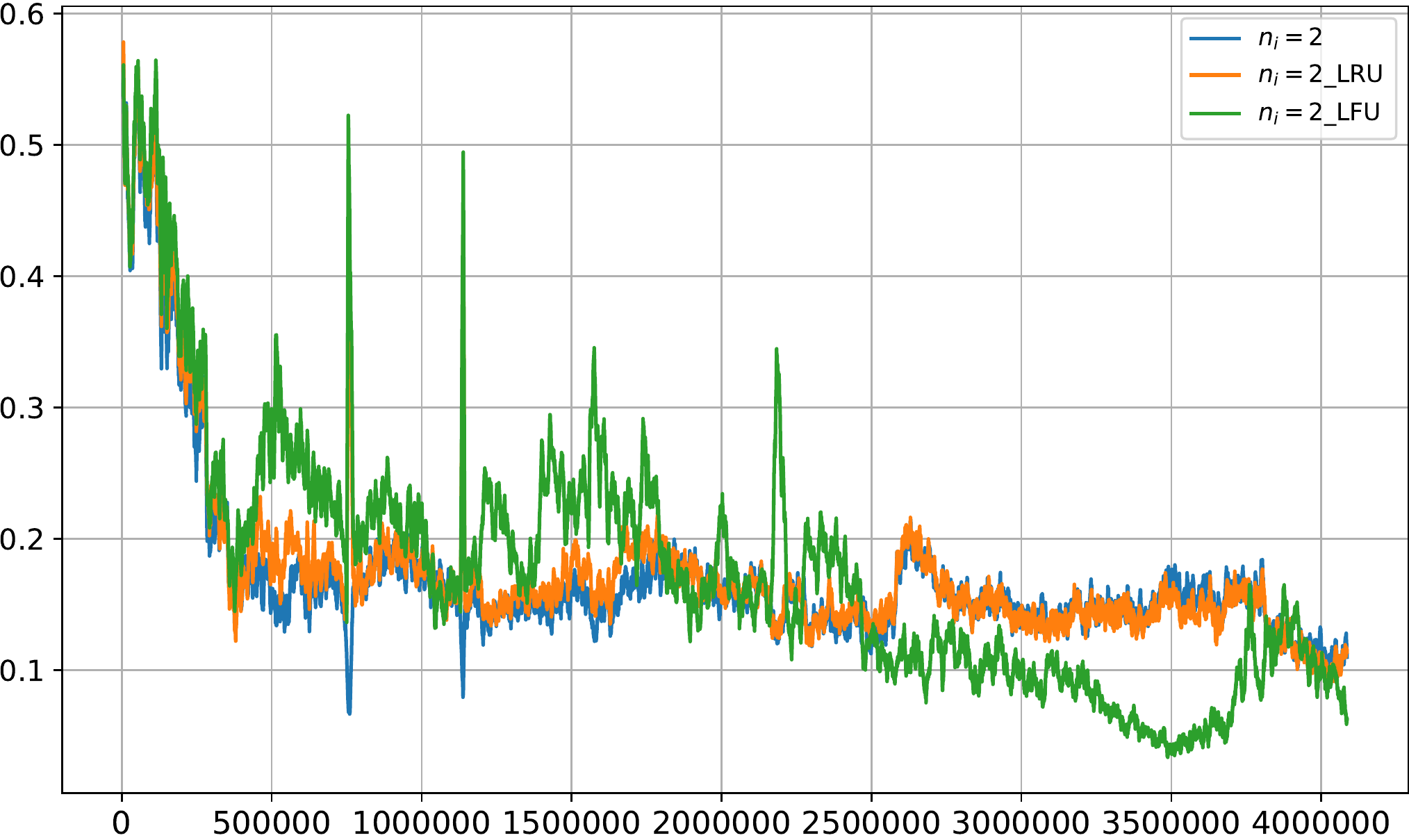}
  \caption{Netflix with $n_{i}=2$}
  \label{NF_using_ni_2}
\end{subfigure}
\qquad 
\begin{subfigure}[t]{0.5\textwidth}
  \includegraphics[width=\linewidth]{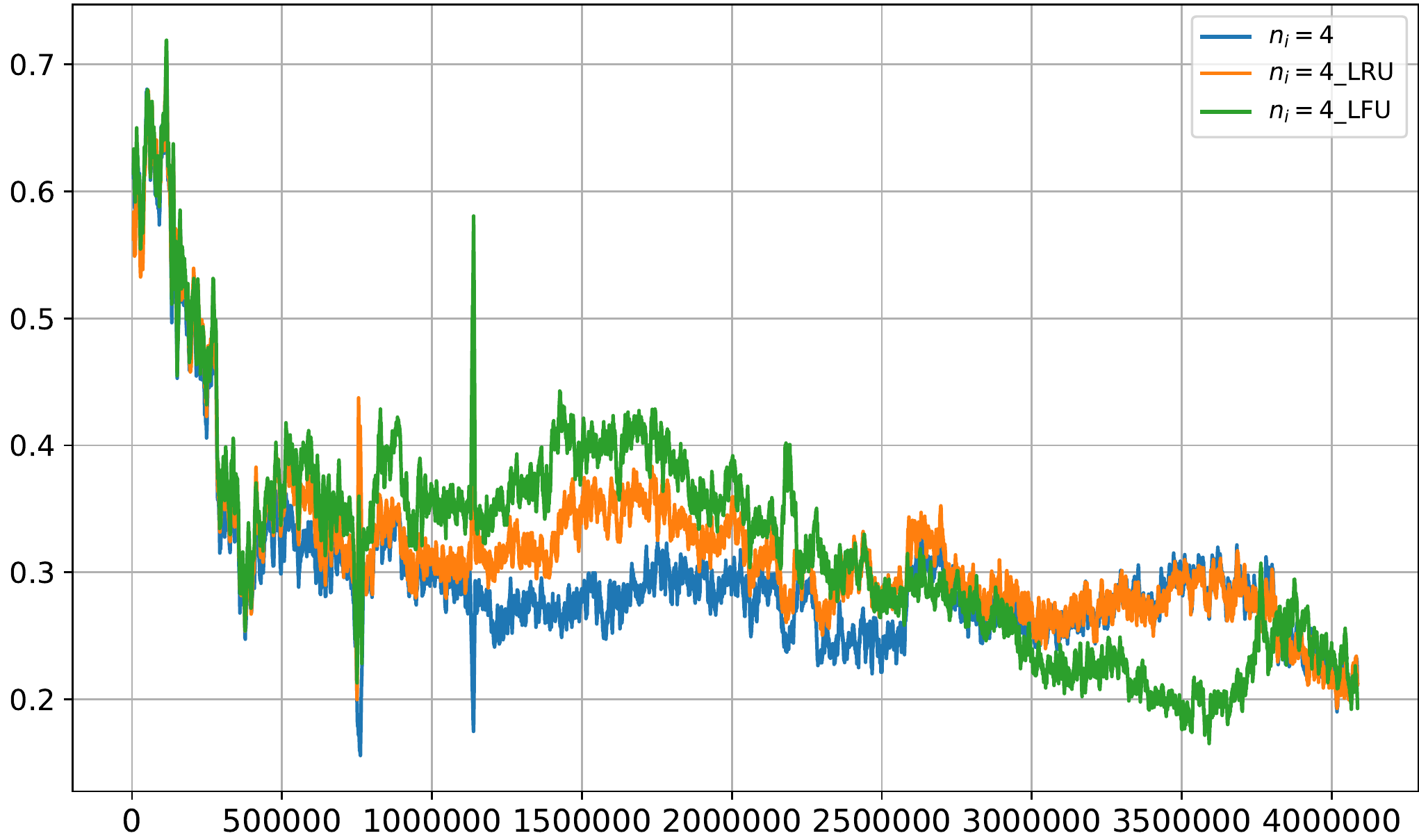}
  \caption{Netflix with $n_{i}=4$}
  \label{NF_using_ni_4}
\end{subfigure}
~ 
\begin{subfigure}[t]{0.5\textwidth}
  \includegraphics[width=\linewidth]{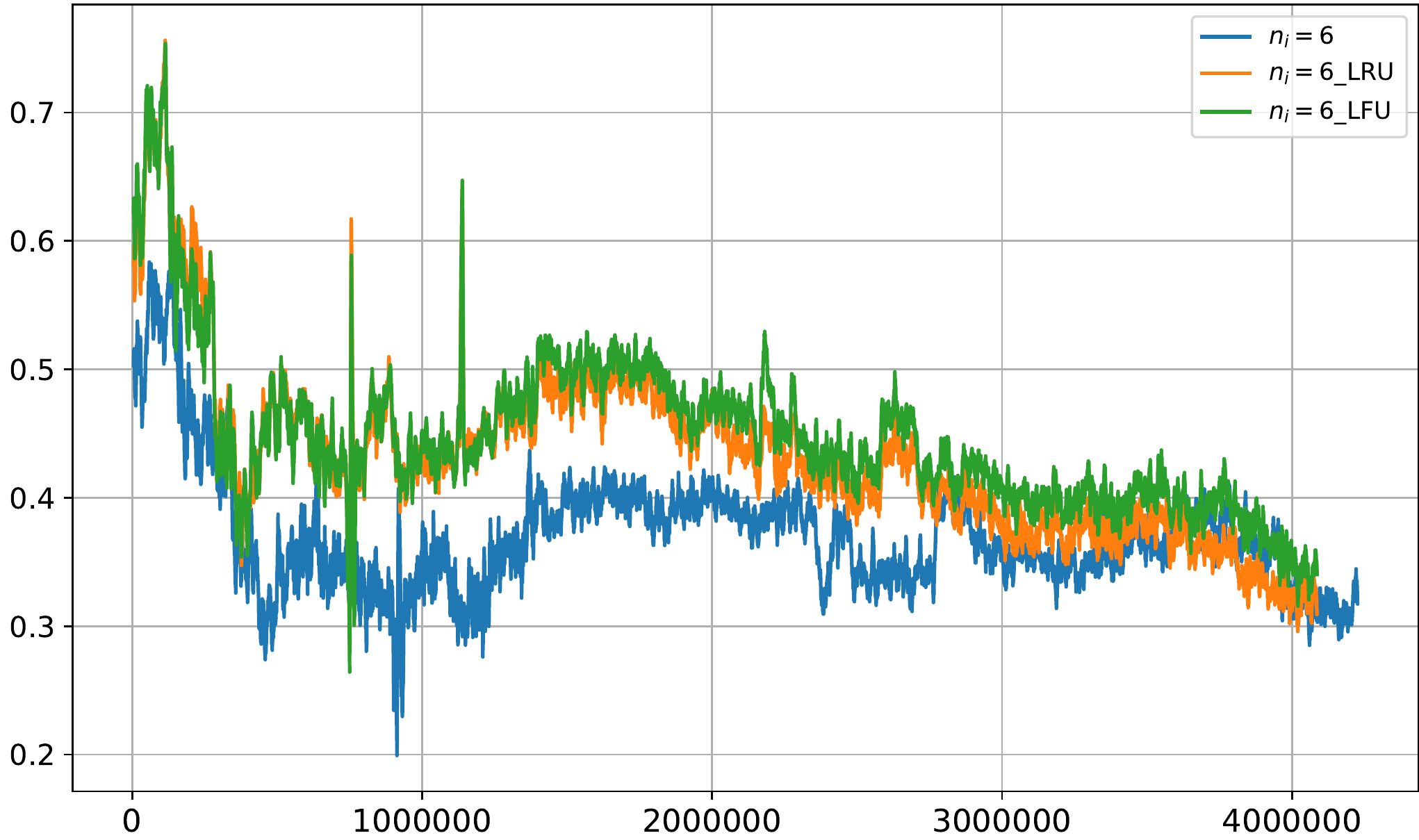}
  \caption{Netflix with $n_{i}=6$}
  \label{NF_using_ni_6}
\end{subfigure}
\caption{Comparison of the effect of LFU and LRU on Recall@10 for ISGD (central) and DISGD for the different $n_i$ values}
\label{fig:images_DISGD_closer_look}
\end{figure}

A closer look comparing the recall performance with and without using forgetting techniques using different $n_i$ configurations is plotted in Figure \ref{fig:images_DISGD_closer_look}. It is clear that LRU performs better than LFU while LFU's recall sometimes drops below central and LRU and this behavior can be observed in  figures~\ref{NF_using_ni_2} and~\ref{NF_using_ni_4} with $n_i=2,4$ for Netflix dataset. This can be explained by the aggressive forgetting parameters tuning applied on LFU. So, there is a trade-off between gaining performance with flushing obsolete data but it is up to limit after that the accuracy starts to drop. 

\begin{figure}[h!]
     \begin{subfigure}[t]{0.5\textwidth}
         \includegraphics[width=\linewidth]{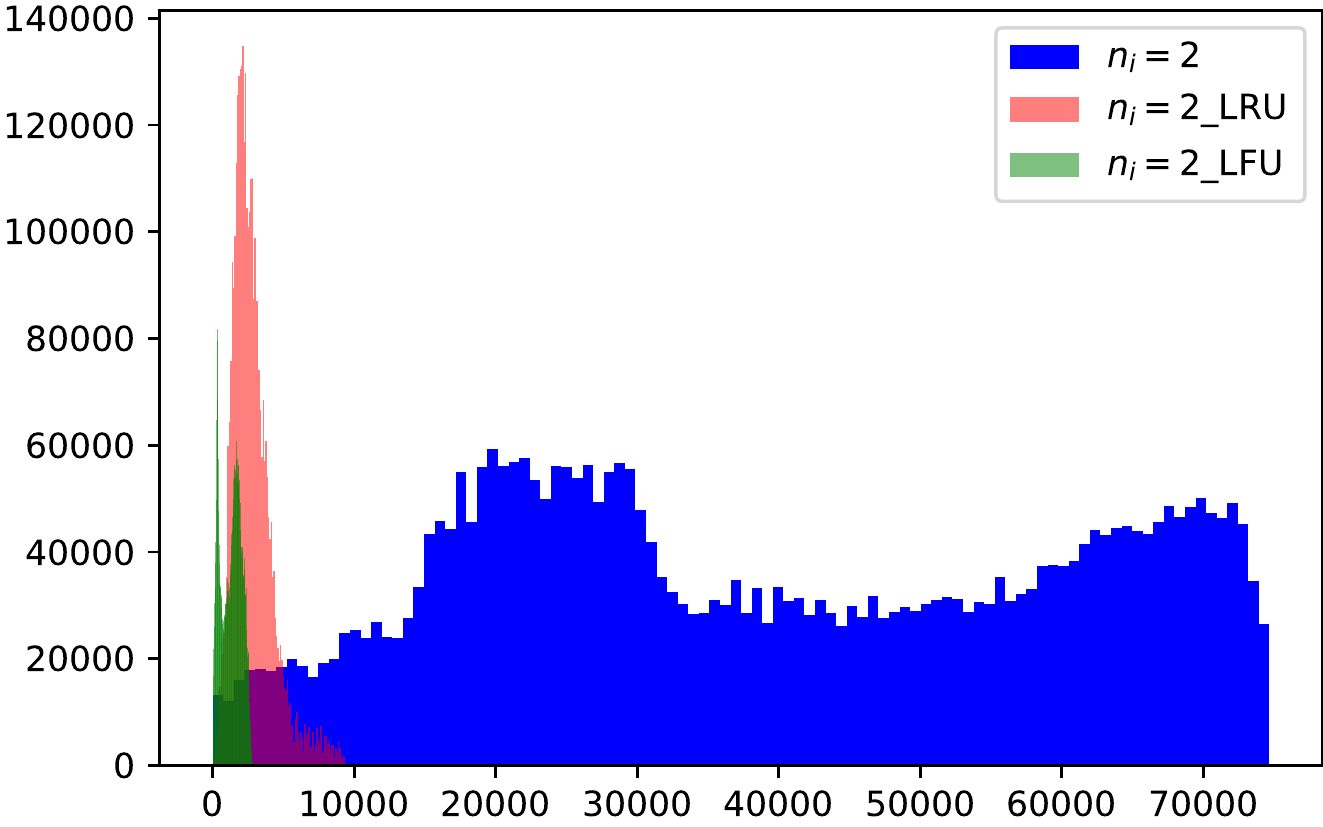}
         \caption{Users' frequency distribution for $n_{i}=2$}
         \label{ISGD_ML_ni_2_user}
     \end{subfigure}
~
     \begin{subfigure}[t]{0.5\textwidth}
         \includegraphics[width=\linewidth]{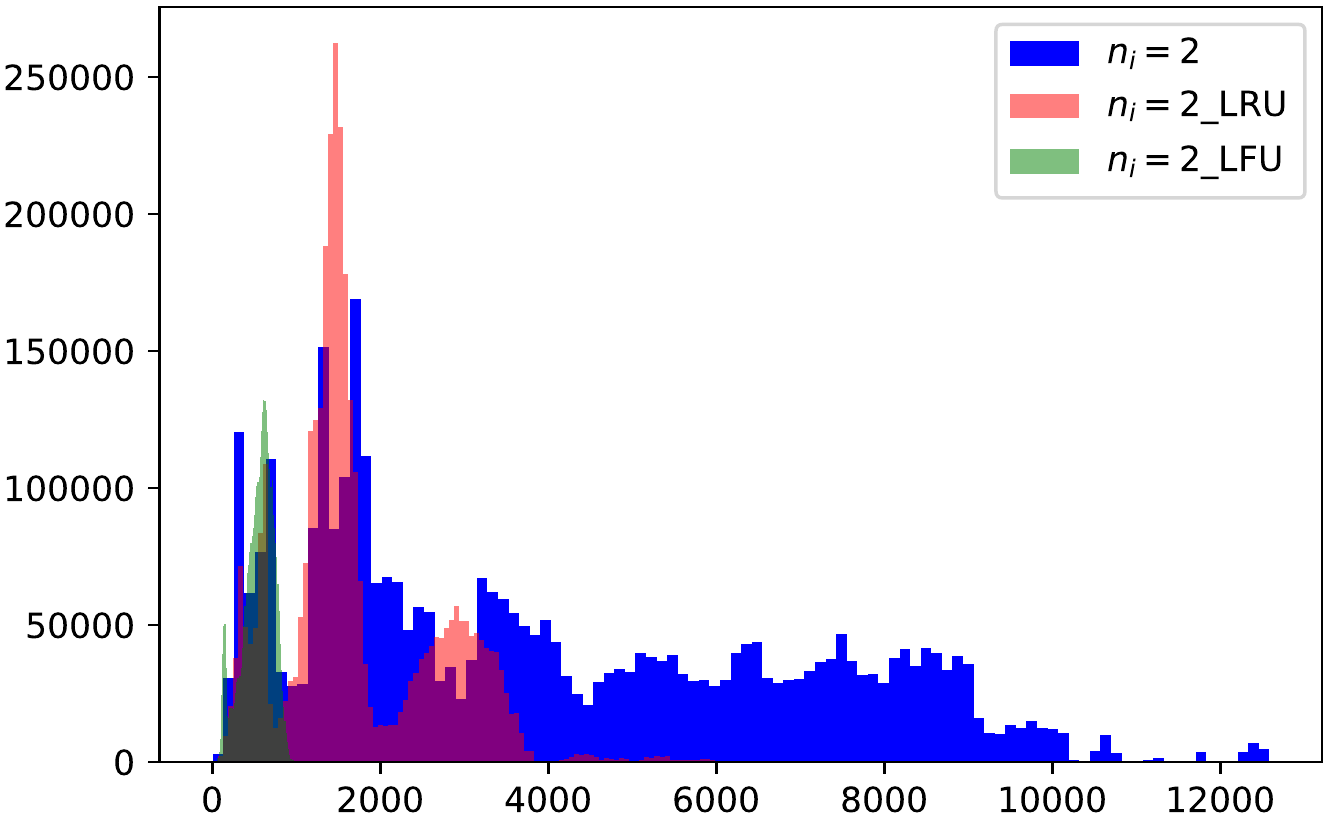}
         \caption{Items' frequency distribution for $n_{i}=2$}
         \label{ISGD_ML_ni_2_item}
     \end{subfigure}
      \caption{The effect of applying forgetting techniques on memory distribution for MovieLens data set}
  \label{Movielens_recall_forgetting_memory_dist}
     \end{figure}
     \begin{figure}[h!]\ContinuedFloat
      \begin{subfigure}[t]{0.5\textwidth}
         \includegraphics[width=\linewidth]{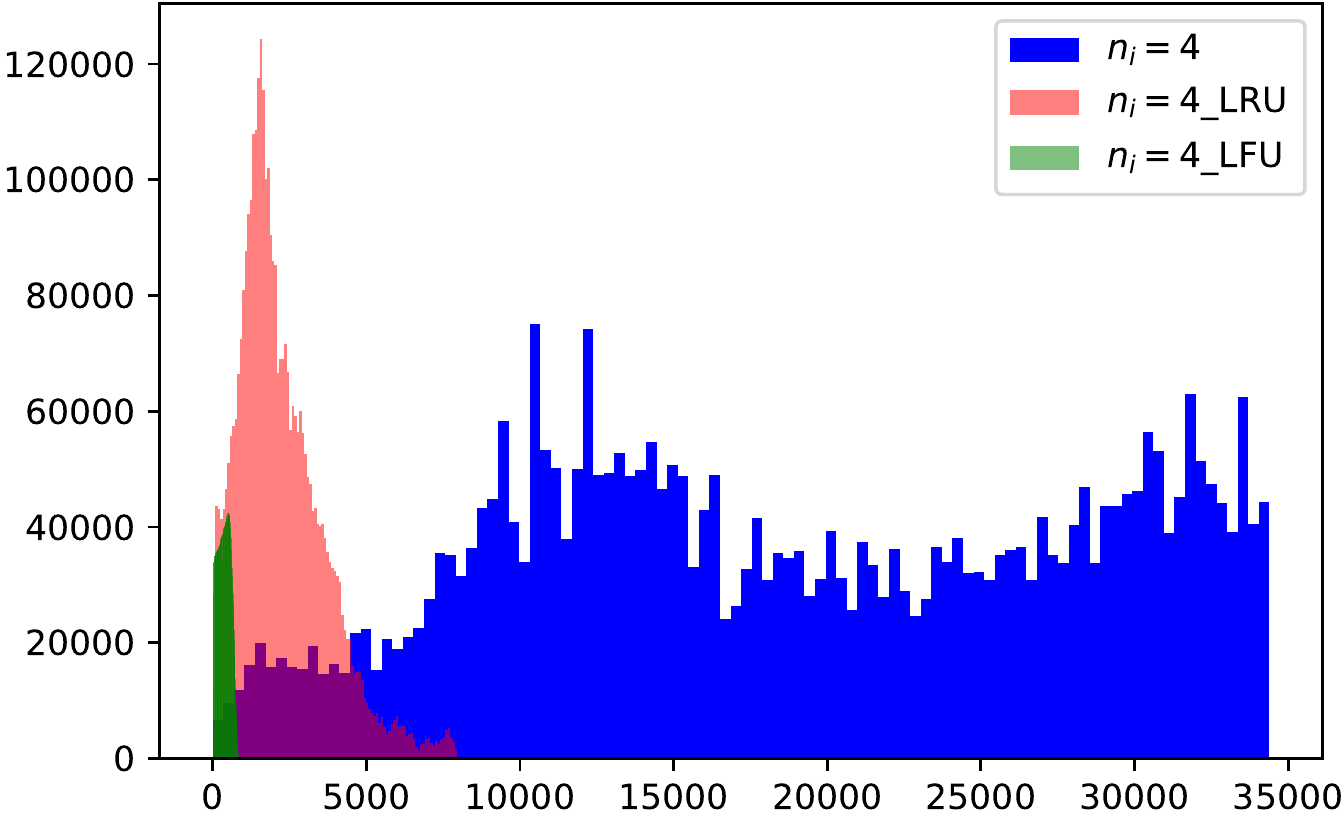}
         \caption{Users' frequency  distribution for $n_{i}=4$}
         \label{ISGD_ML_ni_4_user}
     \end{subfigure}
     ~
     \begin{subfigure}[t]{0.5\textwidth}
         \includegraphics[width=\linewidth]{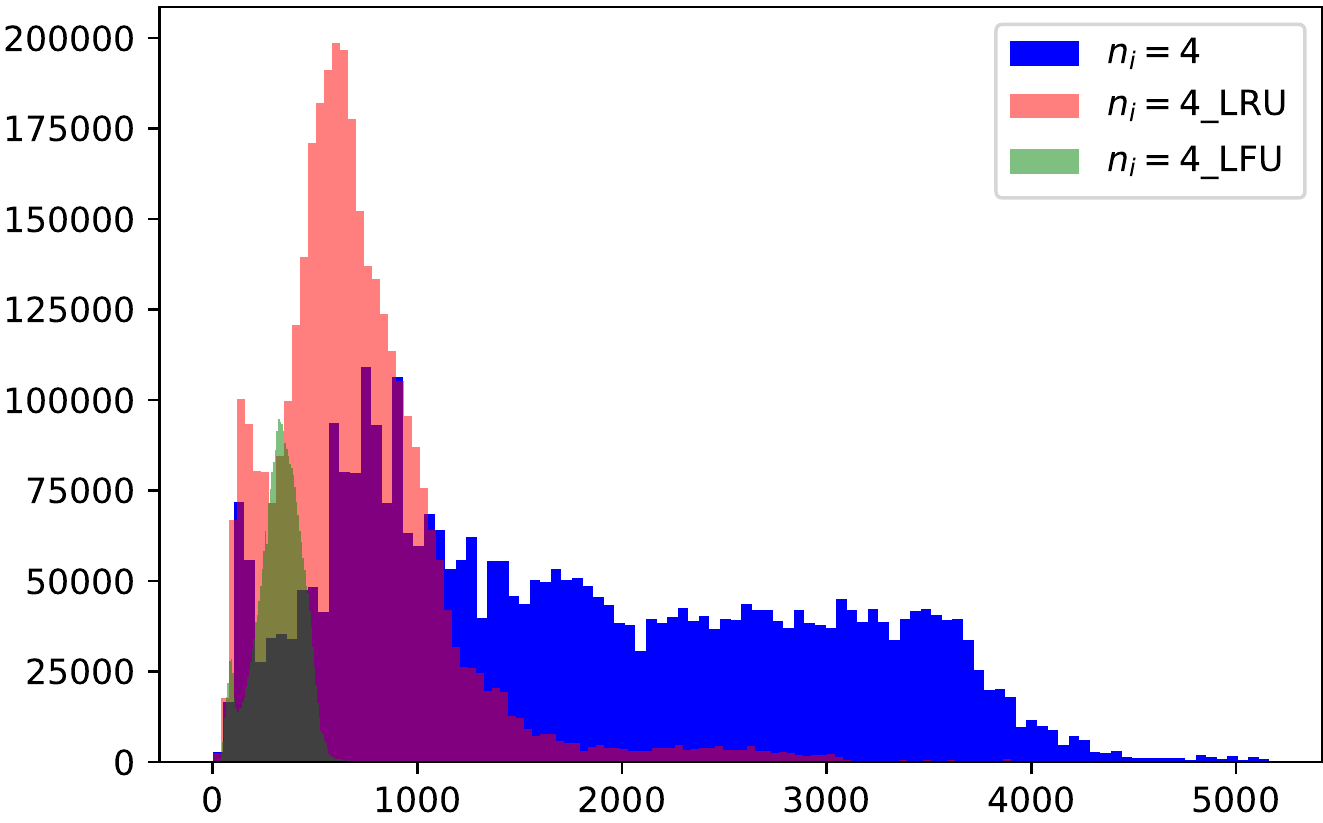}
         \caption{Items' frequency distribution for $n_{i}=4$}
         \label{ISGD_ML_ni_4_item}
     \end{subfigure}
\qquad
     \begin{subfigure}[t]{0.5\textwidth}
         \centering
         \includegraphics[width=\linewidth]{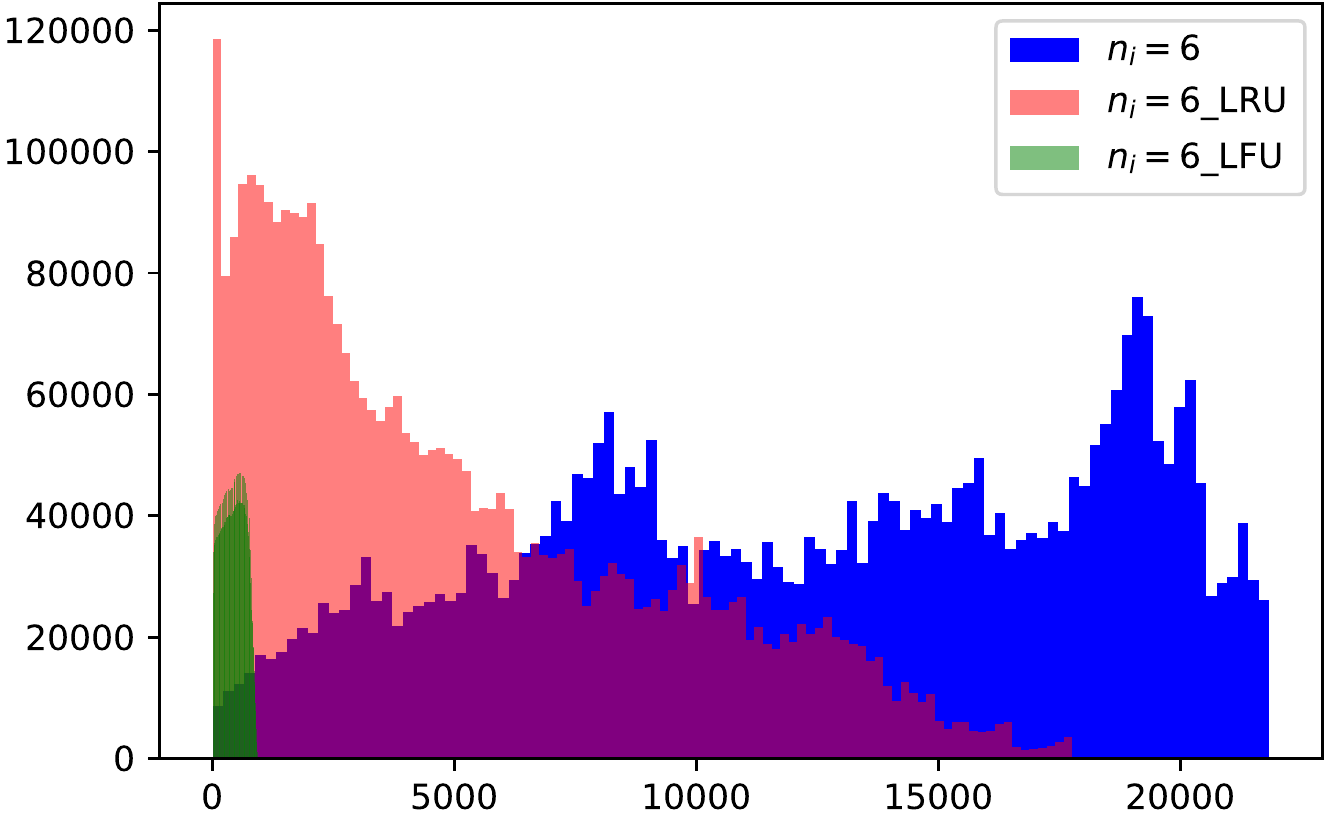}
         \caption{Users' frequency distribution for $n_{i}=6$}
         \label{ISGD_ML_ni_6_user}
     \end{subfigure}
     ~
      \begin{subfigure}[t]{0.5\textwidth}
         \centering
         \includegraphics[width=\linewidth]{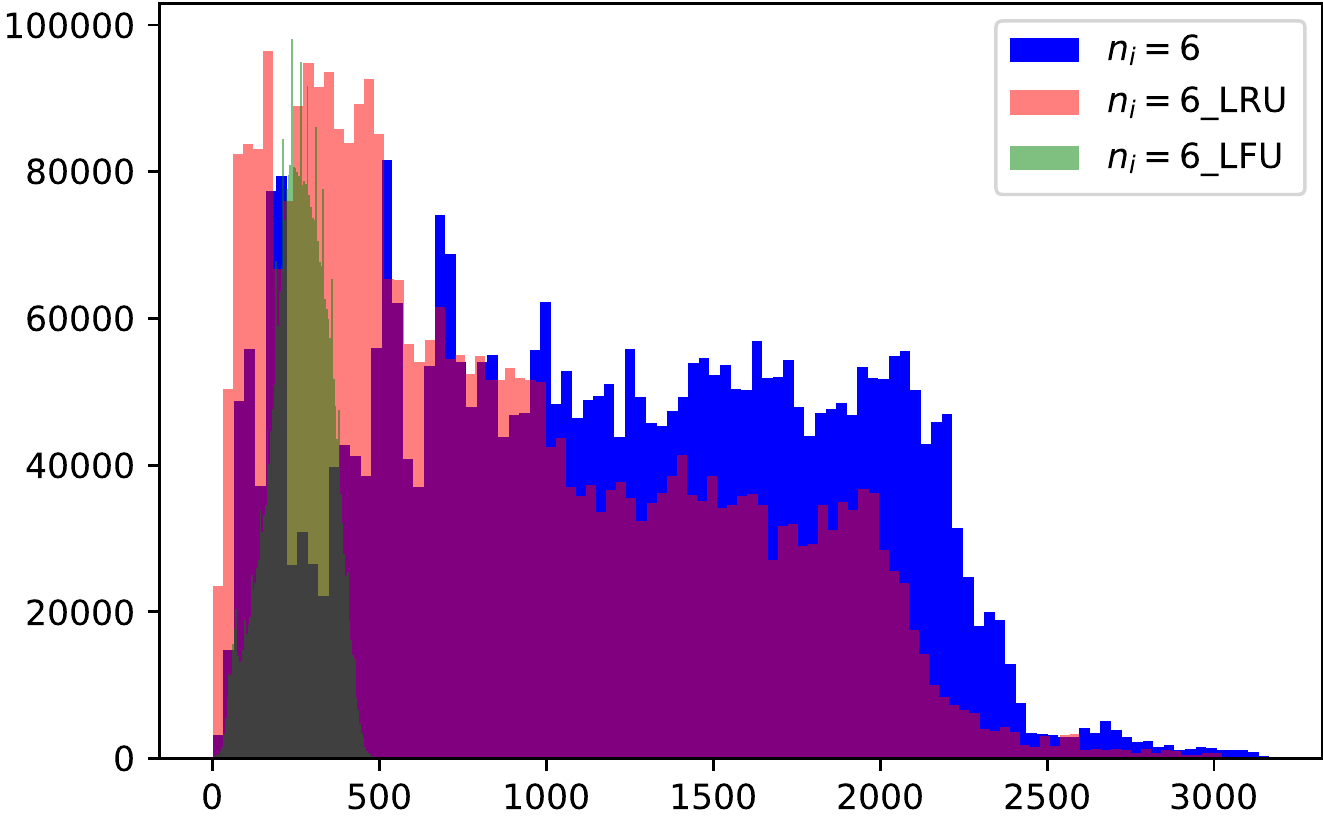}
         \caption{Items' frequency distribution for $n_{i}=6$}
         \label{ISGD_ML_ni_6_item}
     \end{subfigure}
     \caption{The effect of applying forgetting techniques on memory distribution for MovieLens data set}
   \label{Movielens_recall_forgetting_memory_dist}
    
\end{figure}

\begin{figure}[h!]
    \centering
	\includegraphics[width=\linewidth]{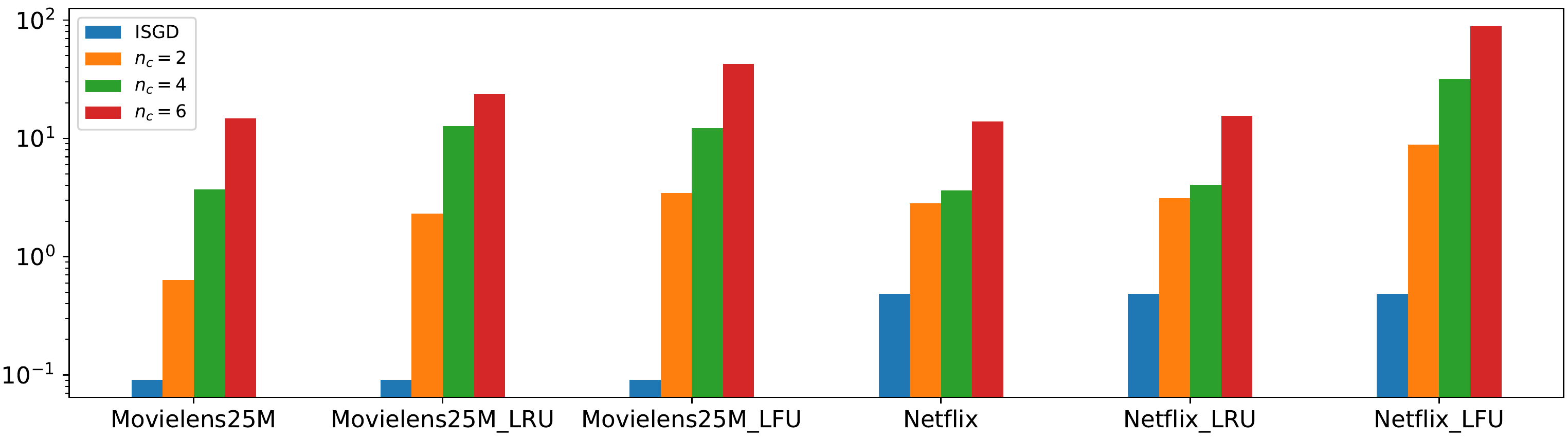}
	\caption{Throughput comparison for D/ISGD with different $n_i$ with and without forgetting technique applied on Movielens, Netflix datasets }
	\label{Pipeline_throughput}
\end{figure}

As forgetting techniques are mainly employed to control the memory consumption growth, Figure~\ref{Movielens_recall_forgetting_memory_dist} shows the effect of LRU and LFU on the memory distribution for the MovieLens dataset. For $n_i=2$,  the mean of the users' distribution falls down to $0.071$  and $0.025$ of the $n_i=2$ users' state mean for LRU and LFU respectively. Likewise, the items' state drops to $0.428$ and $0.13$ of $n_i=2$ items' state mean for LRU and LFU. For replication factor $n_i=4$, the mean of the users' distribution drops to $0.118$  and $0.019$ of the users' state mean for LRU and LFU respectively whereas the items' state drops to $0.39$ and $0.18$ of items' state mean. More gain with $n_i=6$ as the mean of the users' distribution comes down to $0.42$  and $0.034$ of the users' state mean for LRU and LFU respectively and $0.21$ of $n_i=6$ items' state mean.

Figure~\ref{Pipeline_throughput} shows the throughput of centralized ISGD against the DISGD under all configurations discussed for $n_i$ with and without applying forgetting techniques. The DISGD throughput for Movielens has been improved around $38$ times compared to ISGD for $n_i=2$ with LFU. The best improvement for $n_i=4$ occurs with LRU where it has been enhanced $139$ times over ISGD. The best throughput recorded for $n_i=6$ happens with LFU where the throughput has been improved $461$ times against ISGD throughput. The same enhancement behavior can be captured with the Netflix dataset where the best throughput has been enhanced $25,~86$ and $241$ times over ISGD throughput for $n_i=2$, $n_i=4$, and $n_i=6$ with LFU respectively.




\subsubsection{DICS}
\label{section:vector:similarity:results}
In this section, we discuss the evaluation of the incremental Cosine similarity calculation in the central case (the baseline) compared to the distributed version obtained by applying our splitting and replication mechanism. The flow of experiments is similar to that of Section~\ref{sec:disgd:results}.

\begin{figure}[h!]
\begin{subfigure}[t]{0.5\textwidth}
  \includegraphics[width=\linewidth]{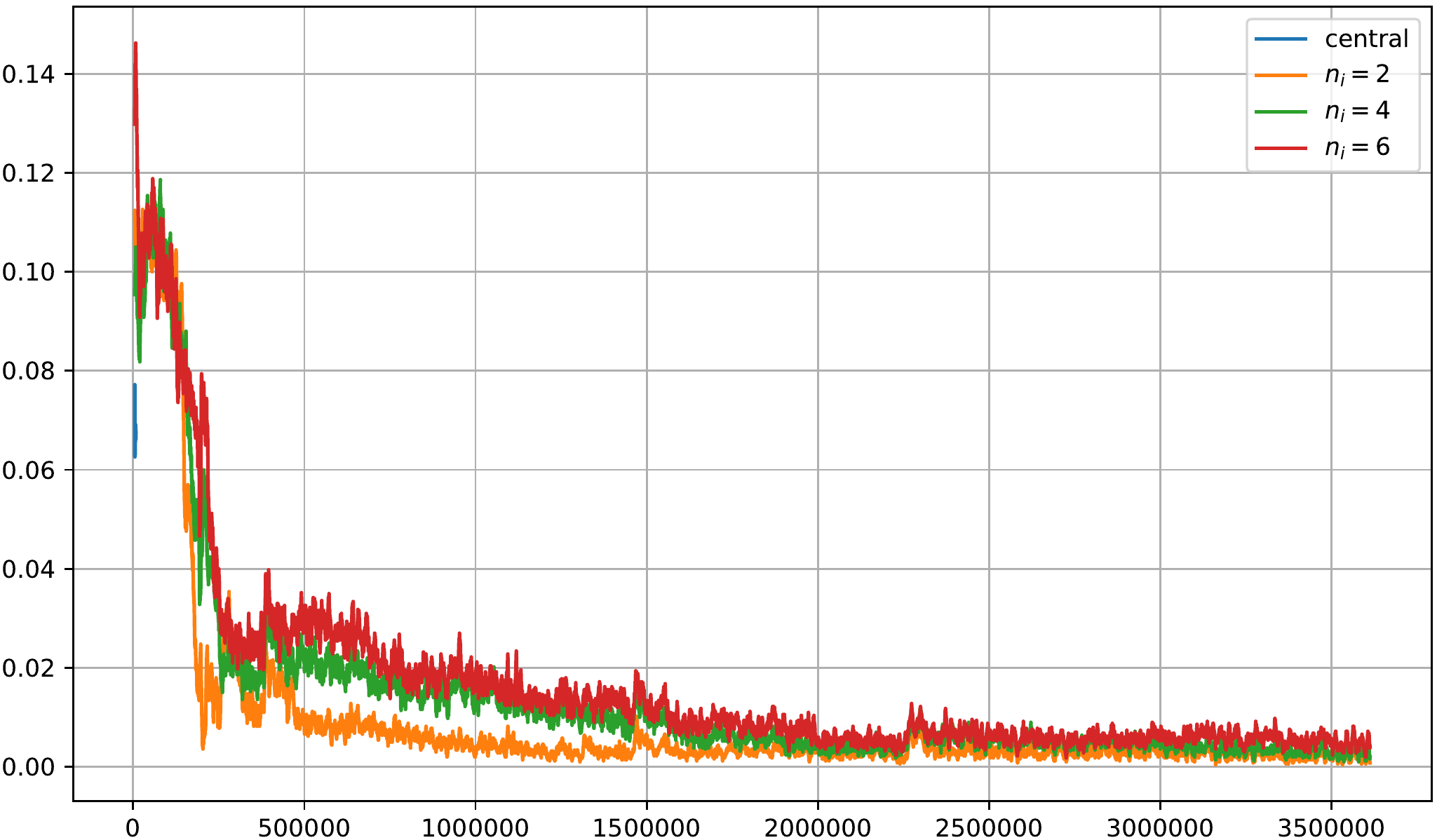}
  \caption{Movielens-25M}
  \label{fig:dtenrec:recall:no:forget:ML25}
\end{subfigure}
~
\begin{subfigure}[t]{0.5\textwidth}
  \includegraphics[width=\linewidth]{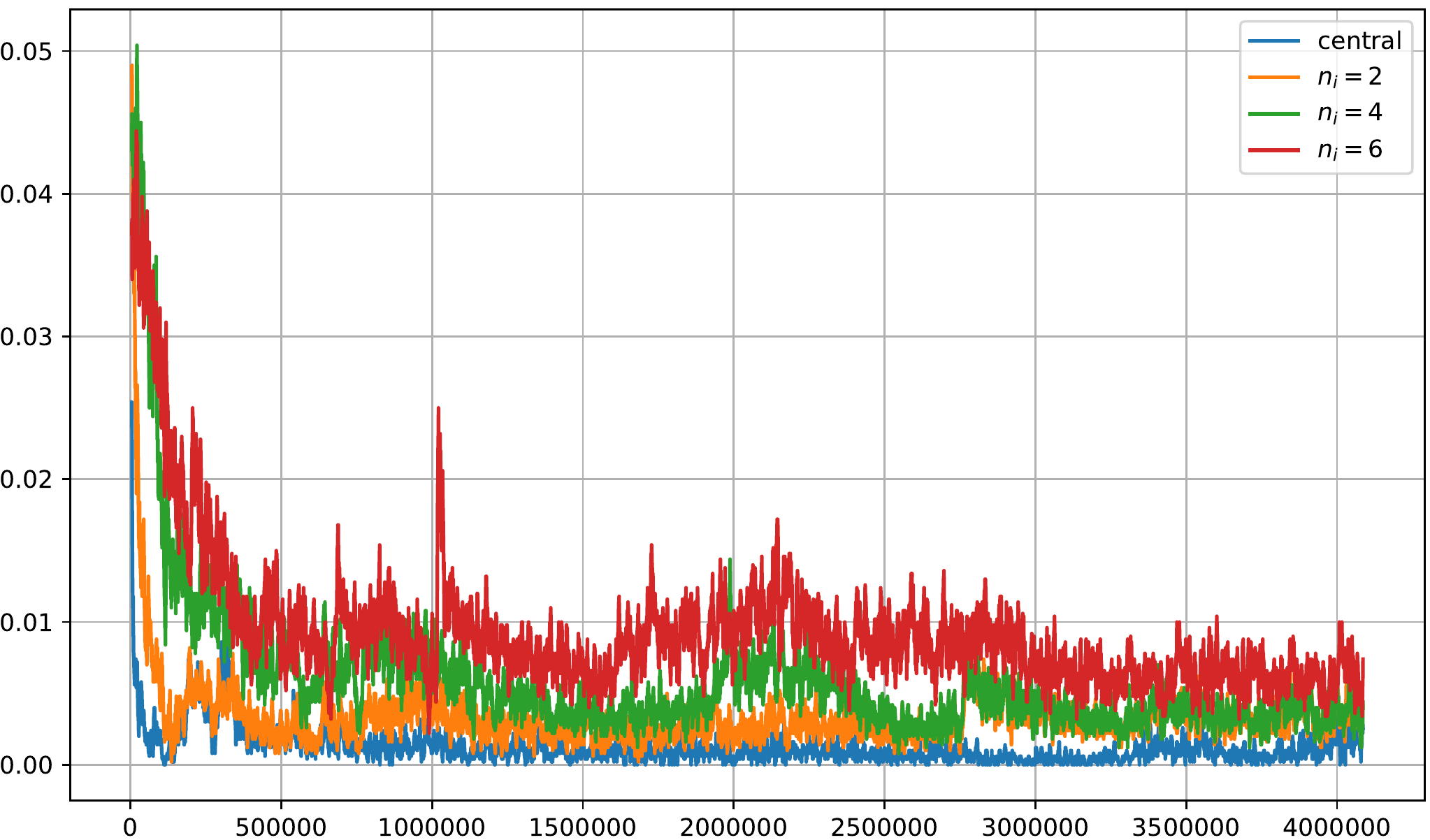}
  \caption{Netflix}
  \label{fig:dtenrec:recall:no:forget:NFlix}
\end{subfigure}
\caption{Moving average Recall@10 for Cosine similarity central/ distributed for the different $n_i$ values}
\label{fig:dtenrec:recall:no:forget}
\end{figure}

Similar to DISGD, the application of the splitting and replication mechanism makes improvements for the recall. This is shown in Figure~\ref{fig:dtenrec:recall:no:forget}. However, the scale of improvement is not the same as in DISGD. This is attributed to the nature of the recommendation approach. For the Movielens dataset (Figure~\ref{fig:dtenrec:recall:no:forget:ML25}), the distributed approach starts to have gains over the baseline with $n_i=6$. For the Netflix data set (Figure~\ref{fig:dtenrec:recall:no:forget:NFlix}), the gain is observed starting from replication factor $n_i=2$. In general, the average recall is enhanced from $3\%$ to more than $40\%$ depending on the replication factor $n_{i}$ and the dataset. Again,  it is demonstrated empirically that there is a positive correlation between the recall and $n_{i}$.

\begin{figure}[h!]
\begin{subfigure}[t]{0.5\textwidth}
  \includegraphics[width=\linewidth]{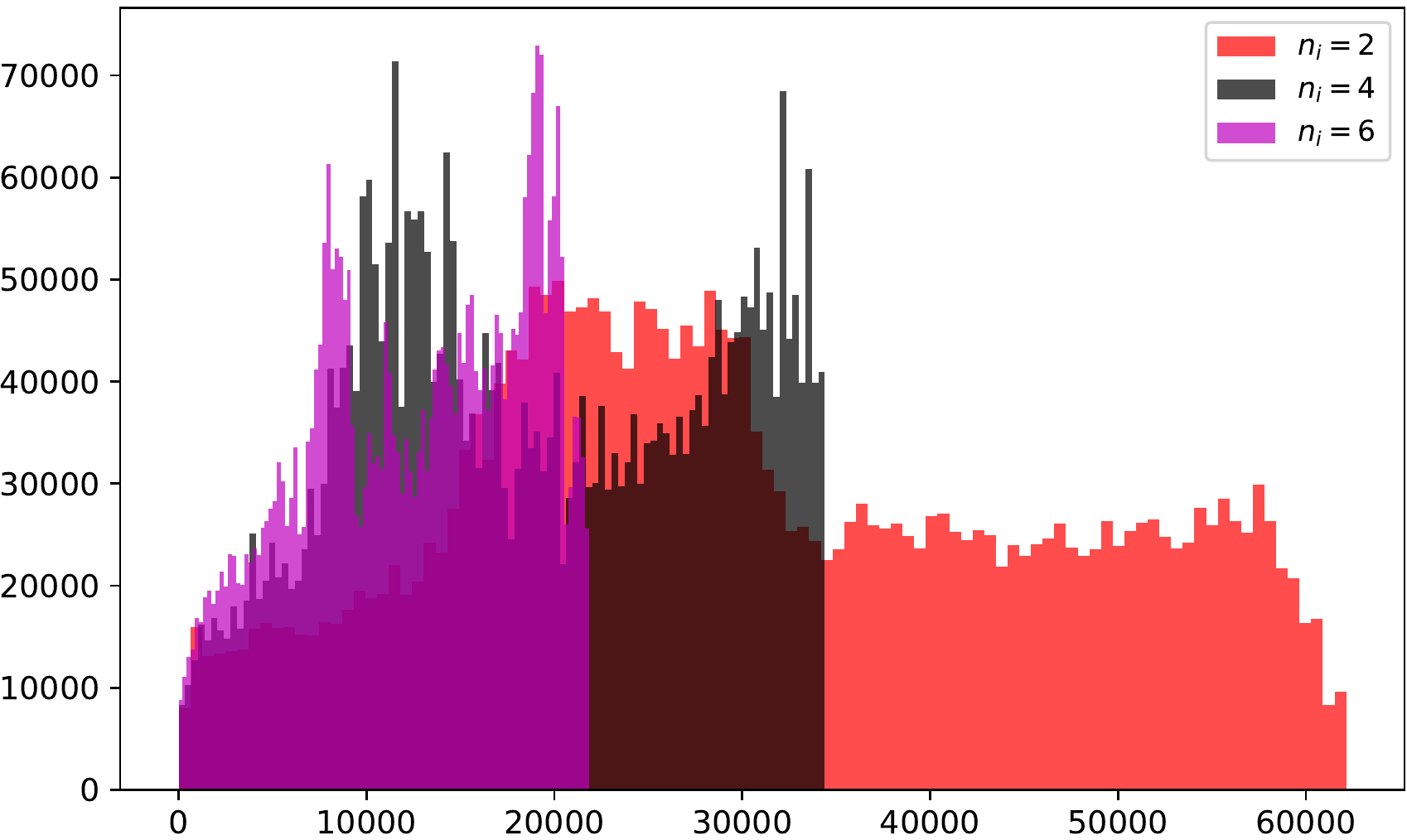}
  \caption{Movielens-25M users' state distribution}
  \label{fig:dtenrec:memory:no:forget:ML25:users}
\end{subfigure}
~
\begin{subfigure}[t]{0.5\textwidth}
  \includegraphics[width=\linewidth]{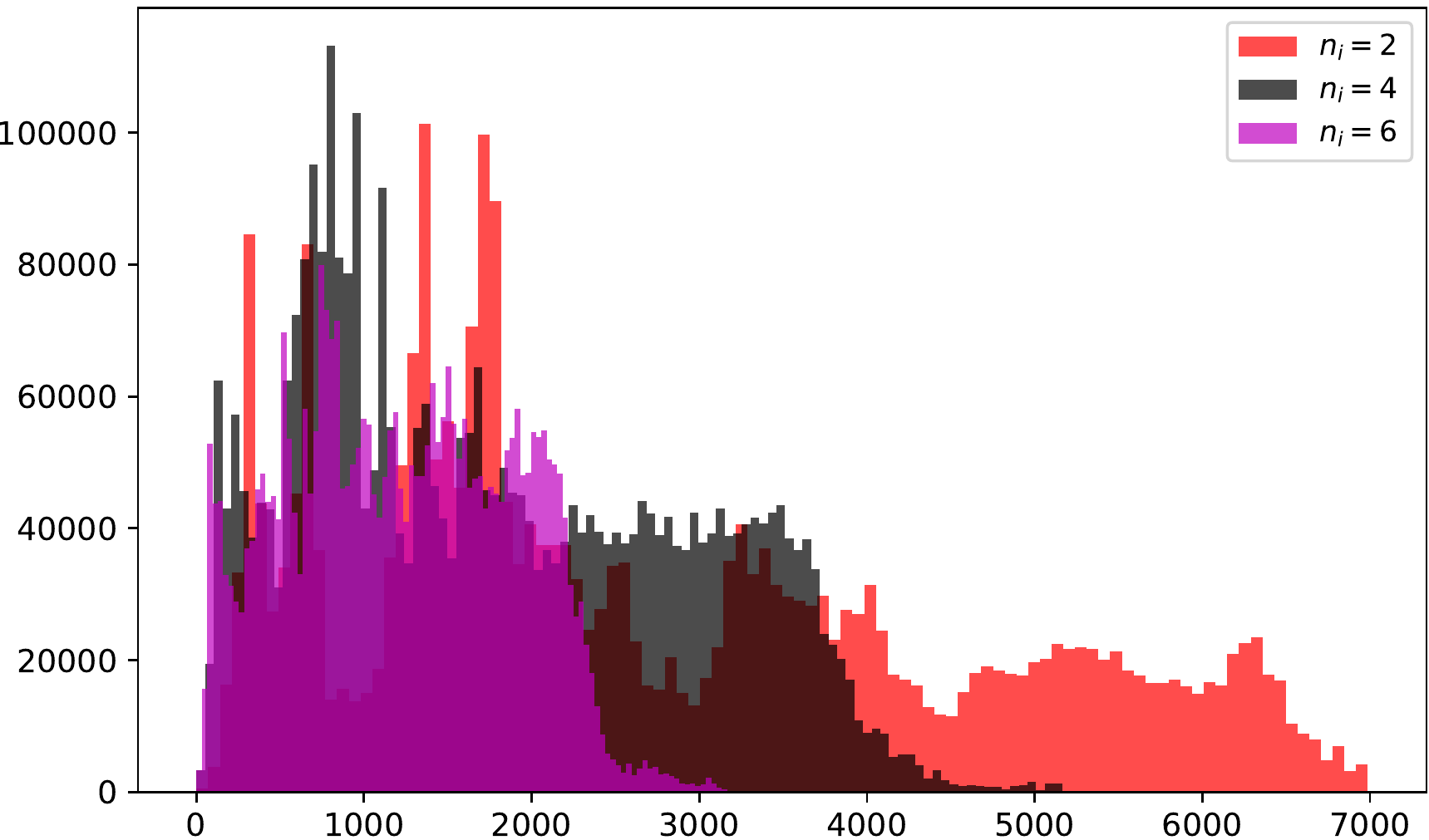}
  \caption{Movielens-25M items' state distribution}
\label{fig:dtenrec:memory:no:forget:ML25:items}
\end{subfigure}
\qquad
\begin{subfigure}[t]{0.5\textwidth}
  \includegraphics[width=\linewidth]{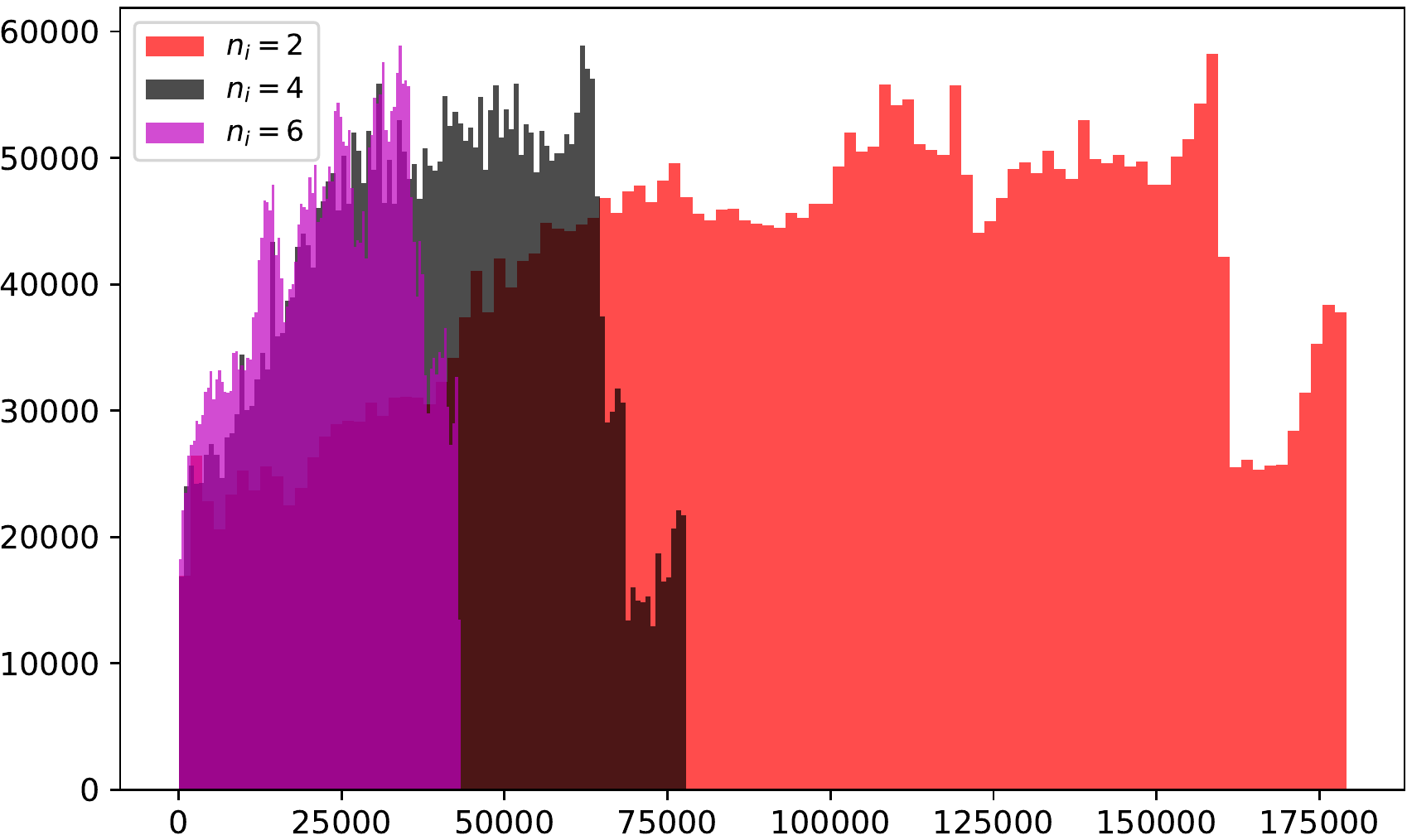}
  \caption{Netflix users' state distribution}
\label{fig:dtenrec:memory:no:forget:NFlix:users}
\end{subfigure}
~
\begin{subfigure}[t]{0.5\textwidth}
  \includegraphics[width=\linewidth]{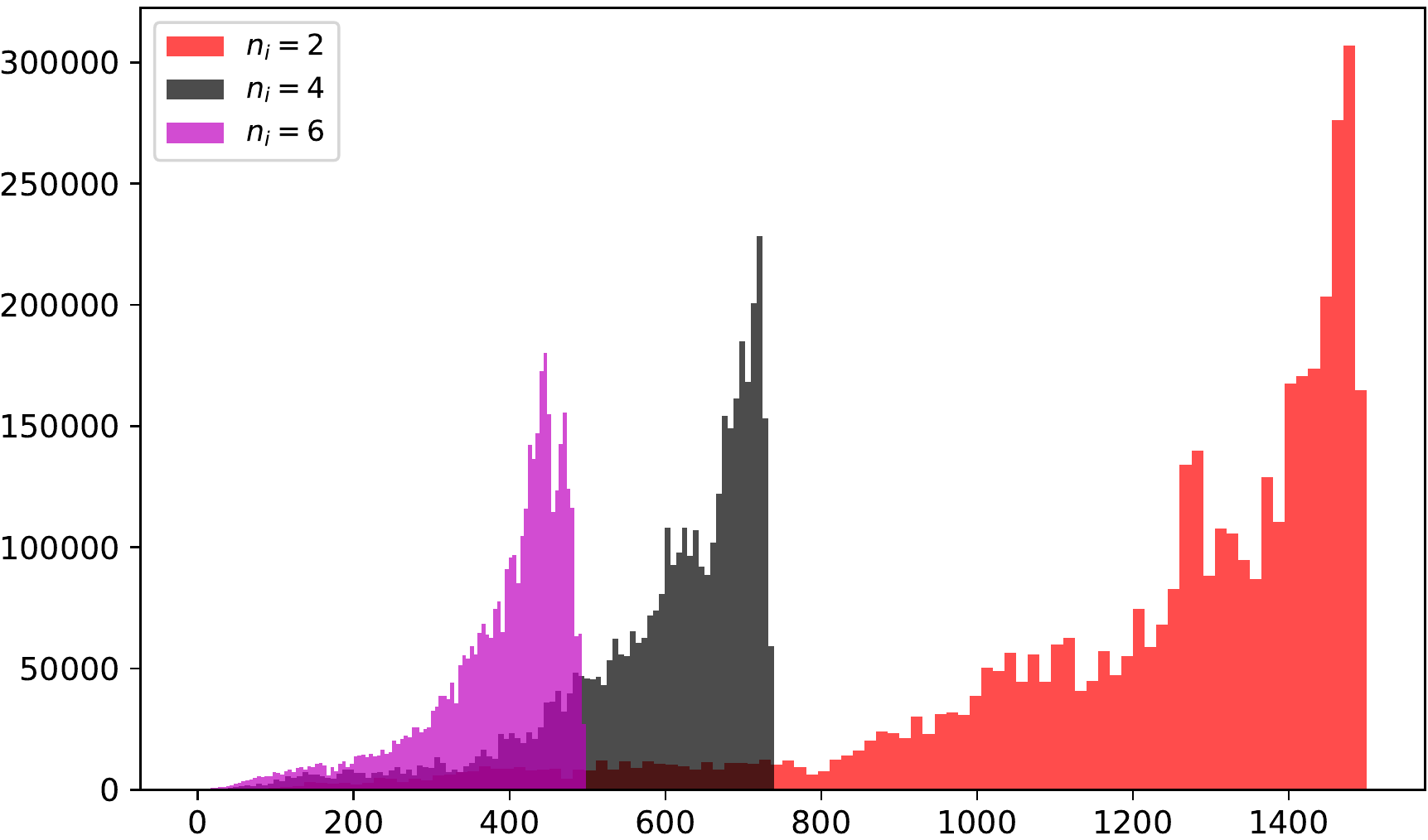}
  \caption{Netflix items' state distribution}
\label{fig:dtenrec:memory:no:forget:NFlix:items}
\end{subfigure}
\caption{Memory distribution for Cosine similarity central and distributed with the different $n_{i}$ values. X-axis shows the size number of entries and Y-axis shows the frequency.}
\label{fig:dtenrec:memory:no:forget}
\end{figure}

The central version with the ML25M dataset has been left to process the data for $11$ days and it has just processed $8356$ records so the job was killed. Applying our splitting and replication mechanism, the jobs completed successfully. As a consequence of incomplete processing of data for the central configuration, we do not report memory consumption and other aspects for the rest of the experiments.

Moving to study how memory consumption distribution looks like, Figure \ref{fig:dtenrec:memory:no:forget} plots these distributions for users and items vectors for the Movielens and Netflix data sets respectively. For the Movielens dataset, the user's state size (Figure~\ref{fig:dtenrec:memory:no:forget:ML25:users}) is considerably reduced as $n_i$ increases. The mean of users' state distribution dropped from $30897.85$ for $n_i=2$ to $18928.42$ for $n_i=4$ and to $12530.09$ for $n_i=6$. The mean of items' state (Figure~\ref{fig:dtenrec:memory:no:forget:ML25:items}) goes from $2765.174$ to $1813.77$ and to $1234.08$ for $n_i=2, 4,6$ respectively. The same reduction behavior can be observed for the Netflix dataset. The mean of users' state distribution (Figure~\ref{fig:dtenrec:memory:no:forget:NFlix:users}) drops from $96115.62$ to $38983.9$ and to $22787.59$ whereas the items' state distributions (Figure~\ref{fig:dtenrec:memory:no:forget:NFlix:items}) means are reduced from $1227.12$ to $591.92$ and to $386.11$ for $n_i = 2,4,6$ respectively.

\begin{figure}[htb]
\begin{subfigure}[t]{0.5\textwidth}
  \includegraphics[width=\linewidth]{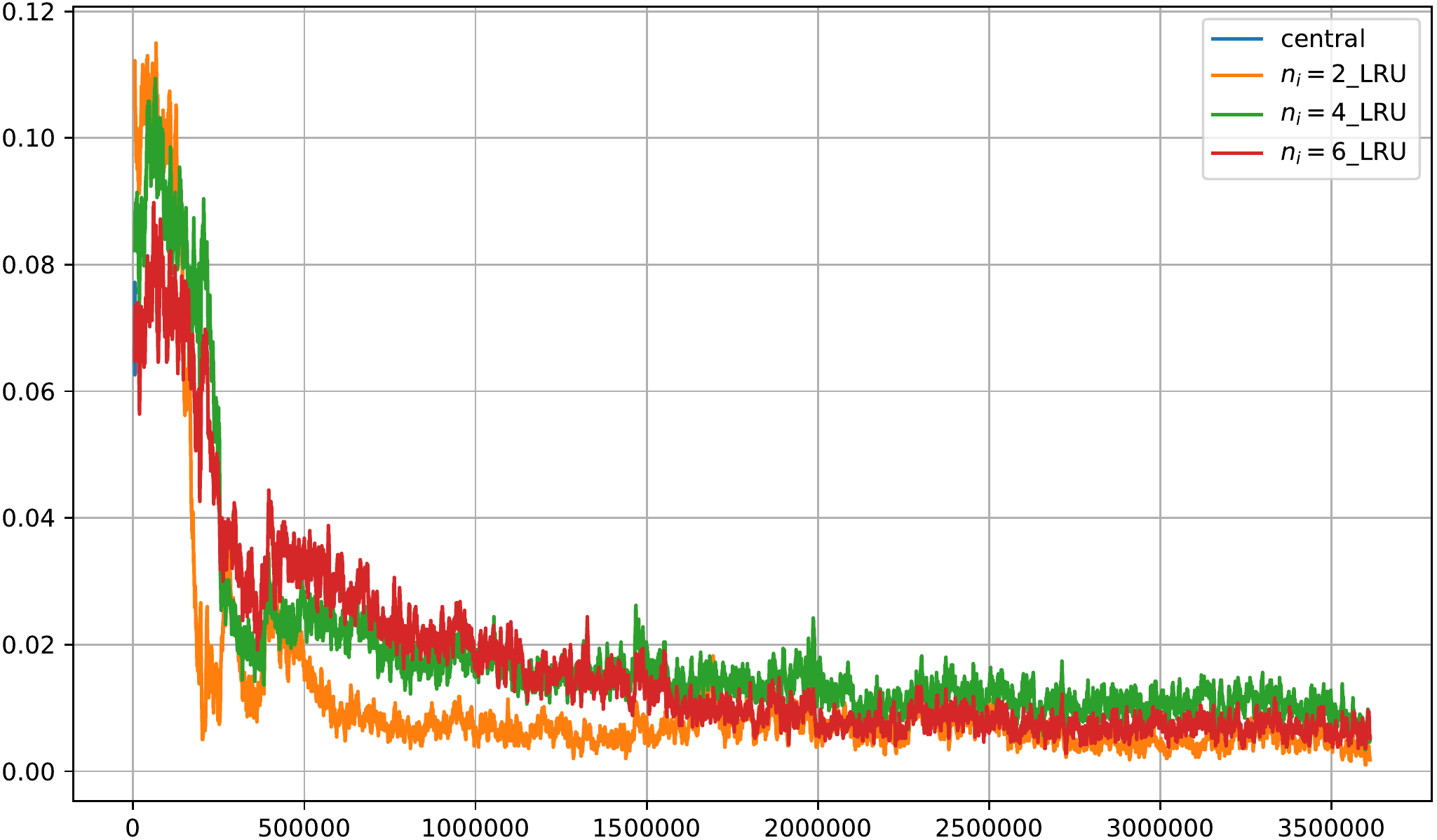}
  \caption{Movielens-25M with LRU Forgetting}
\end{subfigure}
~
\begin{subfigure}[t]{0.5\textwidth}
  \includegraphics[width=\linewidth]{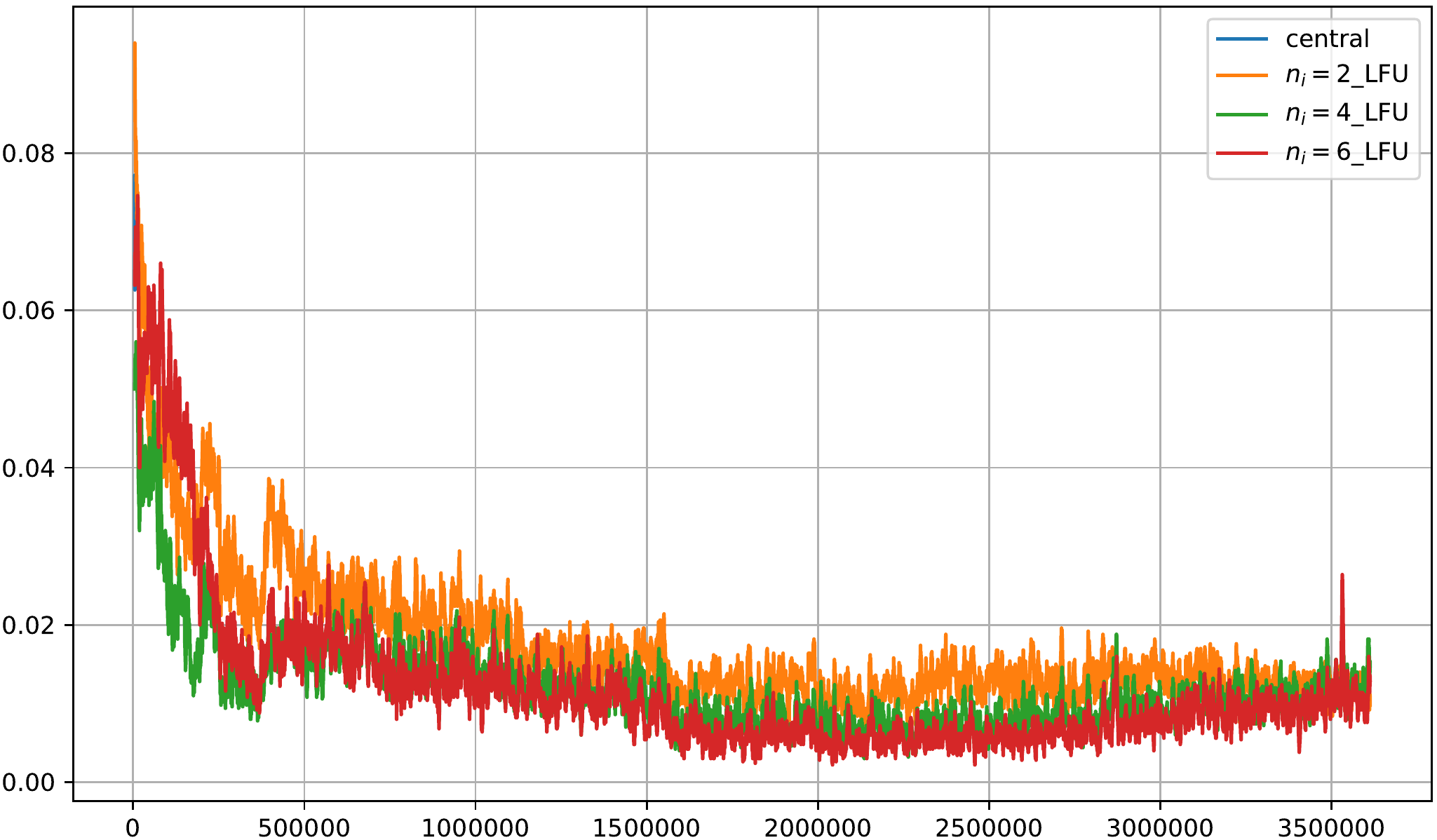}
  \caption{Movielens-25M with LFU Forgetting}
\end{subfigure}
\qquad
\begin{subfigure}[t]{0.5\textwidth}
  \includegraphics[width=\linewidth]{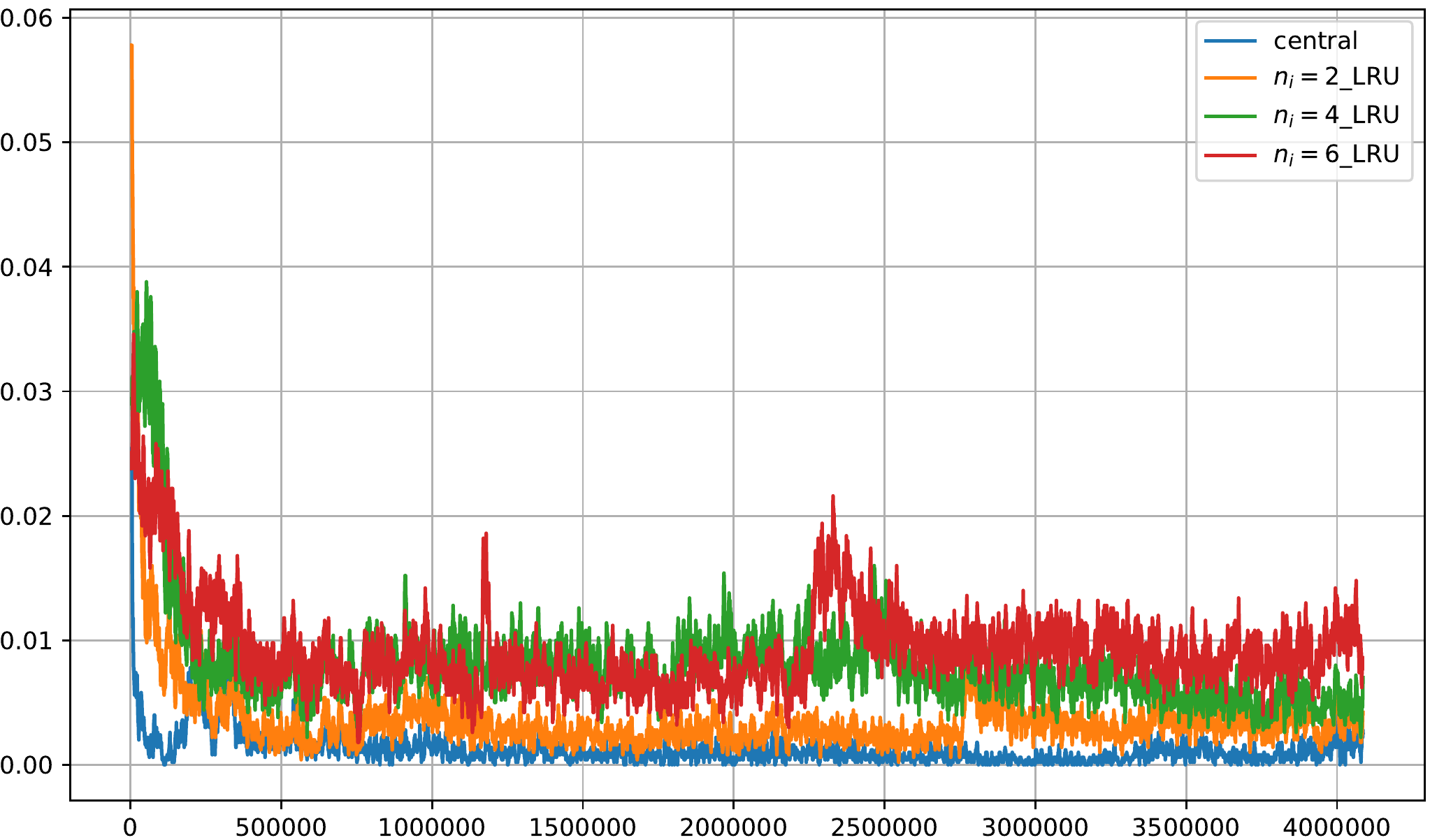}
  \caption{Netflix with LRU Forgetting}
\end{subfigure}
~
\begin{subfigure}[t]{0.5\textwidth}
  \includegraphics[width=\linewidth]{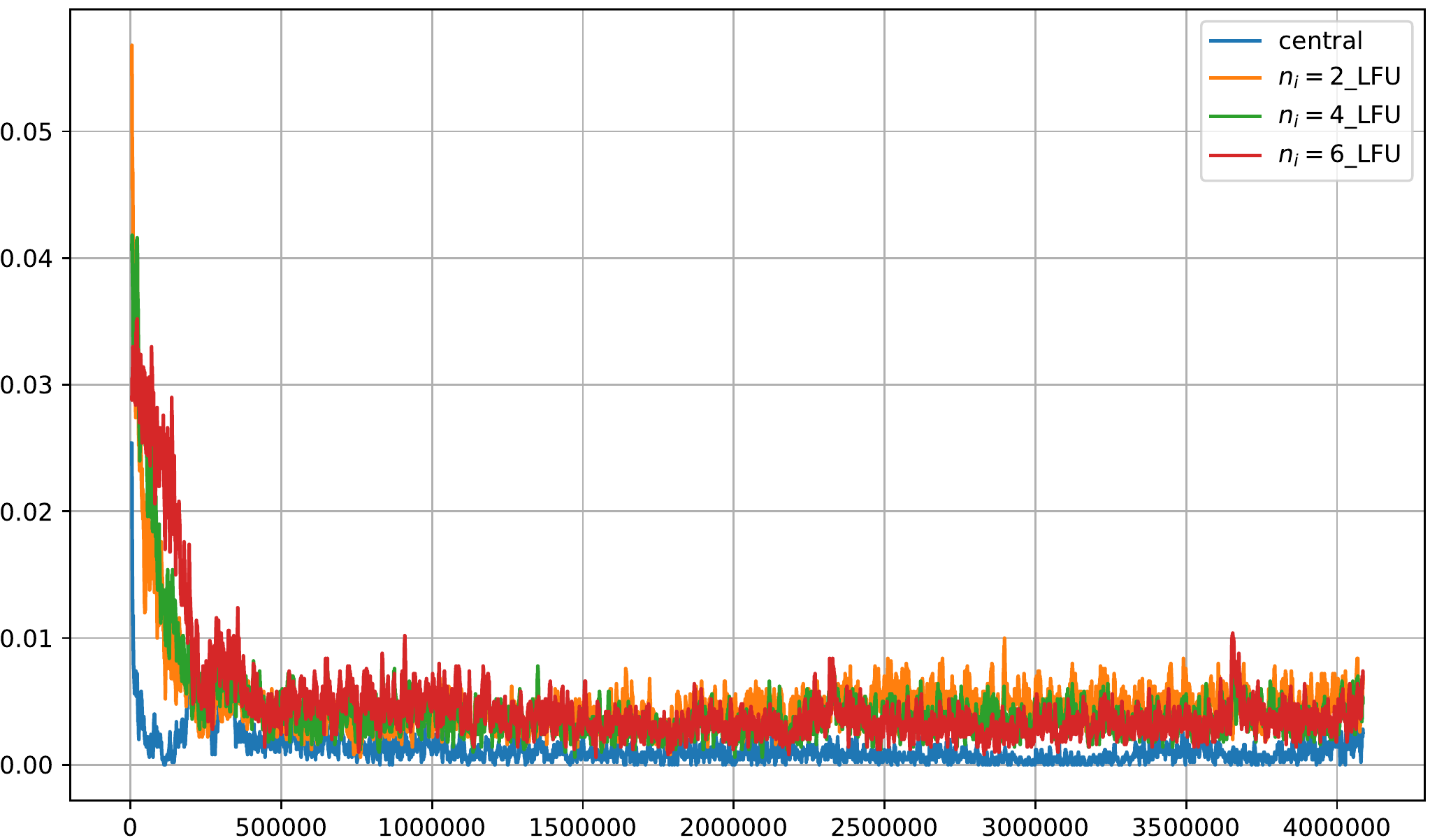}
  \caption{Netflix with LFU Forgetting}
\end{subfigure} 
\caption{The effect of applying forgetting techniques on Recall@10 for Cosine similarity central and distributed for the different $n_i$ values}
\label{fig:dtenrec:recall:forget}
\end{figure}

Next, we report about the effect of applying forgetting techniques on recall and memory consumption. Figure \ref{fig:dtenrec:recall:forget} plots the effect of applying the two different forgetting techniques for the different replication factors. It is important to remind that for the Movielens data set, we do not plot data for the central configuration as it did not complete. Other than that, the forgetting techniques can accomplish acceptable recall compared to the non-forgetting version. 
\clearpage
\begin{figure}[h!]
\begin{subfigure}[t]{0.5\textwidth}
  \includegraphics[width=\linewidth]{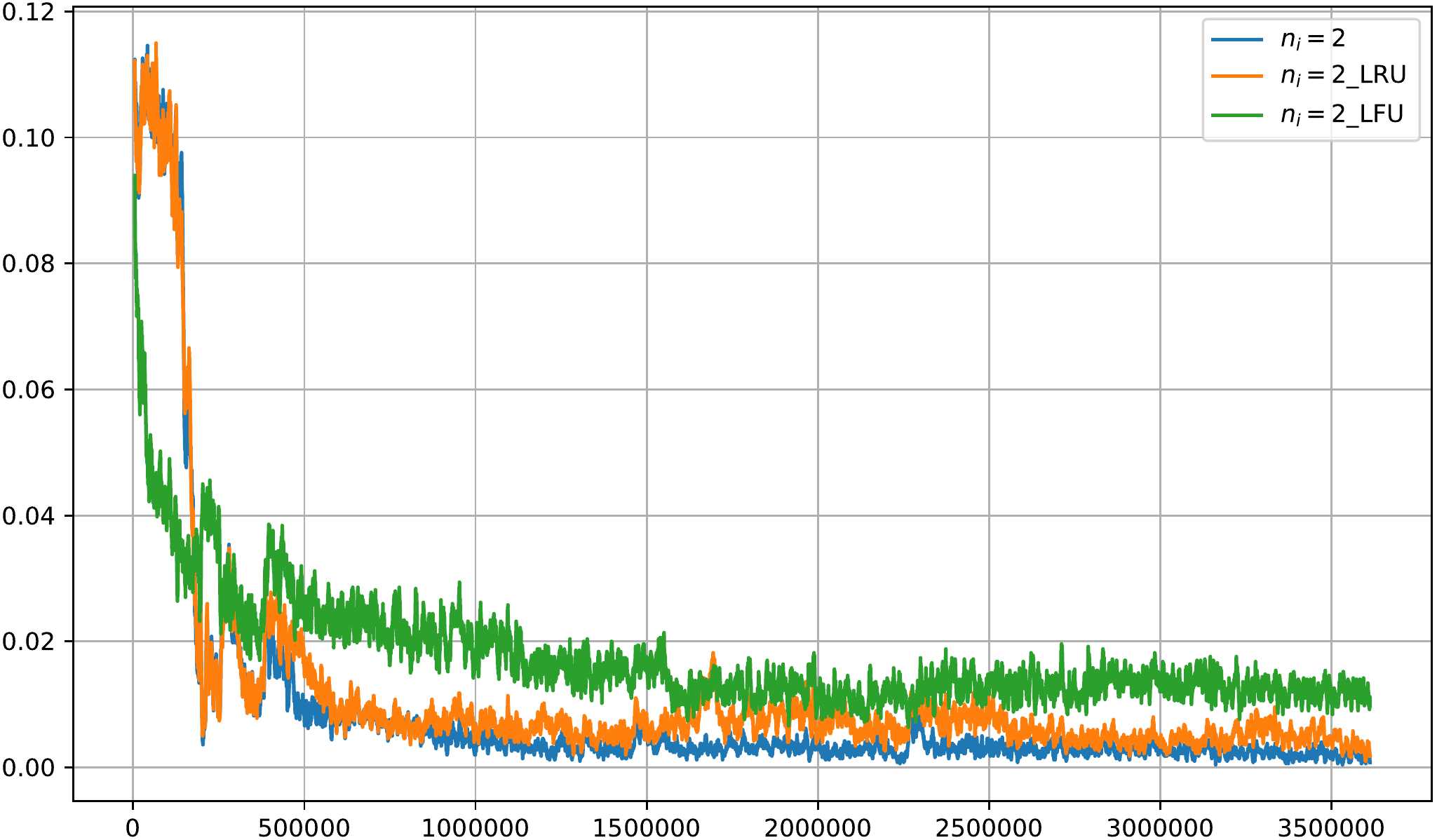}
  \caption{Movielens-25M with $n_{i}=2$}
  \label{TENREC_ML_using_ni_2}
\end{subfigure}
~
\begin{subfigure}[t]{0.5\textwidth}
  \includegraphics[width=\linewidth]{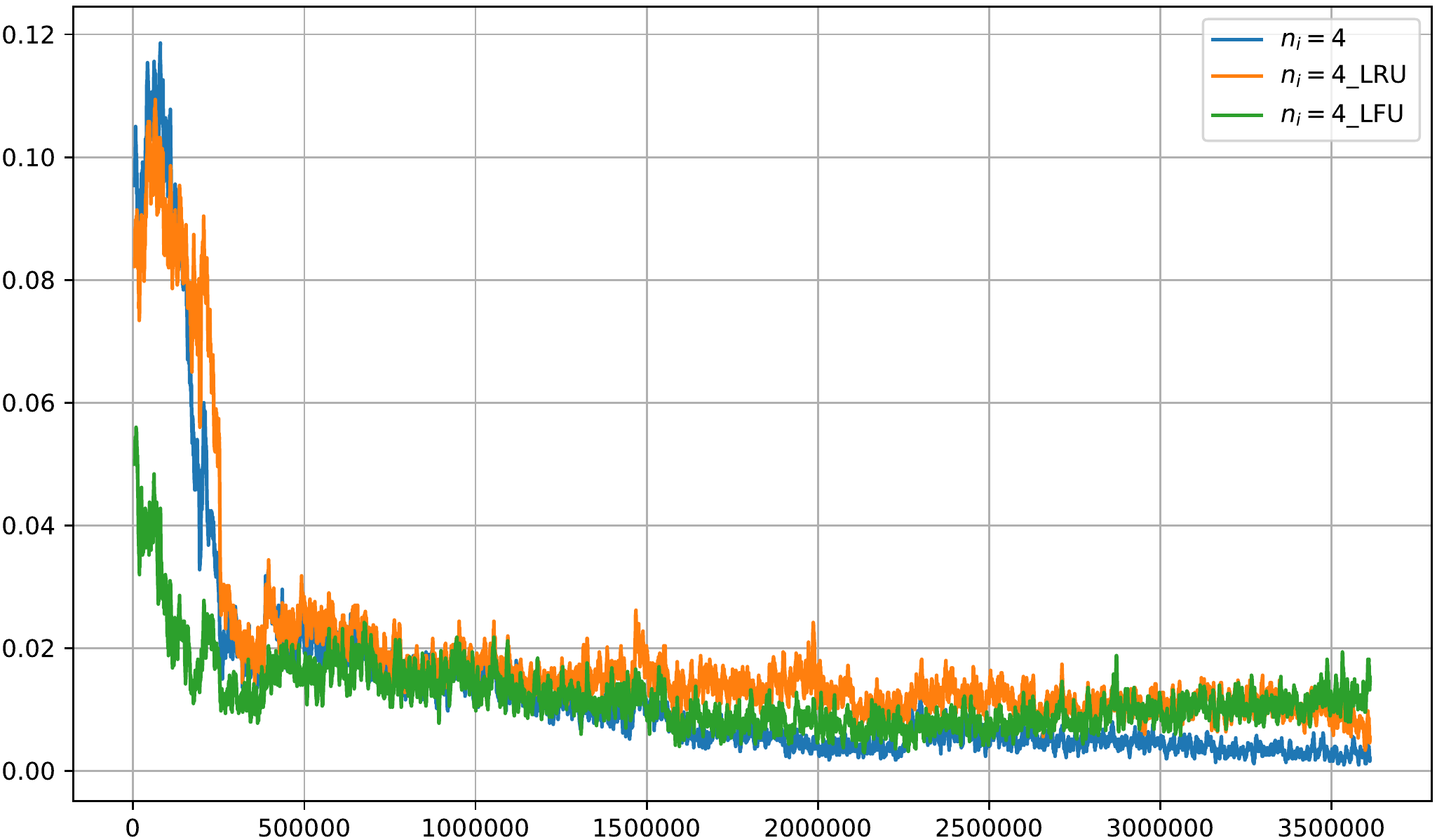}
  \caption{Movielens-25M with $n_{i}=4$}
  \label{TENREC_ML_using_ni_4}
\end{subfigure}
\qquad
\begin{subfigure}[t]{0.5\textwidth}
  \includegraphics[width=\linewidth]{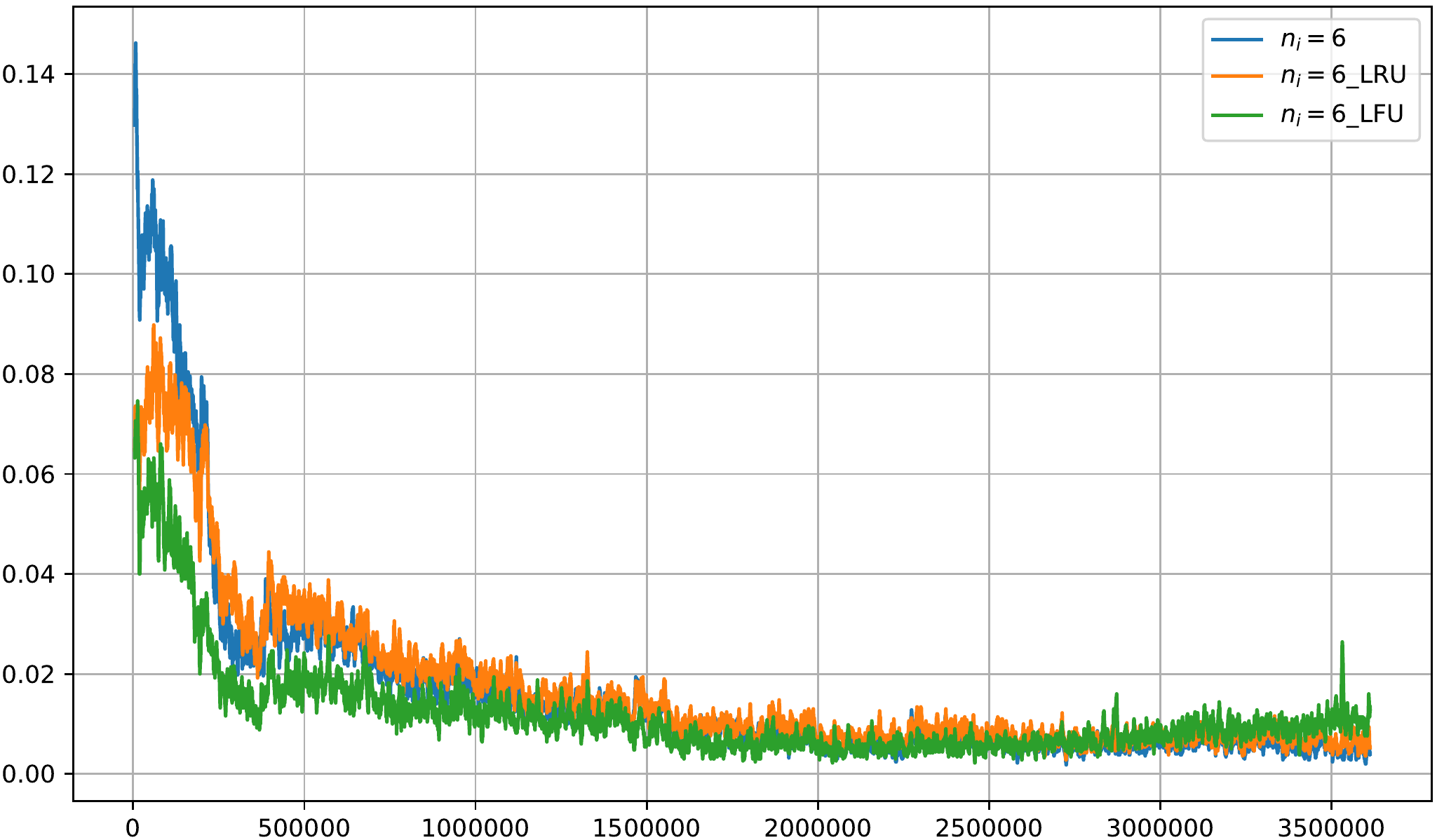}
  \caption{Movielens-25M with $n_{i}=6$}
  \label{TENREC_ML_using_ni_6}
\end{subfigure}~
\begin{subfigure}[t]{0.5\textwidth}
  \includegraphics[width=\linewidth]{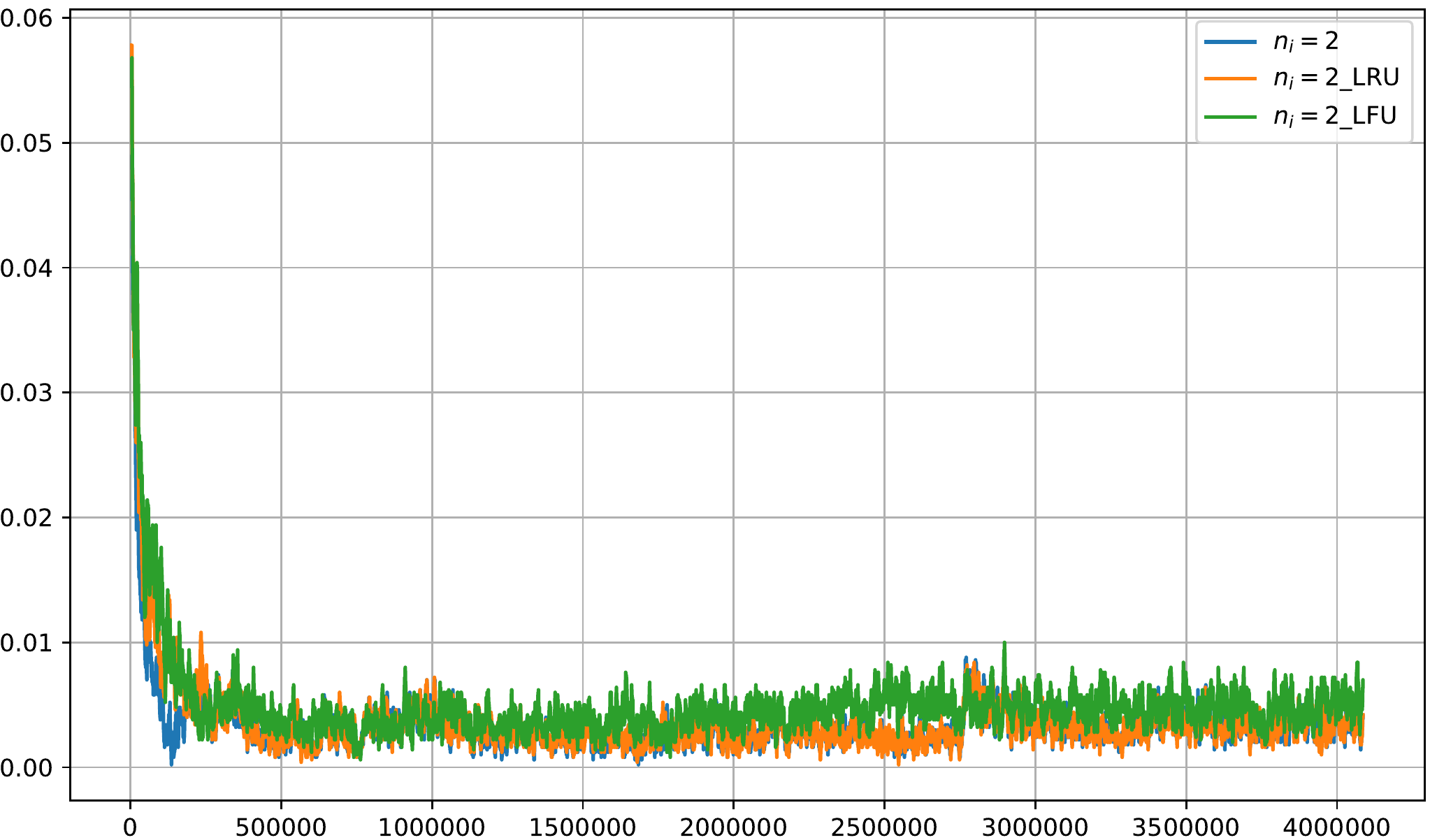}
  \caption{Netflix with $n_{i}=2$}
  \label{TENREC_NF_using_ni_2}
\end{subfigure}
\qquad
\begin{subfigure}[t]{0.5\textwidth}
  \includegraphics[width=\linewidth]{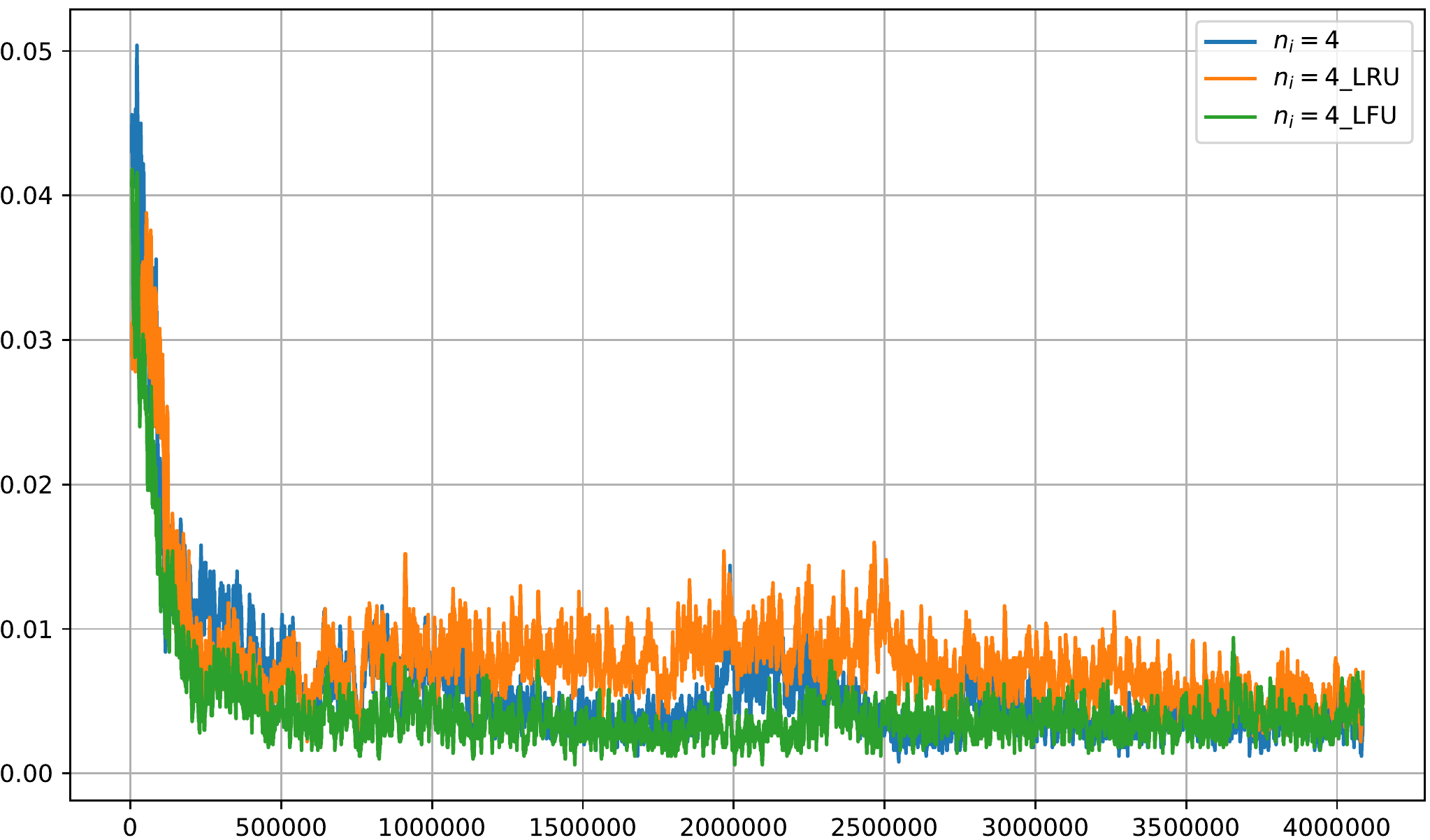}
  \caption{Netflix with $n_{i}=4$}
  \label{TENREC_NF_using_ni_4_e_e}
\end{subfigure}
~ 
\begin{subfigure}[t]{0.5\textwidth}
  \includegraphics[width=\linewidth]{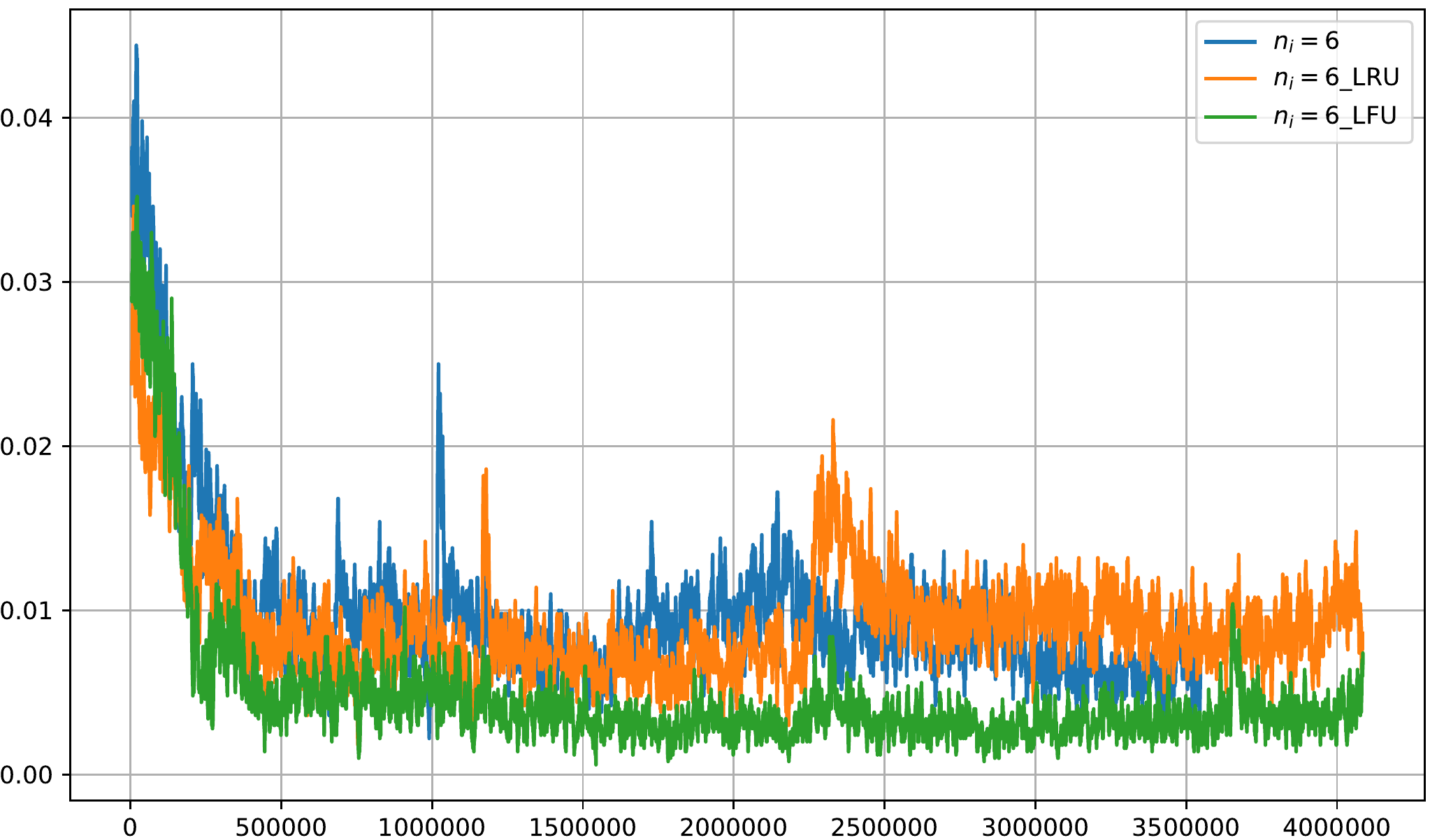}
  \caption{Netflix with $n_{i}=6$}
  \label{TENREC_NF_using_ni_6}
\end{subfigure}
\caption{Comparison of the effect of LFU and LRU on Recall@10 for the distributed Cosine similarity for the different $n_i$ values}
\label{fig:images_TRall}
\end{figure}

Figure~\ref{fig:images_TRall} provides a one-to-one comparison of LFU and LRU for the different replication factor configurations. The impact of the aggressive forgetting represented in LFU led to a considerable loss in the recall like in figures~\ref{TENREC_NF_using_ni_4_e_e} and ~\ref{TENREC_NF_using_ni_6}  although it achieves better results in figures~\ref{TENREC_ML_using_ni_2} and~\ref{TENREC_NF_using_ni_2} for $n_i=2$ for Movielens and Netflix datasets. This can be explained as: increasing the parallelism leads to a smaller share of state for each worker, even with the replication, makes the recall more sensitive to the removal of items from the state by applying LFU.

Overall, splitting and replication along with forgetting techniques have a positive impact on not only the memory consumption but also the recall of the recommender algorithms. However, the magnitude of the gain is different among the algorithms. DISGD benefits more in the improvement of the recall compared to DICS. This is inherent to the nature of the algorithm. It is out of scope for this paper to study why there is difference in recall improvement.

\begin{figure}[h!]
     \begin{subfigure}[t]{0.5\textwidth}
         \includegraphics[width=\textwidth]{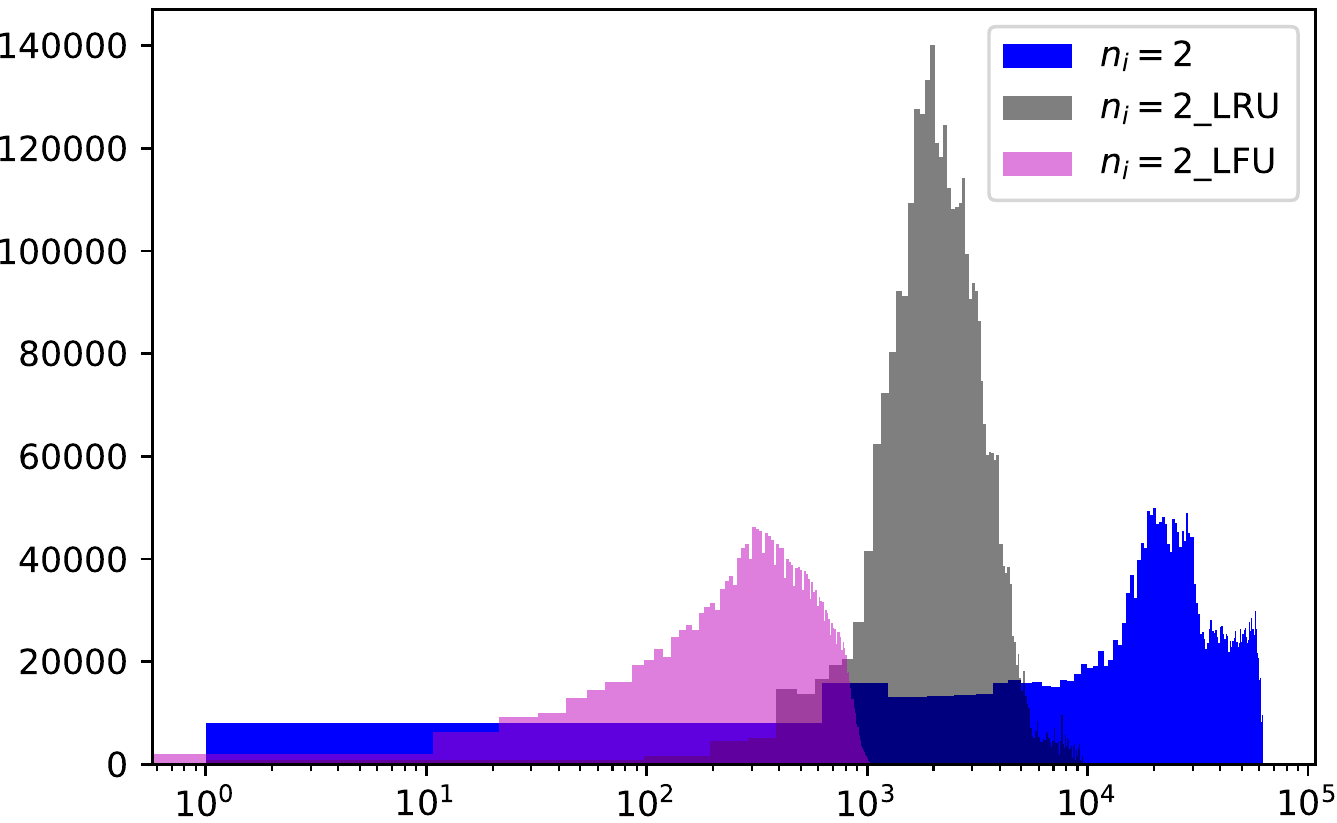}
         \caption{Users' frequency distribution for $n_i=2$}
         \label{NF_ni_2_user}
     \end{subfigure}
     ~
     \begin{subfigure}[t]{0.5\textwidth}
         \includegraphics[width=\textwidth]{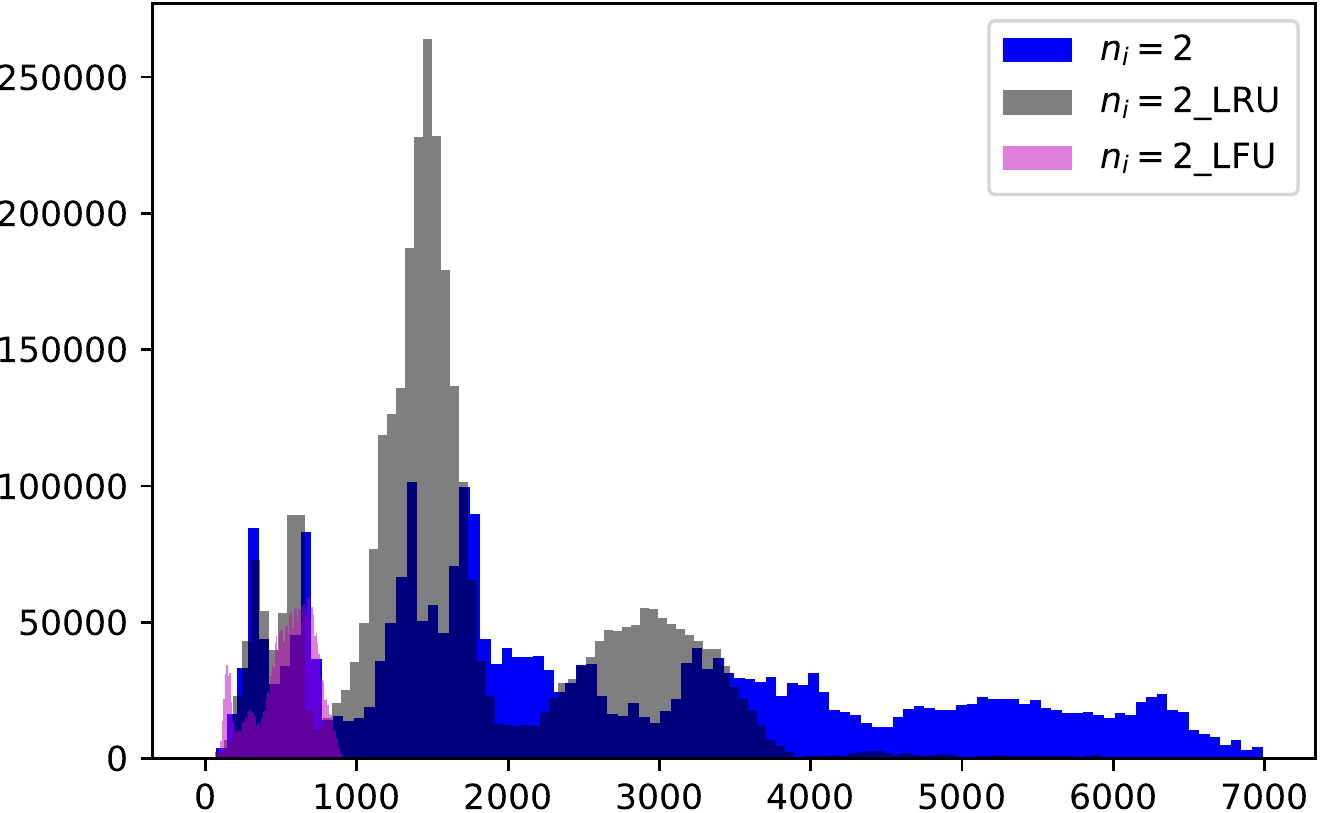}
         \caption{Items' frequency distribution for $n_i=2$}
         \label{NF_ni_2_item}
     \end{subfigure}
     \qquad
    
     \begin{subfigure}[t]{0.5\textwidth}
         \includegraphics[width=\textwidth]{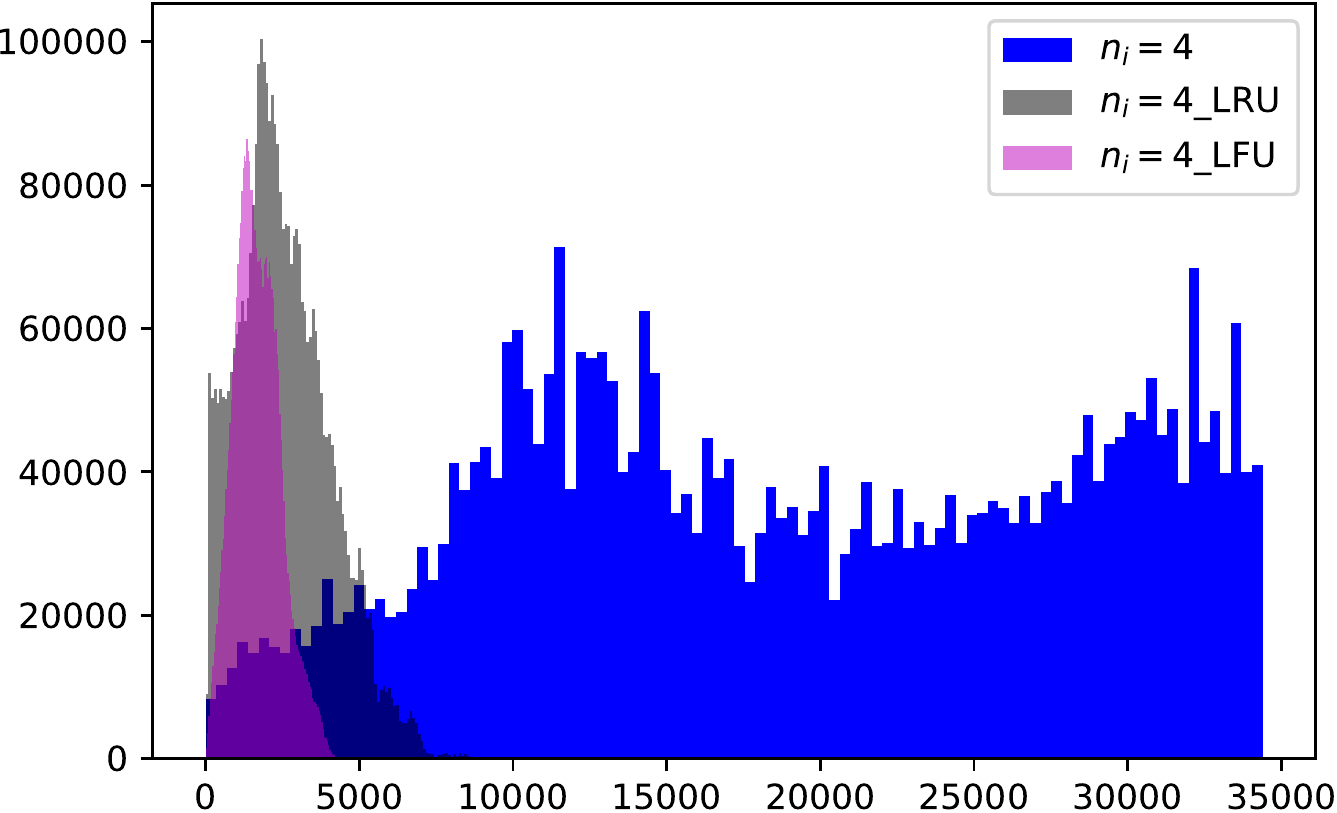}
         \caption{Users' frequency distribution for $n_i=4$}
         \label{NF_ni_4_user}
     \end{subfigure}
     ~
     \begin{subfigure}[t]{0.5\textwidth}
         \includegraphics[width=\textwidth]{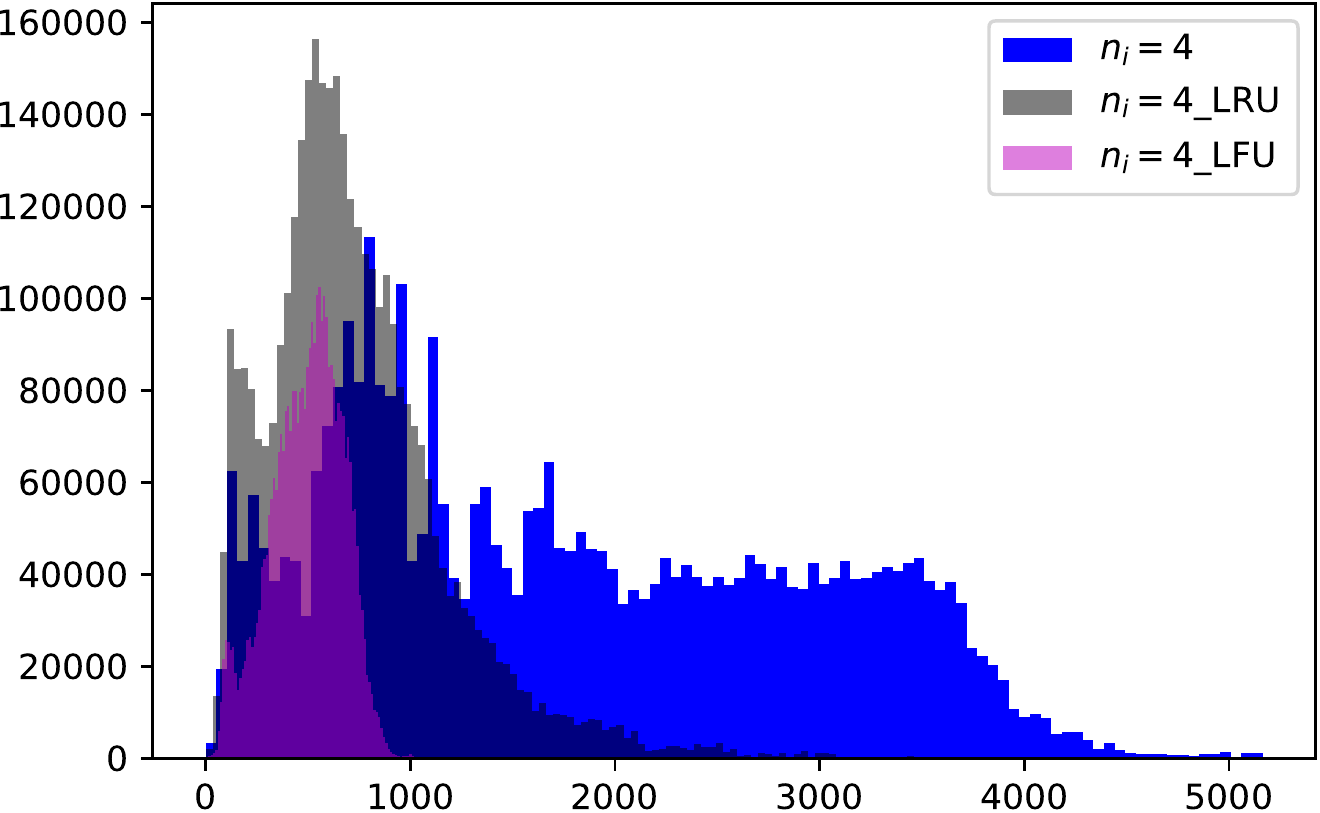}
         \caption{Items' frequency distribution for $n_i=4$}
         \label{NF_ni_4_item}
     \end{subfigure}
    \qquad
     \begin{subfigure}[t]{0.5\textwidth}
         \includegraphics[width=\textwidth]{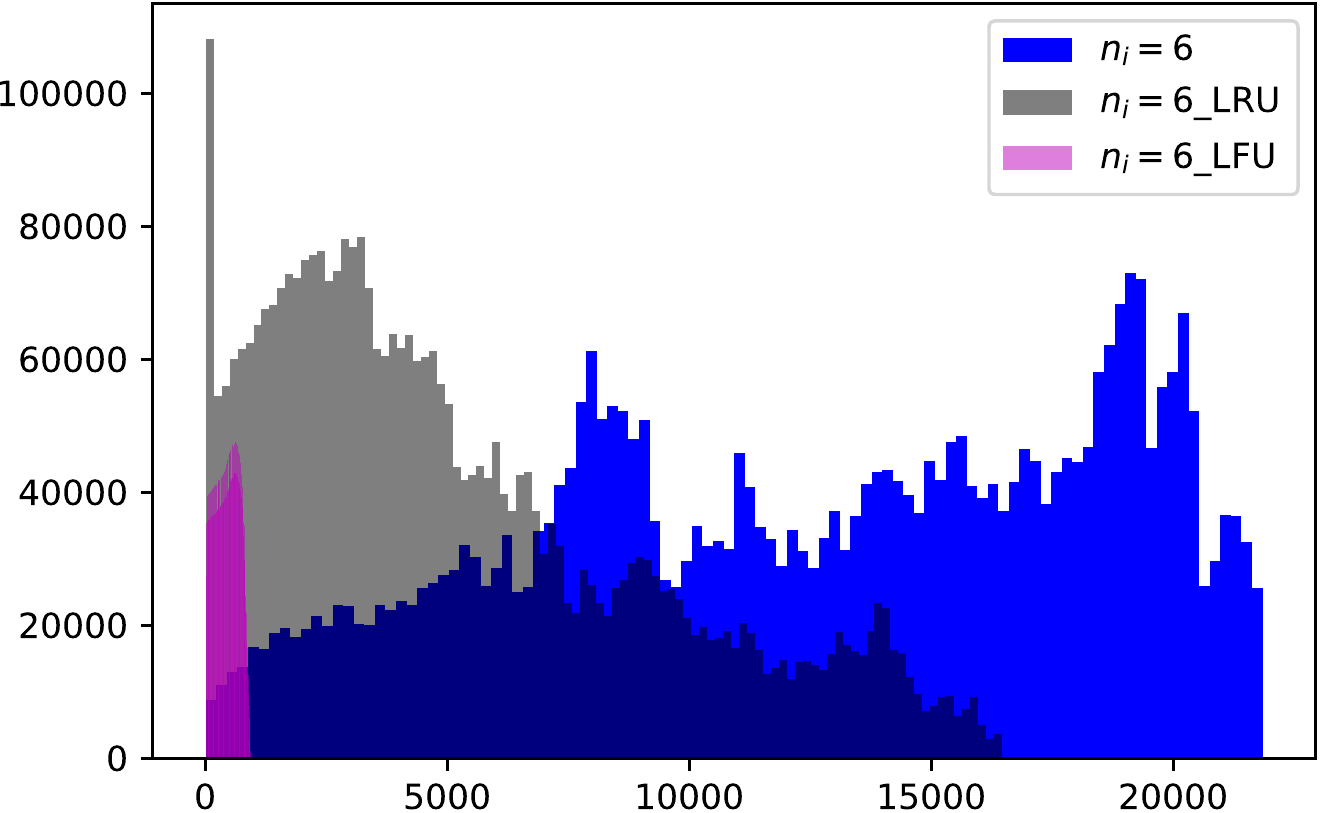}
         \caption{Users' frequency distribution for $n_i=6$}
         \label{NF_ni_6_user}
     \end{subfigure}
     ~
      \begin{subfigure}[t]{0.5\textwidth}
         \includegraphics[width=\textwidth]{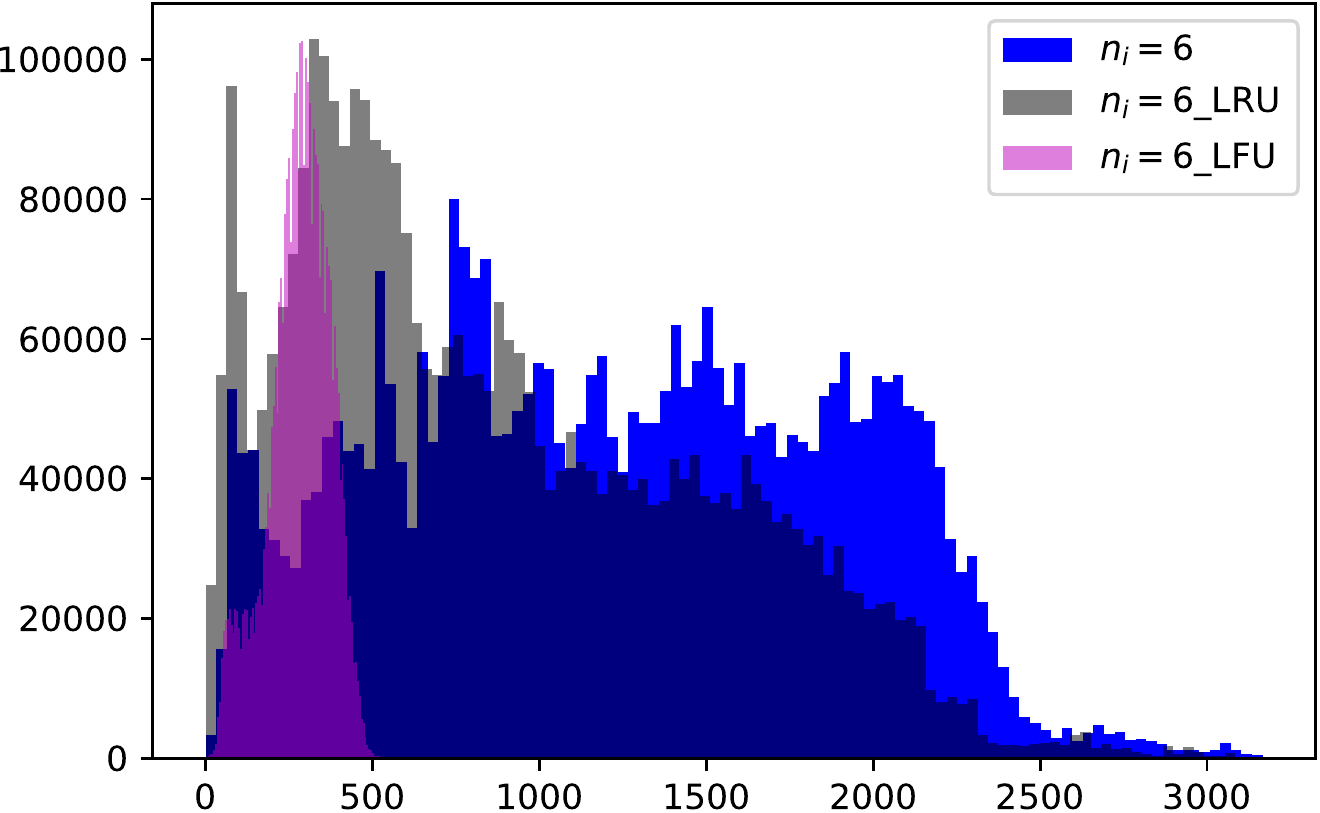}
         \caption{Items' frequency distribution for $n_i=6$}
         \label{NF_ni_6_item}
     \end{subfigure}
     \caption{The effect of applying forgetting techniques on memory distribution for the Netflix data set.}
   \label{Netflix_recall_forgetting_memory_dist3}
\end{figure}

Figure~\ref{Netflix_recall_forgetting_memory_dist3} shows the effect of LRU and LFU on the memory distribution for the Netflix dataset for DISC. For $n_i=2$,  the mean of the users' distribution falls down to $0.071$  and $0.025$ of the $n_i=2$ users' state mean for LRU and LFU respectively. Likewise, the items' state drops to $0.428$ and $0.13$ of i$n_i=2$ items' state mean for LRU and LFU. For replication factor $n_i=4$, the mean of the users' distribution drops to $0.118$  and $0.019$ of the users' state mean for LRU and LFU respectively whereas the items' state drops to $0.39$ and $0.18$ of items' state mean. More gain with $n_i=6$ as the mean of the users' distribution comes down to $0.42$  and $0.034$ of the users' state mean for LRU and LFU respectively and $0.21$ of $n_i=6$ items' state mean. 

\begin{figure}[h!]
    \centering
	\includegraphics[width=\linewidth]{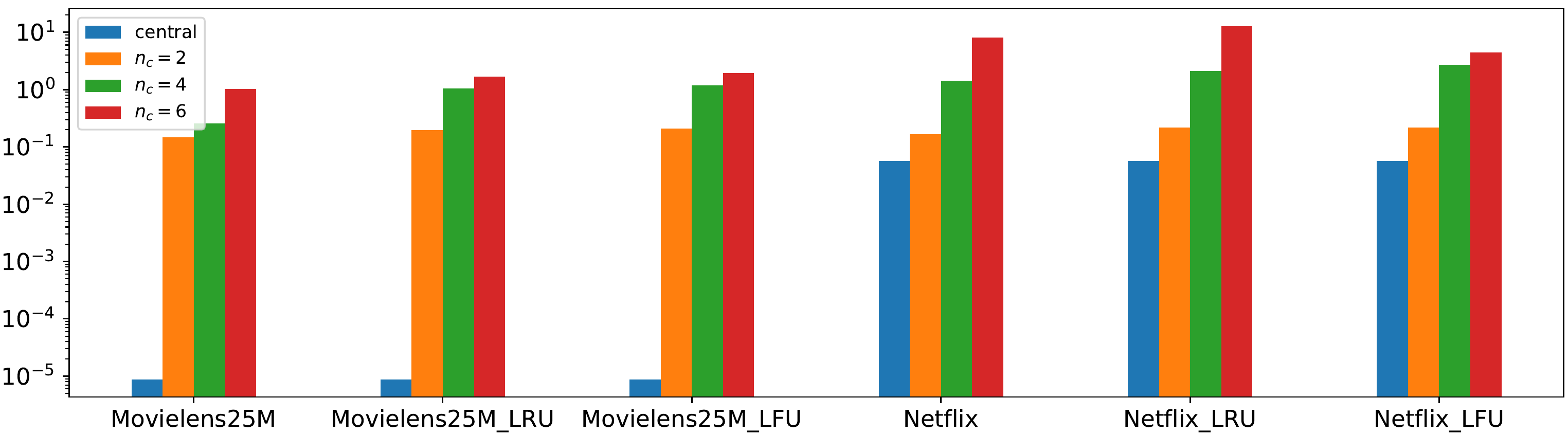}
	\caption{Throughput comparison for DICS with different $n_i$ with and without forgetting technique applied on Movielens and Netflix datasets.}
	\label{Pipeline_throughput2}
\end{figure}

Figure~\ref{Pipeline_throughput2} shows the throughput of centralized incremental Cosine similarity against the DICS under all configurations discussed for $n_i$ with and without applying forgetting techniques. Overall, splitting and replication improves the throughput six orders of magnitude in the case of the Movielens dataset. For the Netflix dataset, there is about one order of magnitude gain. Forgetting techniques still make improvements on the throughput. However, not with the same scale as is achieved with DISGD. This is due to the inherent slowness in the incremental Cosine similarity. The nature of the algorithm stores complex structures in the state, with each item, we store a list of similar items. So, when removing items, we have to iterate and remove relevant items as well. So, the gain of throughput due to splitting is wasted in iterating over the items in memory. The tuning of LFU parameters to forget faster has amplified the loss in throughput due to the frequent invocation of the forgetting procedure.




\section{Conclusion and Future Directions}
\label{sec:conclusion}

In this work, splitting and replication mechanism has been proposed to make streaming recommender systems scalable. The mechanism's capability has been studied with two different streaming algorithms ISGD and incremental Cosine similarity and has been applied on two different datasets Netflix and Movielens25M. To cope with stream unboundedness, forgetting techniques, namely Least Recently Used (LRU) and Least Frequently Used (LFU) have been applied as well. The proposed mechanism has adopted the shared-nothing architecture. It has been implemented using Flink APIs. The proposed mechanism adds an edge to the streaming algorithms to overcome the locking and synchronization problems implied by the need to synchronize model parameters while learning from the stream of data. The scalability of the algorithm has been tested in addition to the quality of the streaming recommender algorithms and the results show that the algorithms operate well in distributed environments. The quality of the distributed recommender systems has been assessed using incremental recall in the context of prequential evaluation. the splitting and replication mechanism is able to scale the streaming recommender systems and the enhancements are not limited to parallelization only but extend to the recommender systems' accuracy in terms of recall. 

Albeit the preliminary results show the power of the mechanism, further improvements can be added in future. During our experiments, we observed skewness of data distribution. The data distribution change might lead to skewness in the load on workers. Load rebalancing techniques already exist in the literature, however, the effect of moving/merging state on the performance of the algorithm is unknown and is an interesting subject for our future work. Additionally, the splitting and replication mechanism can be tested with other streaming recommender algorithms and with different forgetting techniques like sliding window and gradual forgetting. These are also interesting directions for future work.

\section*{Acknowledgment}

The work by Ahmed Awad is funded by the European Regional Development Funds via the Mobilitas Plus programme (grant MOBTT75).

\begin{thebibliography}{10}
\expandafter\ifx\csname url\endcsname\relax
  \def\url#1{\texttt{#1}}\fi
\expandafter\ifx\csname urlprefix\endcsname\relax\def\urlprefix{URL }\fi
\expandafter\ifx\csname href\endcsname\relax
  \def\href#1#2{#2} \def\path#1{#1}\fi

\bibitem{recommenderSystemsBook16}
C.~C. Aggarwal, \href{https://doi.org/10.1007/978-3-319-29659-3}{Recommender
  Systems - The Textbook}, Springer, 2016.
\newblock \href {http://dx.doi.org/10.1007/978-3-319-29659-3}
  {\path{doi:10.1007/978-3-319-29659-3}}.
\newline\urlprefix\url{https://doi.org/10.1007/978-3-319-29659-3}

\bibitem{albertBifetMLDataStreamsBook18}
A.~Bifet, R.~Gavald{\`a}, G.~Holmes, B.~Pfahringer, Machine learning for data
  streams: with practical examples in MOA, MIT press, 2018.

\bibitem{Flink2015}
P.~Carbone, A.~Katsifodimos, S.~Ewen, V.~Markl, S.~Haridi, K.~Tzoumas, Apache
  flink{\texttrademark}: Stream and batch processing in a single engine, {IEEE}
  Data Eng. Bull. 38~(4).

\bibitem{BEAM}
A.~S. Foundation, {Beam}, {https://beam.apache.org/} (2016).

\bibitem{vinagre2014fast}
J.~Vinagre, A.~M. Jorge, J.~Gama, Fast incremental matrix factorization for
  recommendation with positive-only feedback, in: International Conference on
  User Modeling, Adaptation, and Personalization, Springer, 2014, pp. 459--470.

\bibitem{huang2015tencentrec}
Y.~Huang, B.~Cui, W.~Zhang, J.~Jiang, Y.~Xu, Tencentrec: Real-time stream
  recommendation in practice, in: Proceedings of the 2015 ACM SIGMOD
  International Conference on Management of Data, 2015, pp. 227--238.

\bibitem{contentBasedRecommenderSystem19}
P.~Lops, D.~Jannach, C.~Musto, T.~Bogers, M.~Koolen,
  \href{https://doi.org/10.1007/s11257-019-09231-w}{Trends in content-based
  recommendation - preface to the special issue on recommender systems based on
  rich item descriptions}, User Model. User Adapt. Interact. 29~(2) (2019)
  239--249.
\newblock \href {http://dx.doi.org/10.1007/s11257-019-09231-w}
  {\path{doi:10.1007/s11257-019-09231-w}}.
\newline\urlprefix\url{https://doi.org/10.1007/s11257-019-09231-w}

\bibitem{collaborativeFilteringRecommenderSystemDS13}
J.~Liu, C.~Wu, W.~Liu,
  \href{https://doi.org/10.1016/j.dss.2013.04.002}{Bayesian probabilistic
  matrix factorization with social relations and item contents for
  recommendation}, Decis. Support Syst. 55~(3) (2013) 838--850.
\newblock \href {http://dx.doi.org/10.1016/j.dss.2013.04.002}
  {\path{doi:10.1016/j.dss.2013.04.002}}.
\newline\urlprefix\url{https://doi.org/10.1016/j.dss.2013.04.002}

\bibitem{knowledgeBasedRecommenderSystem18}
J.~K. Tarus, Z.~Niu, G.~Mustafa,
  \href{https://doi.org/10.1007/s10462-017-9539-5}{Knowledge-based
  recommendation: a review of ontology-based recommender systems for
  e-learning}, Artif. Intell. Rev. 50~(1) (2018) 21--48.
\newblock \href {http://dx.doi.org/10.1007/s10462-017-9539-5}
  {\path{doi:10.1007/s10462-017-9539-5}}.
\newline\urlprefix\url{https://doi.org/10.1007/s10462-017-9539-5}

\bibitem{ALS08}
Y.~Zhou, D.~M. Wilkinson, R.~Schreiber, R.~Pan,
  \href{https://doi.org/10.1007/978-3-540-68880-8\_32}{Large-scale parallel
  collaborative filtering for the netflix prize}, in: R.~Fleischer, J.~Xu
  (Eds.), Algorithmic Aspects in Information and Management, 4th International
  Conference, {AAIM} 2008, Shanghai, China, June 23-25, 2008. Proceedings, Vol.
  5034 of Lecture Notes in Computer Science, Springer, 2008, pp. 337--348.
\newblock \href {http://dx.doi.org/10.1007/978-3-540-68880-8\_32}
  {\path{doi:10.1007/978-3-540-68880-8\_32}}.
\newline\urlprefix\url{https://doi.org/10.1007/978-3-540-68880-8\_32}

\bibitem{matrixFactorization09}
Y.~Koren, R.~M. Bell, C.~Volinsky,
  \href{https://doi.org/10.1109/MC.2009.263}{Matrix factorization techniques
  for recommender systems}, Computer 42~(8) (2009) 30--37.
\newblock \href {http://dx.doi.org/10.1109/MC.2009.263}
  {\path{doi:10.1109/MC.2009.263}}.
\newline\urlprefix\url{https://doi.org/10.1109/MC.2009.263}

\bibitem{le1998lecun}
Y.~LeCun, L.~Bottou, G.~B. Orr, K.~M{\"{u}}ller,
  \href{https://doi.org/10.1007/978-3-642-35289-8\_3}{Efficient backprop}, in:
  G.~Montavon, G.~B. Orr, K.~M{\"{u}}ller (Eds.), Neural Networks: Tricks of
  the Trade - Second Edition, Vol. 7700 of Lecture Notes in Computer Science,
  Springer, 2012, pp. 9--48.
\newblock \href {http://dx.doi.org/10.1007/978-3-642-35289-8\_3}
  {\path{doi:10.1007/978-3-642-35289-8\_3}}.
\newline\urlprefix\url{https://doi.org/10.1007/978-3-642-35289-8\_3}

\bibitem{gemulla2011large}
R.~Gemulla, E.~Nijkamp, P.~J. Haas, Y.~Sismanis, Large-scale matrix
  factorization with distributed stochastic gradient descent, in: Proceedings
  of the 17th ACM SIGKDD international conference on Knowledge discovery and
  data mining, 2011, pp. 69--77.

\bibitem{schelter2014factorbird}
S.~Schelter, V.~Satuluri, R.~Zadeh, Factorbird-a parameter server approach to
  distributed matrix factorization, arXiv preprint arXiv:1411.0602.

\bibitem{li2014scaling}
M.~Li, D.~G. Andersen, J.~W. Park, A.~J. Smola, A.~Ahmed, V.~Josifovski,
  J.~Long, E.~J. Shekita, B.-Y. Su, Scaling distributed machine learning with
  the parameter server, in: 11th $\{$USENIX$\}$ Symposium on Operating Systems
  Design and Implementation ($\{$OSDI$\}$ 14), 2014, pp. 583--598.

\bibitem{recht2011hogwild}
B.~Recht, C.~Re, S.~Wright, F.~Niu, Hogwild: A lock-free approach to
  parallelizing stochastic gradient descent, in: Advances in neural information
  processing systems, 2011, pp. 693--701.

\bibitem{offlineInitializationOnlineUpdate2010}
D.~Agarwal, B.~Chen, P.~Elango,
  \href{https://doi.org/10.1145/1835804.1835894}{Fast online learning through
  offline initialization for time-sensitive recommendation}, in: B.~Rao,
  B.~Krishnapuram, A.~Tomkins, Q.~Yang (Eds.), Proceedings of the 16th {ACM}
  {SIGKDD} International Conference on Knowledge Discovery and Data Mining,
  Washington, DC, USA, July 25-28, 2010, {ACM}, 2010, pp. 703--712.
\newblock \href {http://dx.doi.org/10.1145/1835804.1835894}
  {\path{doi:10.1145/1835804.1835894}}.
\newline\urlprefix\url{https://doi.org/10.1145/1835804.1835894}

\bibitem{papagelis2005incremental}
M.~Papagelis, I.~Rousidis, D.~Plexousakis, E.~Theoharopoulos, Incremental
  collaborative filtering for highly-scalable recommendation algorithms, in:
  International Symposium on Methodologies for Intelligent Systems, Springer,
  2005, pp. 553--561.

\bibitem{miranda2008incremental}
C.~Miranda, A.~M. Jorge, Incremental collaborative filtering for binary
  ratings, in: 2008 IEEE/WIC/ACM International Conference on Web Intelligence
  and Intelligent Agent Technology, Vol.~1, IEEE, 2008, pp. 389--392.

\bibitem{streamRec2011}
B.~Chandramouli, J.~J. Levandoski, A.~Eldawy, M.~F. Mokbel,
  \href{https://doi.org/10.1145/1989323.1989465}{Streamrec: a real-time
  recommender system}, in: T.~K. Sellis, R.~J. Miller, A.~Kementsietsidis,
  Y.~Velegrakis (Eds.), Proceedings of the {ACM} {SIGMOD} International
  Conference on Management of Data, {SIGMOD} 2011, Athens, Greece, June 12-16,
  2011, {ACM}, 2011, pp. 1243--1246.
\newblock \href {http://dx.doi.org/10.1145/1989323.1989465}
  {\path{doi:10.1145/1989323.1989465}}.
\newline\urlprefix\url{https://doi.org/10.1145/1989323.1989465}

\bibitem{zaouali2018distributed}
K.~Zaouali, M.~R. Haddad, H.~B. Zghal, Distributed collaborative filtering for
  batch and stream processing-based recommendations, in: OTM Confederated
  International Conferences" On the Move to Meaningful Internet Systems",
  Springer, 2018, pp. 243--260.

\bibitem{he2016fast}
X.~He, H.~Zhang, M.-Y. Kan, T.-S. Chua, Fast matrix factorization for online
  recommendation with implicit feedback, in: Proceedings of the 39th
  International ACM SIGIR conference on Research and Development in Information
  Retrieval, 2016, pp. 549--558.

\bibitem{ali2011parallel}
M.~Ali, C.~C. Johnson, A.~K. Tang, Parallel collaborative filtering for
  streaming data, University of Texas Austin, Tech. Rep (2011) 5--7.

\bibitem{Storm2015}
S.~T. Allen, M.~Jankowski, P.~Pathirana, Storm Applied: Strategies for
  Real-time Event Processing, 1st Edition, Manning Publications Co., Greenwich,
  CT, USA, 2015.

\bibitem{sparkStructuredStreaming18}
M.~Armbrust, T.~Das, J.~Torres, B.~Yavuz, S.~Zhu, R.~Xin, A.~Ghodsi, I.~Stoica,
  M.~Zaharia, \href{https://doi.org/10.1145/3183713.3190664}{Structured
  streaming: {A} declarative {API} for real-time applications in apache spark},
  in: G.~Das, C.~M. Jermaine, P.~A. Bernstein (Eds.), Proceedings of the 2018
  International Conference on Management of Data, {SIGMOD} Conference 2018,
  Houston, TX, USA, June 10-15, 2018, {ACM}, 2018, pp. 601--613.
\newblock \href {http://dx.doi.org/10.1145/3183713.3190664}
  {\path{doi:10.1145/3183713.3190664}}.
\newline\urlprefix\url{https://doi.org/10.1145/3183713.3190664}

\bibitem{stateManagementFlink17}
P.~Carbone, S.~Ewen, G.~F{\'{o}}ra, S.~Haridi, S.~Richter, K.~Tzoumas,
  \href{http://www.vldb.org/pvldb/vol10/p1718-carbone.pdf}{State management in
  apache flink{\textregistered}: Consistent stateful distributed stream
  processing}, Proc. {VLDB} Endow. 10~(12) (2017) 1718--1729.
\newblock \href {http://dx.doi.org/10.14778/3137765.3137777}
  {\path{doi:10.14778/3137765.3137777}}.
\newline\urlprefix\url{http://www.vldb.org/pvldb/vol10/p1718-carbone.pdf}

\bibitem{karimov2018benchmarking}
J.~Karimov, T.~Rabl, A.~Katsifodimos, R.~Samarev, H.~Heiskanen, V.~Markl,
  Benchmarking distributed stream data processing systems, in: 2018 IEEE 34th
  International Conference on Data Engineering (ICDE), IEEE, 2018, pp.
  1507--1518.

\bibitem{ShahverdiAS19}
E.~Shahverdi, A.~Awad, S.~Sakr,
  \href{https://doi.org/10.1109/ICDEW.2019.00-35}{Big stream processing
  systems: An experimental evaluation}, in: 35th {IEEE} International
  Conference on Data Engineering Workshops, {ICDE} Workshops 2019, Macao,
  China, April 8-12, 2019, {IEEE}, 2019, pp. 53--60.
\newblock \href {http://dx.doi.org/10.1109/ICDEW.2019.00-35}
  {\path{doi:10.1109/ICDEW.2019.00-35}}.
\newline\urlprefix\url{https://doi.org/10.1109/ICDEW.2019.00-35}

\bibitem{bifet2015efficient}
A.~Bifet, G.~de~Francisci~Morales, J.~Read, G.~Holmes, B.~Pfahringer, Efficient
  online evaluation of big data stream classifiers, in: Proceedings of the 21th
  ACM SIGKDD international conference on knowledge discovery and data mining,
  2015, pp. 59--68.

\bibitem{coldStartup05}
G.~Adomavicius, A.~Tuzhilin, \href{https://doi.org/10.1109/TKDE.2005.99}{Toward
  the next generation of recommender systems: {A} survey of the
  state-of-the-art and possible extensions}, {IEEE} Trans. Knowl. Data Eng.
  17~(6) (2005) 734--749.
\newblock \href {http://dx.doi.org/10.1109/TKDE.2005.99}
  {\path{doi:10.1109/TKDE.2005.99}}.
\newline\urlprefix\url{https://doi.org/10.1109/TKDE.2005.99}

\bibitem{coldStartup2012}
J.~Bobadilla, F.~Ortega, A.~Hernando, J.~Bernal,
  \href{https://www.sciencedirect.com/science/article/pii/S0950705111001882}{A
  collaborative filtering approach to mitigate the new user cold start
  problem}, Knowledge-Based Systems 26 (2012) 225--238.
\newblock \href
  {http://dx.doi.org/https://doi.org/10.1016/j.knosys.2011.07.021}
  {\path{doi:https://doi.org/10.1016/j.knosys.2011.07.021}}.
\newline\urlprefix\url{https://www.sciencedirect.com/science/article/pii/S0950705111001882}

\bibitem{cacheManagement2016}
S.~Kumar, P.~K. Singh, An overview of modern cache memory and performance
  analysis of replacement policies, in: 2016 IEEE International Conference on
  Engineering and Technology (ICETECH), 2016, pp. 210--214.
\newblock \href {http://dx.doi.org/10.1109/ICETECH.2016.7569243}
  {\path{doi:10.1109/ICETECH.2016.7569243}}.

\bibitem{RSBenchmark2012}
A.~Said, D.~Tikk, K.~Stumpf, Y.~Shi, M.~A. Larson, P.~Cremonesi, Recommender
  systems evaluation: A 3d benchmark., in: RUE@ RecSys, 2012, pp. 21--23.

\bibitem{RSBenchmark2020}
D.~Jannach, C.~Bauer, \href{https://doi.org/10.1609/aimag.v41i4.5312}{Escaping
  the mcnamara fallacy: Towards more impactful recommender systems research},
  {AI} Mag. 41~(4) (2020) 79--95.
\newblock \href {http://dx.doi.org/10.1609/aimag.v41i4.5312}
  {\path{doi:10.1609/aimag.v41i4.5312}}.
\newline\urlprefix\url{https://doi.org/10.1609/aimag.v41i4.5312}

\end{thebibliography}

\end{document}